\documentclass[12pt,a4paper]{article}
\usepackage{jheppub}

\usepackage{enumerate}
\usepackage{graphicx}
\usepackage{caption} 
\usepackage{subcaption} 

\usepackage{bm}
\usepackage{amssymb}
\usepackage{amsmath}
\usepackage{cancel}
\usepackage{epigraph}

\newcommand{\captionfonts}{\small}

\makeatletter  % Allow the use of @ in command names
\long\def\@makecaption#1#2{%
  \vskip\abovecaptionskip
  \sbox\@tempboxa{{\captionfonts #1: #2}}%
 \ifdim \wd\@tempboxa >\hsize
    {\captionfonts #1: #2\par}
  \else
    \hbox to\hsize{\hfil\box\@tempboxa\hfil}%
  \fi
  \vskip\belowcaptionskip}
\makeatother   % Cancel the effect of \makeatletter

\usepackage{fancyhdr}
\usepackage{datetime}

\usepackage{mciteplus}

\makeatletter
\renewcommand\section{\@startsection {section}{1}{\z@}%
                                   {-3.5ex \@plus -1ex \@minus -.2ex}%nn
                                   {2.3ex \@plus.2ex}%
                                   {\normalfont\Large\bfseries}}
\renewcommand\subsection{\@startsection{subsection}{2}{\z@}%
                                     {-3.25ex\@plus -1ex \@minus -.2ex}%
                                     {1.5ex \@plus .2ex}%
                                     {\normalfont\bfseries}}
\renewcommand\subsubsection{\@startsection{subsubsection}{3}{\z@}%
                                     {-2.5ex\@plus -1ex \@minus -.2ex}%
                                     {1.25ex \@plus .2ex}%
                                     {\normalfont\textit}}
\makeatother
\newcommand{\ket}[1]{|#1 \rangle}
\def\mathbi#1{\textbf{\em #1}}
\def\tight#1{\! #1 \!}

\def\({\left(}
\def\){\right)}
\def\[{\left[}
\def\]{\right]}

\def\naive{na\"ive}
\def\ads#1{AdS_{#1}}
\def\sltwo{SL(2,\IR)}
\def\sutwo{SU(2)}
\def\uone{U(1)}
\def\slsup{{\it sl}}
\def\susup{{\it su}}
\def\mbar{\bar m}

\def\ie{{i.e.}}
\def\eg{{e.g.}}
\def\cf{{c.f.}}

\def\ext{{\rm ext}}
\def\eff{{\rm eff}}

\def\lstr{\ell_{\textit{str}}}
\def\gstr{g_{\textit s}}

\def\bh{{\sst\rm BH}}
\def\btz{{\sst \rm BTZ}}

\def\lads{\ell_{\sst\rm AdS}}

\def\suplabel{m}
\def\supera{{\sst (\suplabel)}}
\def\eL{e_{\sst L}}
\def\eR{e_{\sst R}}

\def\nfive{{n_5}}
\def\nfivetil{{\tilde n_5}}
\def\nfivehat{{\hat n_5}}
\def\none{{n_1}}

\def\xx{{\mathbi  x}}

\def\avec{{\mathfrak a}}
\def\bvec{{\mathfrak b}}

\def\bX{{\mathbi X}}
\def\bY{{\mathbi Y}}
\def\x1x2{$x^1$-$x^2$}

\def\ytil{{\tilde y}}
\def\Ry{R_y}
\def\Rytil{R_\ytil}

\def\jdual{ \jmath^{\sst\vee} }
\def\Phihat{\widehat\Phi}
\def\Psihat{\widehat\Psi}
\def\zhat{\hat z}
\def\jhat{{\mathfrak j}}
\def\shat{{\mathfrak s}}

\def\sst{\scriptscriptstyle}
\def\half{\frac12}
\def\coeff#1#2{{\textstyle \frac{#1}{#2}}}

\def\tr{{\rm Tr}}
\def\One{{\hbox{ 1\kern-.8mm l}}}

\def\Dbar{{\bar D}}
\def\ch{{\rm ch}}
\def\sh{{\rm sh}}

\def\barray{\begin{array}}
\def\earray{\end{array}}
\def\be{\begin{equation}}
\def\ee{\end{equation}}
\def\bea{\begin{eqnarray}}
\def\eea{\end{eqnarray}}
\def\bal{\begin{align}}
\def\eal{\end{align}}
\def\nn{\nonumber}

\def\MM{{\mathcal{M}}}

\def\IN{\mathbb{N}}

\def\IR{\mathbb{R}}
\def\IS{\mathbb{S}}
\def\IT{\mathbb{T}}
\def\IZ{\mathbb{Z}}
\def\cA{{\cal A}}
\def\cB{{\cal B}}
\def\cC{{\cal C}}
\def\cD{{\cal D}}
\def\cF{{\cal F}}
\def\cG{{\cal G}}
\def\cI{{\cal I}}
\def\cJ{{\cal J}}
\def\cK{{\cal K}}
\def\cL{{\cal L}}
\def\cM{{\cal M}}
\def\cN{{\cal N}}
\def\cO{{\cal O}}
\def\cP{{\cal P}}
\def\cQ{{\cal Q}}
\def\cR{{\cal R}}
\def\cS{{\cal S}}
\def\cT{{\cal T}}
\def\cV{{\cal V}}
\def\cW{{\cal W}}
\def\cX{{\cal X}}

\definecolor{cardinal}{rgb}{0.6,0,0}
\definecolor{darkgreen}{rgb}{0,0.4,0}
\definecolor{golden}{rgb}{0.92, 0.7, 0}
\definecolor{midnight}{rgb}{0, 0, 0.5}
\definecolor{darkblue}{rgb}{0, 0, 0.7}

\newcommand{\Blue}{\color{blue}}

\begin{document}

\numberwithin{equation}{section}

%%%%%%%%%%%%%%%%%%%%%%%%%%%%%%%%%%%%%%%%%%%%%%%%%%%%%%%%%%%%%%%%%%%%%%%%%%%%%
%                       DEFINITIONS

\mathchardef\mhyphen="2D

%%%%%%%%%%%%%%%%%%%%%%%%%%%%%%%%%%%%%%%%%%%%%%%%%%%%%%%%%%%%%%%%%%%%%%%%%%%%%
%                        Commands

%\newcommand{\be}{\begin{equation}}
%\newcommand{\ee}{\end{equation}}
%\newcommand{\bea}{\begin{eqnarray}\displaystyle}
%\newcommand{\eea}{\end{eqnarray}}
%\newcommand{\nn}{\nonumber}

%%%%%%%%%%%%%%%%%%%%%%%%%%%%%%%%%%%%%%%%%%%%%%%%%%%%%%%%%%%%%%%%%%%%%%%%%%%%%
%              Calligraphic & Blackboard letters etc

\def\cA{{\cal A}} \def\cB{{\cal B}} \def\cC{{\cal C}}
\def\cD{{\cal D}} \def\cE{{\cal E}} \def\cF{{\cal F}}
\def\cG{{\cal G}} \def\cH{{\cal H}} \def\cI{{\cal I}}
\def\cJ{{\cal J}} \def\cK{{\cal K}} \def\cL{{\cal L}}
\def\cM{{\cal M}} \def\cN{{\cal N}} \def\cO{{\cal O}}
\def\cP{{\cal P}} \def\cQ{{\cal Q}} \def\cR{{\cal R}}
\def\cS{{\cal S}} \def\cT{{\cal T}} \def\cU{{\cal U}}
\def\cV{{\cal V}} \def\cW{{\cal W}} \def\cX{{\cal X}}
\def\cY{{\cal Y}} \def\cZ{{\cal Z}}

\def\mC{\mathbb{C}} \def\mP{\mathbb{P}}
\def\mR{\mathbb{R}} \def\mZ{\mathbb{Z}}
\def\mT{\mathbb{T}} \def\mN{\mathbb{N}}
\def\mH{\mathbb{H}} \def\mX{\mathbb{X}}

\def\CP{\mathbb{CP}}
\def\RP{\mathbb{RP}}
\def\Z{\mathbb{Z}}
\def\N{\mathbb{N}}
\def\H{\mathbb{H}}

\def\one{{\hbox{\kern+.5mm 1\kern-.8mm l}}}
\def\zero{{\hbox{0\kern-1.5mm 0}}}

%%%%%%%%%%%%%%%%%%%%%%%%%%%%%%%%%%%%%%%%%%%%%%%%%%%%%%%%%%%%%%%%%%%%%%%%%%%%%
%                                 QM

%\newcommand{\bra}[1]{{\langle {#1} |\,}}
%\newcommand{\ket}[1]{{\,| {#1} \rangle}}
\newcommand{\braket}[2]{\ensuremath{\langle #1 | #2 \rangle}}
\newcommand{\Braket}[2]{\ensuremath{\langle\, #1 \,|\, #2 \,\rangle}}
\newcommand{\norm}[1]{\ensuremath{\left\| #1 \right\|}}
\def\corr#1{\left\langle \, #1 \, \right\rangle}
\def\vac{|0\rangle}

%%%%%%%%%%%%%%%%%%%%%%%%%%%%%%%%%%%%%%%%%%%%%%%%%%%%%%%%%%%%%%%%%%%%%%%%%%%%%
%                         General

\def\d{ \partial }

\newcommand{\sq}{\square}

\newcommand{\floor}[1]{\left\lfloor #1 \right\rfloor}
\newcommand{\ceil}[1]{\left\lceil #1 \right\rceil}

\newcommand{\dbyd}[1]{\ensuremath{ \frac{\d}{\d {#1}}}}
\newcommand{\ddbyd}[1]{\ensuremath{ \frac{\d^2}{\d {#1}^2}}}

\newcommand{\T}[3]{\ensuremath{ #1{}^{#2}_{\phantom{#2} \! #3}}}		%general tensor with upper indices displayed first

\newcommand{\sech}{\operatorname{sech}}
\newcommand{\Spin}{\operatorname{Spin}}
\newcommand{\Sym}{\operatorname{Sym}}
\newcommand{\Com}{\operatorname{Com}}
\def\adj{\textrm{adj}}
\def\id{\textrm{id}}

\def\ha{\frac{1}{2}}
\def\tha{\tfrac{1}{2}}
\def\wt{\widetilde}
\def\ra{\rangle}
\def\la{\langle}

\def\ii{{i}}

\def\r{\rightarrow}
\def\Ri{\Rightarrow}

\def\rplus{\rho_{+}}
\def\rminus{\rho_{-}}

\def\rhoone{\varrho_{1}}
\def\rhotwo{\varrho_{2}}
\def\rhoA{\varrho_{A}}
\def\rhofour{\varrho_{4}}
\def\rhohat{\hat\varrho}

\newcommand{\tblue}[1]{\scriptsize{\Blue #1}}

%%%%%%%%%%%%%%%%%%%%%%%%%%%%%%%%%%%%
%%%%%%%%%%%%%%%%%%%%%%%%%%%%%%%%%%

%\vspace{5mm}

%\begin{center}
%{\LARGE \textbf{{String Theory of Supertubes}}}
%
\title{String Theory of Supertubes}

%\vspace{1 cm} {\large Emil J. Martinec {\it and}\, Stefano Massai}\\
\author{Emil J. Martinec {\it and}\, Stefano Massai}

\vspace{0.85 cm}

\affiliation{
Enrico Fermi Institute and Dept. of Physics \\
5640 S. Ellis Ave.,
Chicago, IL 60637-1433, USA \\
}

\vspace{0.7cm} 

%{\small\upshape\ttfamily ejmartin[at]uchicago.edu} \\

%\end{center}
%
%
%\begin{abstract}

%\vspace{3mm}
\abstract{
String theory on $AdS_3$ backgrounds arises as an IR limit of Little String Theory on NS5-branes.
A wide variety of holographic RG flows from the fivebrane theory in the UV to (orbifolds of) $AdS_3$ in the IR is amenable to exact treatment in worldsheet string theory as a class of null-gauged WZW models.  The condensate of stringy winding operators which resolves the near-source structure of fivebranes on the Coulomb branch plays a crucial role in $AdS_3$, revealing stringy structure invisible to the supergravity approximation.  The D-brane sector contains precursors of the long strings which dominate black hole entropy in the dual spacetime CFT. 
}
%\end{abstract}
%
%\newpage

\maketitle
%%%%%%%%%%%%%%%%%%%%%%%%%%%%%%%%%%%%
%%%%%%%%%%%%%%%%%%%%%%%%%%%%%%%%%%%%

\baselineskip=15pt
\parskip=3pt

%\vskip 1cm
%\tableofcontents
%\newpage

\setcounter{footnote}{0}

%%%%%%%%%%%%%%%%%%%%%%%%%%%%%%%%%%%%
%%%%%%%%%%%%%%%%%%%%%%%%%%%%%%%%%%%%

\section{Introduction}

The problem of quantizing general relativity has in principle been with us for as long as both quantum mechanics and general relativity have entered the canon of physics -- almost a century.  However, an application of standard methods of quantization to the gravitational field, and matter fields coupled to gravity, leads to deep conceptual issues -- for instance the meaning of causality in a quantum superposition of geometries, and the nature of time in the vicinity of gravitational singularities, to name but two.  

One of the most profound puzzles to have arisen in quantum gravity is the black hole information problem~\cite{Hawking:1976ra}, which has been with us for more than 40 years.  We begin with a rather discursive presentation of the issues involved, laying the groundwork for subsequent developments; the expert reader may want to skip directly to those developments, beginning with the outline of this paper in section~\ref{sec:outline}.

To illustrate the issues, consider a 4d charged (Reissner-Nordstrom) black hole geometry written in Eddington-Finkelstein coordinates
\begin{align}
\label{dsRN}
ds^2 &= -f(r)\, dv^2 + 2\, dv\, dr + r^2 d\Omega_2^{\, 2}	\nn\\
f(r) &= \frac{(r-r_+)(r-r_-)}{r^2} \quad,\qquad r_\pm = M\pm \sqrt{M^2-Q^2}
\end{align}
In these coordinates, ingoing radial null trajectories are lines of constant $v$ and can be used as a clock by observers anywhere in the geometry.%
\footnote{They thus have an advantage over the use of Schwarzschild time 
$dt = dv + {dr}/{f(r)}$,
which becomes singular at the (outer) horizon $r=r_+$.  }
%
%One sees directly from~\eqref{dsRN} that $r=r_\pm$, $\Omega_2={\textsl const.}$ are null surfaces.
%More generally, 
The horizons $r_\pm$ are accumulation points of outgoing radial null trajectories
\be
\frac{dr}{dv} = \frac{2}{f(r)} ~.
\ee
The residues of the poles in $f(r)$
\be
\frac{dv}{dr}\Bigl|_{r\sim r_\pm} = \Bigl(\frac{2 r_\pm^2}{r_\pm - r_\mp}\Bigr)\, \frac{1}{r-r_\pm}
\equiv \frac{\kappa_\pm}{r-r_\pm}
\ee  
define the surface gravities $\kappa_\pm$ of the two horizons.  In particular, when one quantizes matter fields in this background geometry, one finds Hawking radiation at temperature
\be
T_H = \frac{\kappa_+}{2\pi} = \frac{r_+-r_-}{4\pi r_+^2}
\ee
Charged black holes are in a sense better behaved than Schwarzschild black holes in that the evaporation process doesn't run away -- the radiation temperature falls smoothly to zero as $M\to M_{\ext}=Q$ and thus $r_+\to r_-$.%
\footnote{Here we have assumed that the black hole doesn't radiate away its charge as it evaporates.  There are situations where indeed this does not happen~\cite{Seiberg:1999xz}.}

%%%%%%%%%%%%%
\begin{figure}[ht]
\centerline{\includegraphics[width=3.0in]{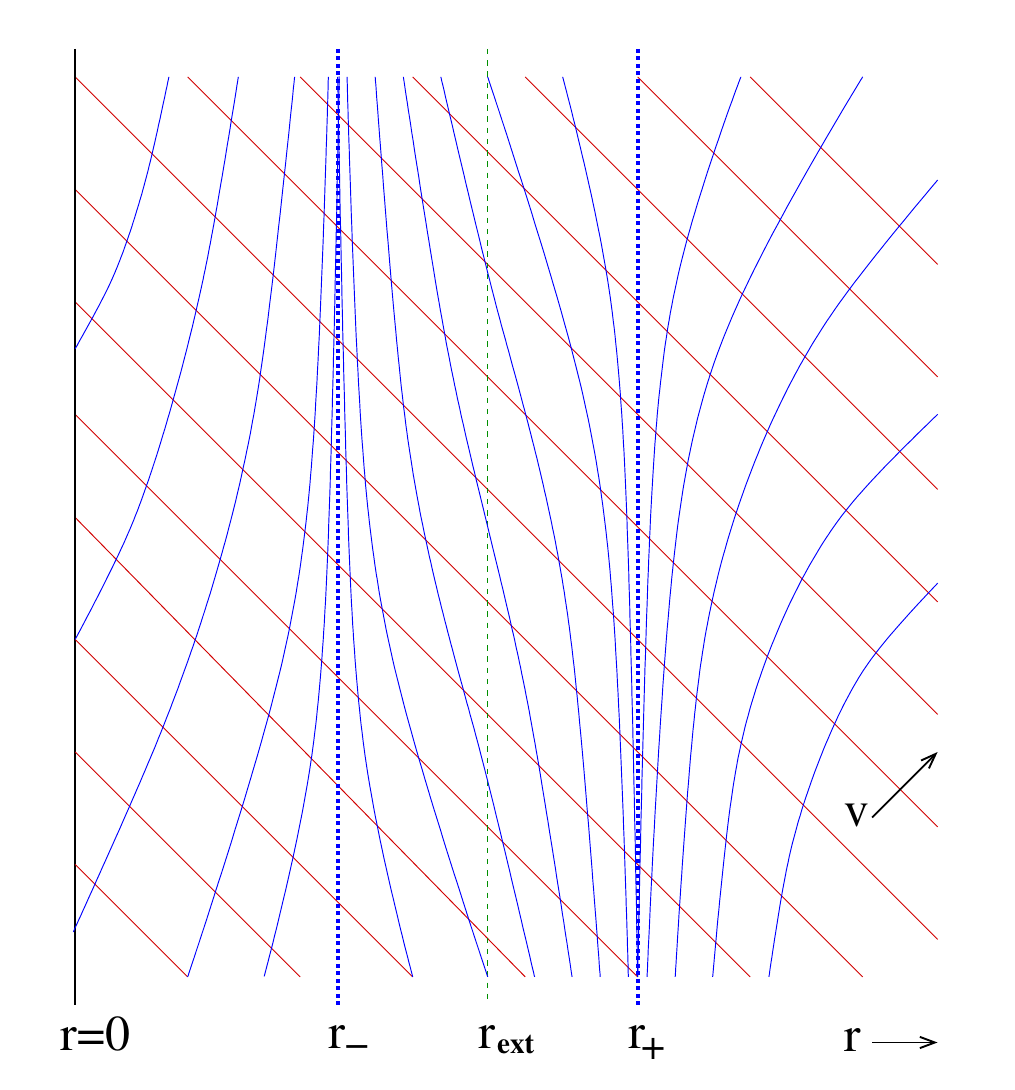}}
\setlength{\unitlength}{0.1\columnwidth}
\caption{\it 
Causal structure of Reissner-Nordstrom black holes depicted in Eddington-Finkelstein coordinates.  Ingoing null trajectories are in red, outgoing in blue.
}
\label{fig:ReissnerNordstrom-alt.pdf}
\end{figure}
%%%%%%%%%%%%%

Although the Hawking temperature vanishes in the limit $r_+\!=\!r_-\!=\!r_\ext$, the Bekenstein-Hawking area law still predicts a nonzero black hole entropy
\be
S_\bh = 4\pi r_\ext^2 = 4\pi Q^2 ~,
\ee
which one can regard as the ground state degeneracy in the superselection sector of fixed charge.

The metric of rotating black holes also has two horizons and much the same structure.  Consider for example the three-dimensional rotating BTZ black hole, with metric
\begin{align}
\label{BTZgeom}
ds^2_\btz &= -f(r) dv^2 + 2 dv\, dr + r^2(d\varphi + \beta(r) dv)^2 
\nn\\[10pt]
f(r) &= \frac{r^2}{\lads^2} - M_3 +\frac{16 G_3^2 J_3^2}{r^2}
=\frac{(r^2-r_+^2)(r^2-r_-^2)}{\lads^2\,r^2}
%\quad,\qquad 
\nn\\[10pt]
\beta(r) &= \frac{4G_3 J_3}{r^2} = \frac{r_+r_-}{\lads\,r^2}  ~~.
\end{align}
One sees again both inner and outer horizons, with a similar structure of outgoing null trajectories.

The structure of outgoing null trajectories reveals a number of essential ingredients of the Hawking calculation of black hole radiance.  Any regular state of a quantum field theory must approach the ground state at asymptotically high frequencies.  In the vicinity of the outer horizon $r\sim r_+$, outgoing null rays diverge as $v$ increases and thus the outgoing modes get stretched as one evolves along $v$.  The vacuum state is an {\it attractor} of evolution near the horizon (in much the same way as in the theory of inflationary perturbations),
%, \cf~\cite{Baumann:2009ds}
and so the state of any sensible field theory rapidly settles down to the vacuum state.  This feature is the root of the slogan ``black holes have no hair''.  Furthermore, the decomposition of the near-horizon vacuum state into the eigenbasis of scattering states at infinity yields a thermal spectrum of outgoing Hawking radiation.

There is thus no structure of a quantum field which can hold a memory of the infall state that formed the black hole at early times, and then coherently release the information in that memory in Hawking radiation at late times.  It is this memory attribute that allows for instance the burning of a lump of coal to maintain unitarity~-- the state of the coal is correlated with that of the radiation it has already emitted, and can imprint those correlations on later radiation.  Thus the resolution of the information puzzle needs some structure to encode memories {\it at the horizon scale}.  But {\it how} can such a structure persist?  Any structure built out of local quantum fields obeys a causal equation of state, and must fall into the black hole.  

To summarize, locality, causality, and regularity of quantum field theory at the outer horizon lead to a loss of unitarity through the Hawking radiation process~\cite{Hawking:1976ra}.  Alternatively, if one insists on unitarity of the outgoing Hawking radiation throughout the evaporation process, one must give up one of these foundational principles of quantum field theory, as has been recently stressed in~\cite{Mathur:2009hf,Almheiri:2012rt}.

While there is as yet no definitive resolution of this puzzle, string theory has provided in several instances a precise accounting of the microstates underlying black hole entropy.  This counting takes place in a weak coupling dual field theory, in a regime of the moduli space far from the regime where gravity is weakly coupled.  We are thus assured that there is something that stores memories and releases them later during a unitary evaporation process, yet we do not gain any further insight into what that something is, and how the Hawking calculation misses it.  Due to the nature of gauge/gravity duality, in the regime where one can count microstates there is no geometrical picture, and where there is a geometrical picture there is no hint of where the microstates are.

If one believes that some structure in string theory encodes memories at the horizon scale, one is led to pose the question: How much of the Reissner-Nordstrom spacetime of figure~\ref{fig:ReissnerNordstrom-alt.pdf} and its properties are reliably embodied in the low-energy effective field theory of gravity and matter?  How much of the spacetime inside the outer horizon even exists, if some violation of effective field theory is happening there?

%%%%%%%%%%%%%%%%%%%%%%%%%%%%%%%%%%%%%%%%%%
%%%%%%%%%%%%%%%%%%%%%%%%%%%%%%%%%%%%%%%%%%

\subsection{The inner horizon}
\label{sec:InnerHorizon}

While it is unclear from the perspective of effective field theory what feature it misses at the {\it outer} horizon, it nevertheless does seem to suggest that nothing dramatic occurs there to local inertial observers.
Let us adopt as a working hypothesis that effective field theory for such observers is valid until it tells us that it is breaking down.  We would then continue to trust the geometry past the outer horizon, into the black hole interior.  However, effective field theory in the vicinity of the {\it inner} horizon $r\sim r_-$ seems suspect, even in the classical domain.  Here we have the opposite problem to that of $r\sim r_+$, in that outgoing null trajectories converge rather than diverge (as reflected in the fact that the surface gravity $\kappa_-$ is {\it negative}).  Infalling matter crossing the inner horizon does in fact leave a nasty memory: The metric backreaction to a local perturbation grows as $e^{-\kappa_- v}$~\cite{Marolf:2011dj,Murata:2013daa}.  We would expect that the geometry relaxes back to the extremal black hole through the Hawking process, but now with a singularity just inside $r=r_-\sim r_+$.  Therefore we should wonder whether the spacetime region $r<r_-$ ever really existed.

\begin{figure}[ht]
\centerline{\includegraphics[width=2.5in]{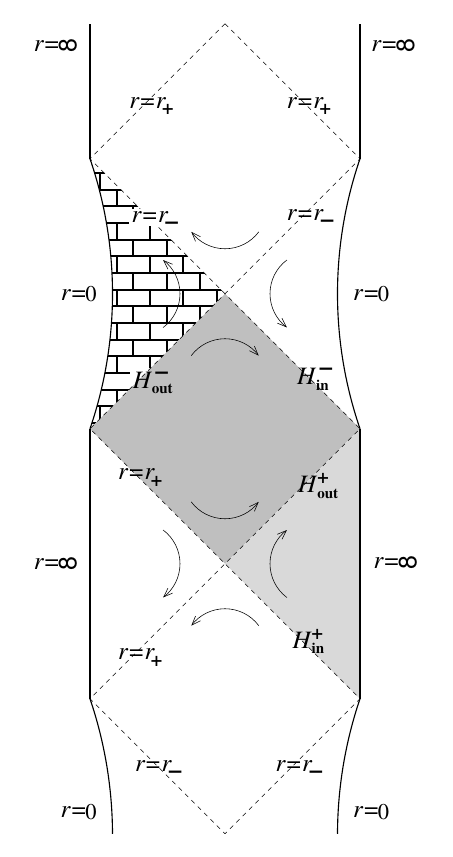}}
\setlength{\unitlength}{0.1\columnwidth}
\caption{\it 
Penrose diagram of the causal structure of the analytically continued vacuum geometry of charged and/or rotating black holes in asymptotically AdS spacetime.  The shaded portion is the region covered by the Eddington-Finkelstein diagram of figure~\ref{fig:ReissnerNordstrom-alt.pdf}.  The flow of the asymptotically timelike Killing vector field is indicated by the arrows.  In physically realized situations, most of the extended diagram is absent.
}
\label{fig:Penrose}
\end{figure}

In the BTZ geometry~\eqref{BTZgeom}, the wave equation of scalar perturbations is separable, and the radial component is given analytically in terms of hypergeometric functions%
~\cite{Maldacena:1997ih,
Cvetic:1997uw,
Cvetic:1998xh,
KeskiVakkuri:1998nw}.
The behavior of infalling wavepackets was analyzed in detail in~\cite{Balasubramanian:2004zu}, with the result that ingoing wave profiles having regular stress-energy on the outer horizon lead to divergent stress-energy on the inner horizon.  If one starts at the outer horizon with a localized, purely ingoing wave profile (\ie\ having support on $\cH^+_{\rm out}$ and vanishing on $\cH^+_{\rm in}$, see figure~\ref{fig:Penrose}), the modular transformation properties of the hypergeometric function map it into a linear combination of ingoing and outgoing waves at the inner horizon.  The trace of the stress tensor blue-shifts as one travels along the inner horizon, leading to a divergence as one moves upward along $\cH^-_{\rm out}$.

%%%%%%%%%%%%%
%%%%%%%%%%%%%

This singularity of perturbations at the inner horizon is of the more standard, short-distance variety which we expect string theory to be able to resolve, but we should not necessarily expect the restoration of the interior region $r<r_-$; instead we might expect some sort of stringy wall ends the geometry at $r\approx r_-$ (which coincides with the outer horizon $r_-=r_+$ for the extremal ground states).

Thus, while the causal structure of the vacuum geometry can be analytically extended to yield the Penrose diagram of figure~\ref{fig:Penrose}, much of this diagram is a fiction.  In physically realized situations, (1) the region below the shaded region is replaced by the formation process of the black hole; (2) the region above the shaded region is replaced by the final state (in particular, if the geometry is asymptotically flat rather than AdS, so that the the black hole evaporates); and (3) the region $r<r_-$ is replaced by whatever regulates the instability toward singularity formation at the inner horizon.  One might say that, just as there is a {\it swampland} of low-energy effective theories that are not realized in string theory~\cite{Ooguri:2006in}, there is also a swampland of Penrose diagrams that are not realized in string theory.

%%%%%%%%%%%%%%%%%%%%%%%%%%%%%%%%%%%%%%%%%%
%%%%%%%%%%%%%%%%%%%%%%%%%%%%%%%%%%%%%%%%%%

\subsection{Extremal (BPS) black holes in string theory}
\label{sec:BPS_BHs}

The BTZ geometry~\eqref{BTZgeom} arises in one of the central examples of AdS/CFT duality in string theory~-- the low-energy limit of type IIB string theory compactified on $\MM\!\times\!\IS^1$, where $\cM=\IT^4$ or $K3$, with $n_5$ Neveu-Schwarz fivebranes (NS5) wrapped on the compactification, bound to $n_1$ fundamental strings (F1) wrapping $\IS^1$, and carrying $n_p$ units of momentum (P) along $\IS^1$, is dual to a two-dimensional CFT.%
\footnote{As we are mostly dealing with two-dimensional CFT's describing perturbative string worldsheets, we will refer to the non-gravitational dual of the full string theory as the {\it spacetime} CFT.}
The supergravity fields sourced by the branes include the metric
\begin{align}
\label{ds-3chg}
ds^2 &= Z_1^{-1}\bigl[ -dt^2 + dy^2 + Z_P(dt+dy)^2 \bigr] + Z_5 ds_\perp^2 + ds_{\cM}^2 \nn\\
\\
Z_5 &= 1+\frac{Q_5}{r^2}
\quad,\qquad
Z_1 = 1+\frac{Q_1}{r^2}
\quad,\qquad
Z_P = \frac{Q_P}{r^2} \quad, \nn
\end{align}
as well as the dilaton $\Phi$ and NS two-form gauge field strength $H^{\sst (3)}=dB$
\be
\label{PhiB}
e^{2\Phi} = \gstr^2\,Z_5/Z_1 
\quad,\qquad 
H^{\sst (3)}_{tyr} = \partial_r Z_1^{-1}
\quad,\qquad
H^{\sst (3)}_{\theta\phi\psi} = \epsilon_{\theta\phi\psi}^{~~~~r}\,\partial_r Z_5 ~.
\ee
Here the space transverse to the branes is written in spherical coordinates $(r,\theta,\phi,\psi)$.  The geometry is entirely determined by the harmonic functions $Z_{1,5,P}$.  The conserved charges $Q_1, Q_5,Q_P$ in these harmonic functions are in turn specified in terms of the integer quanta $n_1, n_5, n_p$ and the compactification data as
\be
\label{chargequant}
Q_1 = \frac{\gstr^2\lstr^6}{V_4} \, n_1
\quad,\qquad
Q_5 = \lstr^2\, n_5
\quad,\qquad
Q_P = \frac{\gstr^2\lstr^4}{V_4 R_y^2}\, n_p
\ee
where $R_y$ is the radius of $\IS^1$ and $V_4$ is the volume of $\cM$.

The low-energy limit decouples near-horizon physics from that of the ambient spacetime.  This decoupled theory provides one of the central examples of AdS/CFT duality~-- the near-horizon geometry is ${\textsl BTZ}\times \IS^3\times \MM$.%
\footnote{The $\IS^1$ parametrized by $y$, wrapped by both the F1 and NS5, is the angular direction $\varphi$ of the BTZ space.}  
As usual, low energy is related to the large redshift near the brane sources for $r^2 \ll Q_1,Q_5$.  In this limit the geometry reduces to
\be
\label{nearhor}
ds^2 \sim \frac{r^2}{Q_1}\Bigl[ -dt^2+dy^2 + \frac{Q_P}{r^2}(dt+dy)^2\Bigr]
	+ Q_5\Bigl(\frac{dr^2}{r^2} + d\Omega_3^{\,2}\Bigr) + ds_\cM^2  ~.
\ee
The radial coordinate $r_\btz$ in the BTZ geometry~\eqref{BTZgeom} is related to the radial coordinate $r_{\mathbi{brane}}$ in the brane geometry above via
\be
\label{coordmap}
(r_\btz)^2 = r_-^2 + {(r_{\mathbi{brane}})^2}
~~,~~~~
t_\btz = {\textstyle \sqrt{\frac{Q_5}{Q_1}}}\; t_{\mathbi{brane}} 
~~,~~~~
\lads \, \varphi_\btz = {\textstyle \sqrt{\frac{Q_5}{Q_1}}}\, {y^{~}_{\mathbi{brane}}}
%\bigl({y^{~}_{\mathbi{brane}}}/{R_y} \bigr)
\ee
while the parameters are related via
\be
\frac{\lads}{4 G_3} = n_1n_5
%~~,~~~~ \frac{\lads^2}{\lstr^2} = n_5
~~,~~~~ \lads^2 = Q_5
~~,~~~~ r_+r_- = r_\ext^2 = Q_p  \quad.
\ee
The extremal horizon at $r\to 0$ (in the brane coordinates; $r=r_-$ in BTZ Schwarzschild coordinates%
\footnote{Note that the natural coordinates for the branes exclude the region $r_\btz<r_-$~\cite{Skenderis:1999bs}.}) 
is associated to a gravitational entropy; in the extremal limit, this entropy takes the form
\be
\label{Sbh}
S_{\bh} = \frac{A_{\rm hor}}{4 G_{10}} = 2\pi \sqrt{n_1n_5n_p-J^2} ~~,
\ee
%where the ten-dimensional gravitational coupling is $G_{10}=8\pi^6 \gstr^2\lstr^8???$, 
where we have included the effect of rotational angular momentum in the space transverse to the branes~\cite{Cvetic:1997uw,Cvetic:1998xh}.

Putting aside for the moment the thornier issue of the information paradox, a perhaps more tractable question is what resolves the horizon singularity of BPS extremal black holes of finite horizon area, such as the extremal BTZ geometry above.  This problem is something of a stepping stone toward the much murkier issue of unitarity preservation during radiation from non-extremal geometries.  Nevertheless it should reveal some of the essential ingredients, because it forces us to discover coherent and stable horizon-scale structures that resolve the apparent singularity of effective field theory at the extremal horizon.  The hope is that these structures provide the needed data storage facility for later information retrieval via Hawking radiation from non-extremal (or at least near-extremal) black holes.

%%%%%%%%%%%%%%%%%%%%%%%%%%%%%%%%%%%%%%%%%%
%%%%%%%%%%%%%%%%%%%%%%%%%%%%%%%%%%%%%%%%%%

\subsection{The Fuzzball conjecture}
\label{sec:Fuzzball}

Beneath the black hole threshold where $S_\bh$ in~\eqref{Sbh} vanishes, there are still many states in the Hilbert space, but they cannot be thought of as semiclassical black holes.  However, the generic state near the threshold might provide clues to the nature of the microstates in the black hole regime.  By approaching the threshold as a limit of states which are not black holes, we can hope to reveal the structures that arise at the extremal horizon scale to resolve its singularity, and wall off the region $r<r_-$ of the naive vacuum geometry from forming.

For instance one can consider the ensemble of BPS states carrying two of the three charges of BTZ black holes.  The generic two-charge BPS solutions are nonsingular, capping off in a region of topological bubbles supporting the brane charges via electric and magnetic fluxes, as is familiar from many examples of the AdS/CFT correspondence, \eg~\cite{Lunin:2001fv,Lunin:2002iz,Polchinski:2000uf,Lin:2004nb}.  The entropy of these two-charge {\it microstate geometries} grows with the charges as $\sqrt{n_1n_5}$; these features might lead one to hope that the effect of adding the third charge is sufficiently benign that the geometry remains nonsingular in the presence of all three charges.  If there are enough different ways of adding momentum excitations, one would then account for the three-charge entropy via an ensemble of geometries that look like the black hole from far away, but differ substantially from it sufficiently close to the brane sources (where the horizon would have been a priori).  Since a microstate has no entropy, each such three-charge microstate geometry should be smooth and have no horizon since horizons imply macroscopic gravitational entropy.  The idea that black hole entropy can be described this way is known as the {\it fuzzball conjecture} (see~\cite{Mathur:2005zp,Bena:2007kg,Skenderis:2008qn,Bena:2013dka} for reviews).

However, there is no guarantee that the generic three-charge state can be arrived at this way; indeed this question is at the heart of the issue of what is breaking down at the scale of the extremal horizon, and what physics resolves the singularity we expect to find there according to the logic of section~\ref{sec:InnerHorizon}.  More recently, a refinement of this picture%
~\cite{Martinec:2014gka,Martinec:2015pfa}
posits that the resolution of the extremal horizon results from the appearance of a horizon-scale condensate of branes wrapping topological cycles at the bottom of the AdS throat.  Indeed, we will argue below that something akin to a Coulomb-Higgs phase transition occurs precisely in the regime where one expects a horizon to form, resulting in a significant modification to our picture of the fuzzball.

For BPS black holes, the microstate geometries program is an attempt to resolve the inner horizon singularity entirely within the context of higher-dimensional supergravity, or perhaps an extrapolation of supergravity to the regime where the bubbles and fluxes that cap off the geometry become small -- perhaps of order the Planck scale -- and the location of the cap approaches the location of the horizon of the \naive\ black hole solution carrying the same conserved charges.  This is the picture advocated in~\cite{Mathur:2005zp,Bena:2007kg,Bena:2013dka}.  The proposal is that in string theory, black holes do have hair at the horizon~-- a fuzz of KK monopole bubbles and antisymmetric tensor flux.  In this scenario, the singularity at the extremal horizon is resolved geometrically, by capping off the AdS throat region {\it before} one reaches the horizon.  

\begin{figure}[ht]
\centerline{\includegraphics[width=3.5in]{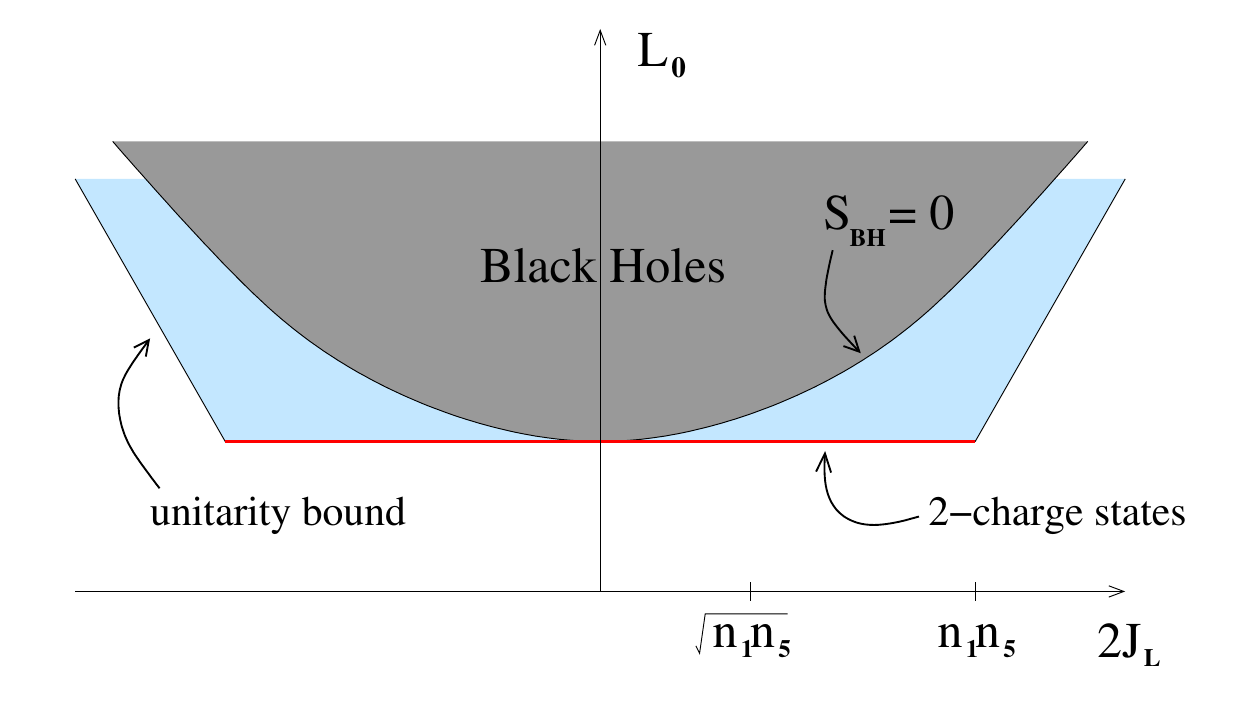}}
\setlength{\unitlength}{0.1\columnwidth}
\caption{\it 
Phase diagram for the spectrum.  Generic states with $c_\eff L_0/6\gg J_L^2$ are BTZ black holes with rotation on the $\IS^3$; states below this bound (depicted in blue) are not.  The 1/4-BPS states (on the red line) all lie at or below the bound.
}
\label{fig:Spectrum-alt}
\end{figure}

It should be emphasized that the two-charge solutions discussed above, while exhibiting macroscopic entropy, are nevertheless outside the regime where generic states are macroscopic black holes.  The structure of the Hilbert space of the spacetime CFT is sketched in figure~\ref{fig:Spectrum-alt}.%
\footnote{A more detailed analysis leads to a slightly more elaborate phase diagram~\cite{Bena:2011zw} with additional black hole phases between the BPS bound and the threshold of vanishing geometric entropy~\eqref{Sbh}.} 
%
%%%%%%%%%%%%%
%%%%%%%%%%%%%
Supersymmetric ground states have zero energy, regardless of their $\IS^3$ angular momenta $J_L$, $J_R$.
The asymptotic density of spacetime CFT states which are in a supersymmetric ground state for the right-movers but in an excited state carrying left-moving momentum $L_0$ and nonzero angular momentum $J_L$, reproduce the BPS black hole entropy:
\be
\label{CardyJne0}
S \sim 2\pi \sqrt{\frac{c_\eff}{6} L_0-J_L^2} = 2\pi \sqrt{\strut n_5 n_1 n_p -J_L^2} ~.
\ee
The requirement that the state carries angular momentum subtracts from the free energy available for entropic oscillator excitations of the spacetime CFT.

A vanishing or imaginary value of this expression for the entropy does not mean there are no states, rather it simply means that these states are not macroscopic black holes, and so their density of states is not approximated by the usual Bekenstein-Hawking formula.%  
\footnote{For some observables these states may nevertheless approximate the results obtained in black hole states~\cite{Balasubramanian:2007qv,Balasubramanian:2008da}. }
The BPS three-charge states are in a right-moving ground state while being excited on the left.  The BPS two-charge states have zero left-moving energy $L_0$ as well, and are at best extremal black holes when $J_L=0$.  They are nonsingular (if stringy) geometries known as {\it supertubes}~\cite{Mateos:2001qs,Emparan:2001ux}, as we review below in section~\ref{sec:Supertubes}.  These supertubes have macroscopic rotation in the transverse space in order to keep the fivebrane sources separated, and the smoother they are, the further they are below the black hole threshold.  In a sense, the geometry is over-rotating, and that rotation supports the brane constituents against collapsing in on one another to form a black hole.

Within the fuzzball program, two approaches have been taken to adding the third charge to move upward in $L_0$ into the macroscopic black hole regime:
\begin{enumerate}
\item
Construct more general geometries, where the brane/flux transition is applied to all three charges to generate smooth geometries whose quantum numbers place them solidly in the black hole regime (for a review, see~\cite{Bena:2007kg,Gibbons:2013tqa}).
\item
Add momentum waves to a seed two-charge NS5-F1 supertube solution in such a way that the geometry remains smooth.  If one can add enough momentum, the state will be pushed upward from the two-charge line in figure~\ref{fig:Spectrum-alt} into the black hole regime%
~\cite{Bena:2015bea,Giusto:2015dfa,Bena:2016agb,Bena:2016ypk}.
Such smooth, horizonless geometries built by adding waves to a two-charge solution are known as {\it superstrata}.
\end{enumerate}
In the first approach, it was shown recently in~\cite{Martinec:2015pfa} that precisely at the point where the bubbles and fluxes descend down a deep AdS throat to approximate the extremal black hole, a condensate of branes wrapping the bubbles develops and dominates the density of states (at least in a crude truncation of the collective dynamics to a quantum-mechanics approximation).  In this admittedly over-simplified approximation scheme, the region where the geometry is smoothly capped off by bubbles and fluxes is the Coulomb branch of the quantum mechanics, while the brane condensate is the Higgs branch, suggesting the process of horizon formation is something akin to a Coulomb-Higgs phase transition.

These results motivate a search for similar phenomena in the second approach.  Since the superstrata are built on a supertube base, we would like to understand the topology of supertube states, what sort of branes we might wrap around that topology, and what novel features might arise as a result.  However, the typical supertube state is rather stringy because the curvature near the sources is of order the string scale~\cite{Mathur:2005zp}, and so to make progress one needs to understand the string theory of supertubes.  In this work, we will construct an exact worldsheet CFT for a  particular class of supertube states, and illustrate features which we expect to be generic.  We will indeed find stringy topology at the bottom of the AdS throat related to the configuration of background branes sourcing the supertube geometry; we will also find indications that once again, branes wrapping this topology play an increasingly important role as one approaches the black hole threshold.

%%%%%%%%%%%%%%%%%%%%%%%%%%%%%%%%%%%%%%%%%%
%%%%%%%%%%%%%%%%%%%%%%%%%%%%%%%%%%%%%%%%%%

\subsection{Outline of the paper}
\label{sec:outline}

We begin in section~\ref{sec:Supertubes} with a review of two-charge NS5-F1 supertubes, which comprise the supersymmetric two-charge ground states of fivebranes compactified on $\cM\!\times\!\IS^1$, and the related NS5-P supertubes that arise upon T-duality along $\IS^1$.
A particularly well-studied class of supertubes are those with a circular source profile in the transverse $\IR^4$, which preserve axial symmetry in the transverse spacetime.  

On the covering space of $\IS^1$, the NS5-P supertube consists of $\nfive$ fivebranes separated on their Coulomb branch and spiraling around one another.  This leads us to review in section~\ref{sec:NS5_CFTs} the variety of known worldsheet descriptions of fivebranes on the Coulomb branch: (i) the \naive\ nonlinear sigma model; (ii) the $\bigl(\frac{\sltwo}{\uone}\!\times\!\frac{\sutwo}{\uone}\bigr)/\IZ_\nfive$ coset orbifold; (iii) the non-compact Landau-Ginsburg orbifold; and (iv) the null-gauged $\sltwo\!\times\!\sutwo$ WZW model.  Each of these worldsheet descriptions exhibits different features of fivebrane physics.  The one that turns out to be most suited to our purpose is the null-gauged WZW model; as we discuss in section~\ref{sec:supertube}, it turns out that a simple modification of the gauging~-- a tilting of the null current being gauged into the $\IR^{1,1}$ directions along the fivebrane~-- transforms the worldsheet description of static fivebranes into that of stationary NS5-P supertubes.  This modification is at the same time an utterly simple, and yet seemingly quite powerful tool.  Having an exact worldsheet CFT for the NS5-P supertube helix, one can compactify along the helix direction in a manner that respects its discrete periodicity.  Then the T-dual is an NS5-F1 supertube, whose properties follow trivially as a mere relabelling of the CFT spectrum; several generalizations and extensions are discussed as well.  In all these models, the radius $R_y$ of the circle $\IS^1$ in the compactification is a control parameter, as it enters the constant in the harmonic function $Z_1$ that controls the crossover between the linear dilaton regime of the fivebranes and the $\ads3$ regime of the spacetime CFT in the infrared.

Section~\ref{sec:spectrum} works out the spectrum of closed string excitations for the NS5-P and NS5-F1 supertubes, and section~\ref{sec:dbranes} discusses features of the D-brane spectrum that exhibit aspects of the long string sector of the spacetime CFT, and at the same time are the analogues of the wrapped brane states considered in~\cite{Martinec:2015pfa}.  Section~\ref{sec:discussion} concludes with a summary and discussion of next steps.  The Appendix lays out our conventions for $\sltwo$ and $\sutwo$ current algebra.

%%%%%%%%%%%%%%%%%%%%%%%%%%%%%%%%%%%%%%%%%%
%%%%%%%%%%%%%%%%%%%%%%%%%%%%%%%%%%%%%%%%%%

\section{Supertubes and Fuzzballs}
\label{sec:Supertubes}

If the extremal horizon is singular under any perturbation, one may expect that whatever stringy or quantum gravitational effects resolve the classical BTZ geometry into individual microstates, also perturb the \naive\ classical solution~\eqref{ds-3chg} and at the same time resolve the inner horizon singularity predicted by effective field theory.

In the context of the fuzzall program, the two-charge supertube geometries were proposed as prototypes of  a mechanism to cap off the throat of the extremal black hole before a horizon could form~\cite{Mathur:2005zp}.  One might entertain the notion that some aspects of the extremal horizon singularity resolution mechanism arise already in the supergravity approximation, perhaps for some non-generic microstates.  This idea originated with the explicit construction of generic supertube geometries~\cite{Lunin:2001fv,Kanitscheider:2007wq} carrying two of the three charges above, plus angular momentum; these geometries were shown to be generically nonsingular supergravity solutions~\cite{Lunin:2002iz}.  

We will begin at an even more basic level, with the supergravity solution sourced by coincident fivebranes.  We can put a chiral wave profile on the fivebranes that wiggles the center of mass of the branes as well as the compact scalar $Y$ and self-dual tensor $C_{ab}$ on the fivebrane worldvolume (here $a,b$ are $\sutwo$ spinor indices).  The general result for the supergravity solution~\cite{Lunin:2001fv,Kanitscheider:2007wq} is again written in terms of harmonic functions.  Specializing to $\cM=\IT^4$, the metric takes the form
\begin{align}
\label{NS5Pmetric}
ds^2 &= -du\,dv +\frac{\cP}{Z_5} dv^2 +2A_i \,dx^i \, dv
	+ Z_5\, d\xx\!\cdot\!d\xx + ds^2_\cM  \nn\\
\cP &= Z_p Z_5 - Z_{ab}Z^{ab}  
~~,~~~~
\end{align}
in terms of null coordinates $u=t-\ytil$, $v=t+\ytil$ 
the harmonic functions $Z_*$ and harmonic one-form $A$ are given by
\begin{align}
\label{NS5Pharmfns}
Z_5 = 1 + \frac{Q_5}{|\xx-\bX(v)|^2 }
~~&,~~~~
Z_p = {Q_5}\,\frac{|\partial_v\bX|^2 + |\partial_v \!\bY|^2}{|\xx-\bX(v)|^2}  \nn\\
Z_{ab} = - {Q_5}\, \frac{ \partial_v Y^{ab}}{|\xx-\bX(v)|^2 }
~~&,~~~~
A_{i} =  -{Q_5}\, \frac{ \partial_v X^{i}}{|\xx-\bX(v)|^2 }
\end{align}
where the singlet part $\epsilon^{ab}Y_{ab}(v)$ represents the profile of the scalar $Y$, and the triplet part $Y^{(ab)}(v)$ corresponds to that of the self-dual antisymmetric tensor $C_{ab}$.
%%%
For $\cM=\IT^4$, the possible polarizations follow from dimensional reduction of the fivebrane worldvolume along $\cM$, yielding eight bosonic and eight fermionic modes of an effective type II string (which is on the U-duality orbit of the background charges).  For $\cM=K3$~\cite{Harvey:1995rn}, there is no consistent dimensional reduction of the fermion fields on the fivebrane, instead all the light momentum-carrying excitations are bosonic.  The self-dual tensor dimensionally reduces on $\cM$ according to the ansatz
\be
C_{ ab} (v) = \sum_I  f^{\strut}_{I}(v) \,\omega_{ ab}^{\sst(I)}
\ee
where $\omega_{ab}^{\sst(I)}$ comprise a complete set of self-dual two-forms on $\cM$ (the $ab$ indices are a symmetric bispinor on $\cM$), of which there are 19 for $\cM=K3$.  These 19 modes, plus the transverse scalars $X^{\alpha\dot\alpha}$ and the scalar $Y$, comprise a set of 24 bosonic excitations of an effective bosonic string (as follows from heterotic-type II duality~\cite{Harvey:1995rn}).

%%%%%%%%%%%%%%%%%%%%%%%%%%%%%%%%%%%%%%%%%%
%%%%%%%%%%%%%%%%%%%%%%%%%%%%%%%%%%%%%%%%%%

\subsection{Aside on the IR CFT dual}
\label{sec:CFTdual}

The spectrum of 1/4-BPS states arising from the various choices of wave profile on the fivebranes is independent of where we are on the moduli space of the compactification; therefore one can also match it to the states of the solvable spacetime CFT dual which appears in the IR limit after T-duality along $\ytil$, namely the symmetric product orbifold $(\cM)^N/S_N$, $N=n_1n_5$, which arises in a particular weak-coupling limit of the spacetime CFT moduli space.  But since neither the T-duality nor the IR limit affect the spectrum of ground states, we can count the ground states of the orbifold and they are guaranteed to match the count of waves on the fivebrane.  The supersymmetric ground states of the orbifold CFT are labelled by conjugacy classes of the symmetric group, which in turn are labelled by a partition of $N$
\be
\Bigl\{ N_\kappa \: \Bigl |\; \sum_\kappa \kappa N_\kappa = N \Bigr\} ~.
\ee 
This partition specifies that the twisted sector contains $N_\kappa$ copies of the cyclic permutation of order $\kappa$, \ie\ the twisted boundary condition on the fields whose monodromy cyclically permutes the $\kappa$ copies of the sigma model CFT on $\cM$.  The effect of the $\kappa$-cyclic twist can be described via single copy of the CFT on $\cM$ living on a spatial circle $\kappa$ times longer.  Because of this simple covering-space interpretation of the effect of the twist, the Ramond ground states of the $\kappa$-cycle CFT are in one-to-one correspondence with those of the CFT on $\cM$ itself.  For the $\cM=\IT^4$ CFT, the Ramond ground states consist of
\begin{align}
\label{polarizations}
{\rm bosonic} ~ &:~~~~\ket{\alpha\dot\alpha} ~~,~~~~ \ket{ab} \nn\\
{\rm fermionic} ~ &:~~~~\ket{\alpha b} ~~,~~~~ \ket{a\dot\alpha} ~;
\end{align}
these states are related by the action of the zero modes of the fermions $\psi_{\alpha a}$, $\tilde\psi_{\dot\alpha b}$.  Note that these ground states map one-to-one onto the polarizations of the type IIB string on $\IT^4\times\IS^1$, in accord with the discussion above.  The eight transverse bosonic polarizations of the string winding along $\IS^1$ split into those of the noncompact $\IR^4$, with the vector index $X^i$ rewritten as an $SO(4)$ bispinor $X^{\alpha\dot\alpha}$ corresponding to the states $\ket{\alpha\dot\alpha}$; and those along $\IT^4$, rewritten in terms of an $SO(3)$ bi-spinor $\ket{ab}$ that decomposes into a singlet corresponding to $Y$ and a triplet corresponding to $C_{ab}$.
For $\cM=K3$ the fermionic ground states are absent, and there are sixteen more bosonic ground states (which are the twisted sector ground states in a realization of $K3$ as $\IT^4/\IZ_2$).%
\footnote{Both of these results are special cases of the fact that the Ramond ground states are in one-to-one correspondence with the cohomology of the sigma model target space~\cite{Alvarez:1987wg,Pilch:1986en}.}  
In other words, one has the polarizations of the bosonic chirality of the heterotic string on $\IT^4$, again in accord with the previous discussion.  

Assembling the components, the 1/4-BPS states consist of a partition of $N$, and for each element of that partition, a choice of Ramond ground state whose counting is the same as the polarizations of waves on a fivebrane wrapped on $\cM\times \IS^1$ and carrying momentum along~$\IS^1$.  The partition of $N$ is the counting of the number of quanta of different wavenumbers in the wave.  Thus, there is a one-to-one match between 1/4-BPS states of the supergravity theory and its dual CFT.
The resulting two-charge entropy amounts to 
\be
\label{S2chg}
S_{\rm 2\;chg} = 2\pi\sqrt{\strut \frac{c_\eff}{6}n_5 n_1}
\ee
where $c_\eff=12$ for $\cM=\IT^4$ and $c_\eff=24$ for $\cM=K3$.

%%%%%%%%%%%%%%%%%%%%%%%%%%%%%%%%%%%%%%%%%%
%%%%%%%%%%%%%%%%%%%%%%%%%%%%%%%%%%%%%%%%%%

\subsection{Multicenter configurations and monodromy}
\label{sec:monodromy}

\begin{figure}[ht]
\centerline{\includegraphics[width=2.5in]{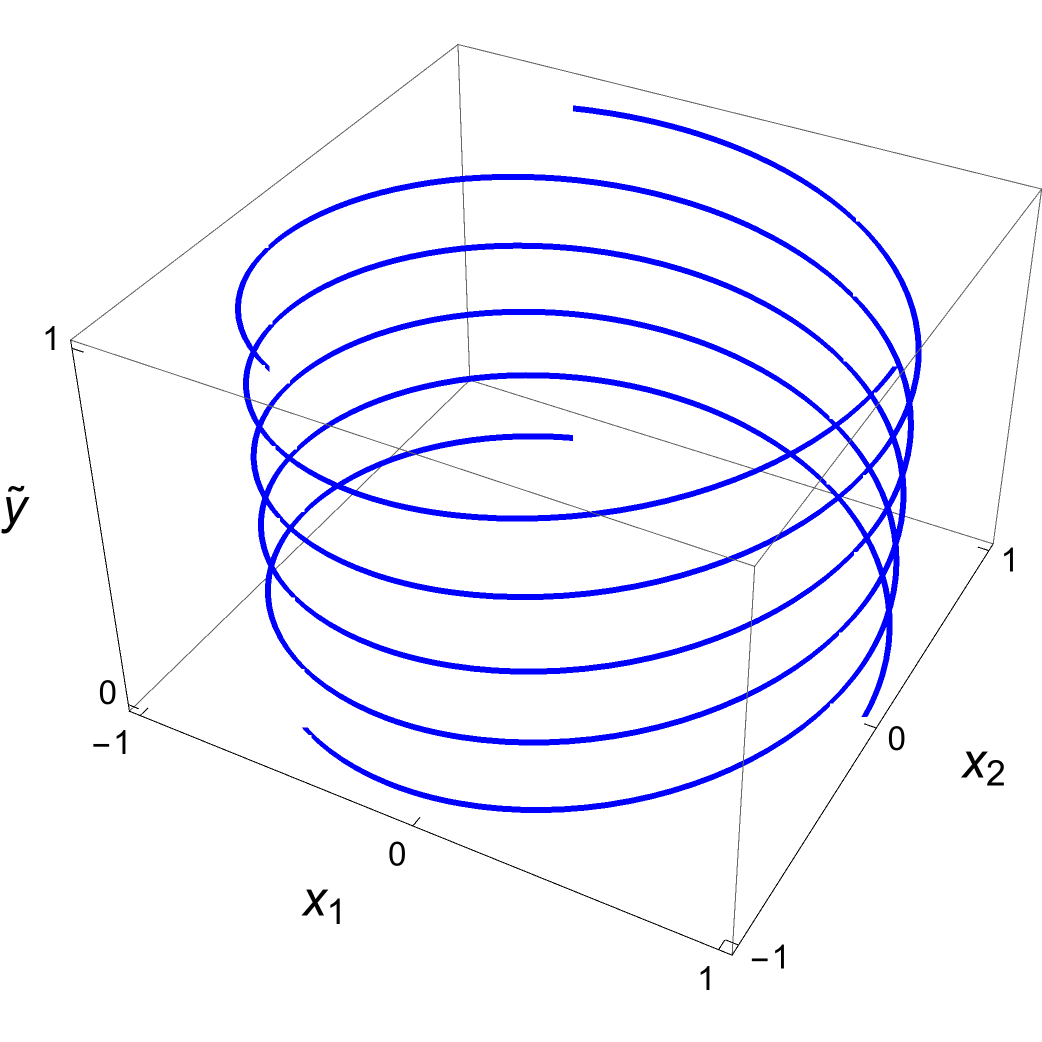}}
\setlength{\unitlength}{0.1\columnwidth}
\caption{\it 
Multiple fivebrane centers wind together around the $\ytil$ circle and lift the relative moduli.
}
\label{fig:NS5-P_1strand-k5n3.pdf}
\end{figure}

One can also consider multi-center solutions, corresponding to the Coulomb branch of the fivebrane gauge theory.  Each center is allowed its own set of source profiles $\bX_{\!\supera}(v)$, $\bY_{\!\supera}(v)$, where $\suplabel$ labels the centers.  Due to the linearity of harmonic functions, the supergravity solution follows from the superposition of the harmonic functions of the individual sources.  When two or more clusters have the same number of fivebranes, a compact $\ytil$ direction allows the option of implementing a twisted boundary condition
\be 
\bX_{\!\supera}(v+2\pi R_\ytil) = \bX_{\sigma\supera}(v)
~~,~~~~
\bY_{\!\supera}(v+2\pi R_\ytil) = \bY_{\!\sigma\supera}(v)
\ee
where $\sigma$ is a permutation.  Any such permutation can be written as a word in the symmetric group $S_N$, where $N$ is the number of identical centers.  The word consists of a product of non-overlapping cyclic permutations.  The twisted boundary condition lifts some of the Coulomb branch moduli -- the number of independent centers reduces from $N$ to the number of cycles in the word $\sigma$.  An example with $N=3$ clusters and the profile
\be
\label{profile1}
X^1_{\!\supera}+iX^2_{\!\supera} 
\equiv |X_{\!\supera}| \, e^{i\phi_{\!\supera}} = a \, \exp\Bigl[ i\frac{k}{N} \frac{(t+\tilde y )}{\Rytil} + \frac{2\pi i \suplabel}{N}\Bigr]
\ee
with $k=5$, and $\sigma$ a single cycle of length three, is shown in figure~\ref{fig:NS5-P_1strand-k5n3.pdf}.

%%%%%%%%%%%%%
%%%%%%%%%%%%%

As it stands, these solutions are na\"ively singular in their cores near the sources.  One can however consider the T-dual frame where one performs a T-duality along $\ytil$.  Since the fivebrane profile is generically not aligned with the $\ytil$ direction, the T-duality is along a cycle partially parallel to and partly orthogonal to the fivebrane worldvolume.  T-duality along an NS5 worldvolume yields again an NS5-brane, while T-duality along a cycle transverse to an NS5-brane yields a KK monopole.  Thus the T-dual of the wiggling type IIA NS5 branes, carrying momentum along $\ytil$ and angular momentum in the transverse space, is a set of NS5-branes in type IIB, carrying fundamental string charge, angular momentum, and KK magnetic dipole charge.

It is not clear how to perform this T-duality transformation directly on the geometry~\eqref{NS5Pmetric}.  If however one smears the source along $\ytil$, then one obtains a set of harmonic functions that are $\ytil$-independent and one can use the standard Buscher rules for T-duality, \cf~\cite{Giveon:1994fu}.  Carrying out this exercise leads to the geometry~\cite{Lunin:2001fv,Kanitscheider:2007wq}
\begin{align}
\label{NS5F1metric}
ds^2 &= \frac{Z_5}{\cP} \Bigl[ -\bigl(dt + \avec\bigr)^2 + \bigl(dy + \bvec\bigr)^2\Bigr] 
	+ Z_5\, d\xx\!\cdot\!d\xx + ds^2_\cM  \nn\\
\cP &= Z_1 Z_5 - Z_{ab}Z^{ab}  
~~,~~~~
d\bvec = - *^{\strut}_4 d\avec
\end{align}
where the smeared harmonic functions $Z_*$ and harmonic one-form $\avec$ are given by
\begin{align}
\label{NS5F1harmfns}
Z_5 = 1 + \frac{\lstr^2}{L_y}\sum_\suplabel \int_0^{L_y}\frac{dv}{|\xx-\bX_{\!\supera}(v)|^2 }
~~&,~~~~
Z_1 = 1+ \frac{\lstr^2}{L_y}\sum_\suplabel \int_0^{L_y} \!\!dv\, \frac{|\partial_v\bX_{\!\supera}|^2 + |\partial_v \!\bY_{\!\supera}|^2}{|\xx-\bX_{\!\supera}(v)|^2}  \nn\\[5pt]
Z_{ab} = - \;\frac{\lstr^2}{L_y}\sum_\suplabel \int_0^{L_y} \frac{dv\; \partial_v Y_{\!\supera}^{ab}}{|\xx-\bX_{\!\supera}(v)|^2 }
~~&,~~~~
\avec_{i} =  \frac{\lstr^2}{L_y}\sum_\suplabel \int_0^{L_y} \frac{dv\; \partial_v X_{\!\supera}^{i}}{|\xx-\bX_{\!\supera}(v)|^2 } ~,
\end{align}
where $L_y=2\pi \Ry$.
An analysis of the local structure near the source~\cite{Lunin:2002iz} shows that the NS5-brane charge has disappeared into flux through the loop of KK monopoles, and the geometry is completely nonsingular (up to orbifold points where the loop self-intersects in the transverse space).

One of the lessons we will learn below is that this smearing is not so innocent.  String theory remembers where the branes are, in the expectation values of stringy winding operators near the source.

%%%%%%%%%%%%%%%%%%%%%%%%%%%%%%%%%%%%%%%%%%
%%%%%%%%%%%%%%%%%%%%%%%%%%%%%%%%%%%%%%%%%%

\subsection{The round supertube}
\label{sec:roundtube}

Perhaps the simplest supertube profile is the circular spiral of equation~\eqref{profile1} and figure~\ref{fig:NS5-P_1strand-k5n3.pdf}, with $N=\nfive$.  
%
%??? We will build it in two stages.  Consider $n_5$ NS5-branes wrapped on $\cM\times \IR^{1,1}$ in type IIA string theory.  The low-energy dynamics of coincident branes is governed by the strongly coupled {\it Little String Theory}.  Without the string sources and the momentum profile (so that $Z_1=1$ and $Z_P=0$ in~\eqref{PhiB}), the dilaton blows up at the brane source.  However, separating the branes sufficiently from one another in the transverse $\IR^4$, the low-energy dynamics is regular in string theory~\cite{Giveon:1999px,Giveon:1999tq}. 
%
This configuration at fixed $\ytil$ places the branes in a $\IZ_{\nfive}$ symmetric arrangement around a circle of radius $a$ in the transverse $x^1$-$x^2$ plane
\be
\label{Z5coulexact}
Z_5 = 1 + \sum_{m=1}^{n_5} \frac{\lstr^2}{|x_1+ix_2  - a e^{i\phi_\supera} |^2 + |x_3+ix_4|^2} 
~~,~~~ \phi_\supera = \frac{k}{\nfive}\frac{(t+\tilde y)}{ \Rytil} + \frac{2\pi  \suplabel}{\nfive} ~,
\ee
and similarly for the other harmonic functions and forms in equation~\eqref{NS5Pmetric}.
We will be interested in the fivebrane {\it decoupling limit} where we simultaneously take the low-energy, near-source limit of the geometry, while bringing the branes together to a separation much smaller than $\sqrt{Q_5}$.  In this limit, the constant term in $Z_5$ is effectively scaled away relative to the source term, and the effect is to drop the constant term from $Z_5$.

\begin{figure}[ht]
\centerline{\includegraphics[width=2.5in]{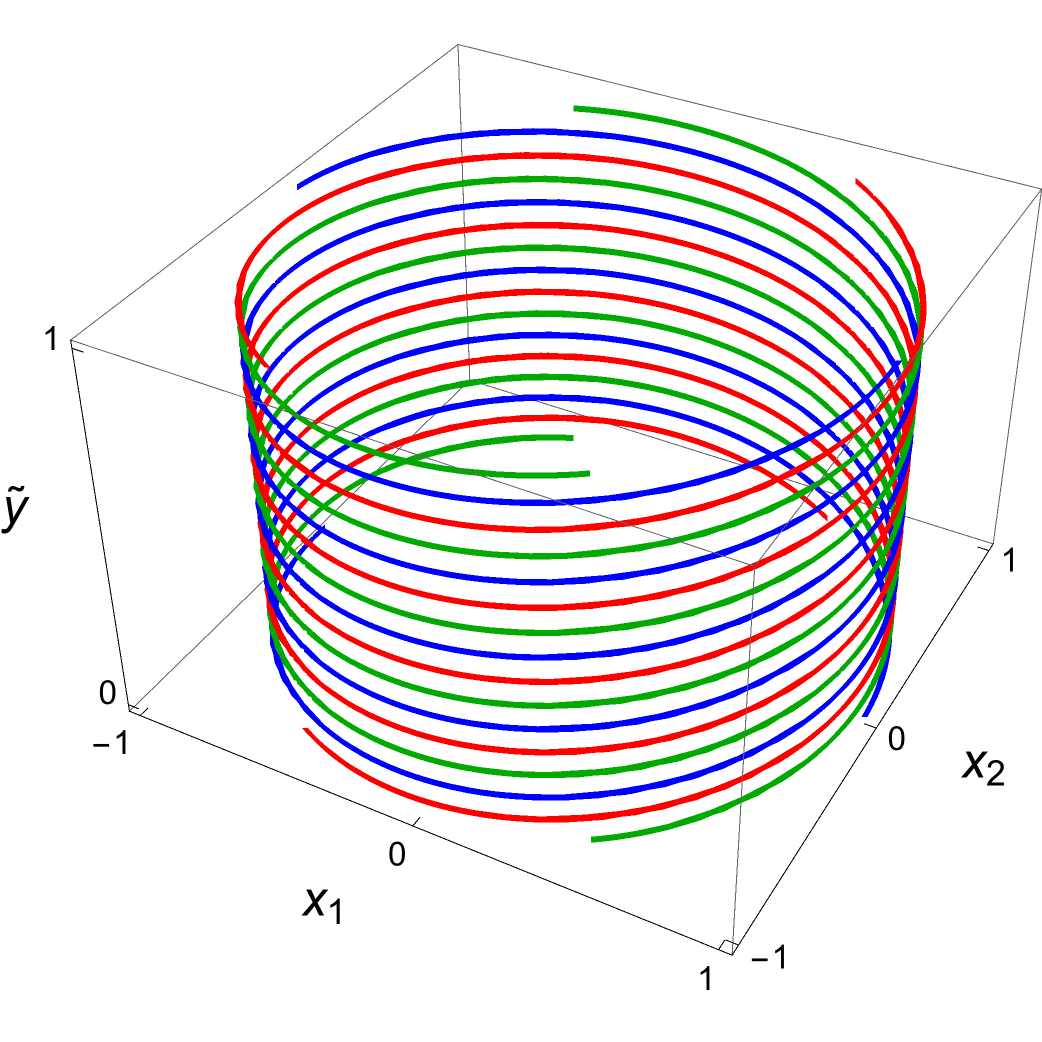}}
\setlength{\unitlength}{0.1\columnwidth}
\caption{\it 
Interleaved fivebrane supertubes; six fivebranes make three supertubes, each of which spirals around the $x_1$-$x_2$ plane five times (so $k=15, n_5=6$).
}
\label{fig:NS5-P_3strand-k15n6.pdf}
\end{figure}

The integer $k$ specifies the permutation $\sigma$, namely a shift by $k$ positions around the circle of radius $a$ in the $x^1$-$x^2$ plane as one passes once around the $\ytil$ circle.   If $k$ and $n_5$ are relatively prime, this monodromy winds the fivebranes into a single fivebrane wrapped $n_5$ times around the $\ytil$ circle as in figure~\ref{fig:NS5-P_1strand-k5n3.pdf}; if $gcd(k,n_5)=p$, then there are $p$ strands of fivebrane interleaved around the circle; see figure~\ref{fig:NS5-P_3strand-k15n6.pdf}.

%%%%%%%%%%%%%
%%%%%%%%%%%%%

After smearing and T-duality along $\ytil$, one obtains a round NS5-F1 supertube.  In the dual spacetime CFT that arises in the IR limit $\Ry=1/\Rytil\to\infty$, this supertube corresponds to a CFT ground state labelled by the symmetric group conjugacy class consisting of cycles of length $k$, all in the $|\alpha\dot\alpha\rangle=|\!+\!+\rangle$ polarization state~\cite{Kanitscheider:2006zf,Giusto:2015dfa}.  This is the class of circular supertube states for which we will now build an exactly solvable worldsheet CFT.

%%%%%%%%%%%%%%%%%%%%%%%%%%%%%%%%%%%%%%%%%%
%%%%%%%%%%%%%%%%%%%%%%%%%%%%%%%%%%%%%%%%%%

\section{Four constructions of fivebranes}
\label{sec:NS5_CFTs}

The harmonic function~\eqref{Z5coulexact} describes round supertubes when $k\tight\ne 0$.  For ${k\tight=0}$, however, it describes NS5-branes on the Coulomb branch.%
\footnote{More precisely, the fivebranes are compactified on $\IS^1\tight\times\cM$, and rather than the branes having fixed locations on a moduli space, one should quantize the branes' collective coordinates.  Alternatively, since our focus is on the configuration in the transverse space, for the present purpose we can decompactify $\cM$ so that the brane locations are indeed moduli.}
In the decoupling limit, NS5-branes in this $\IZ_\nfive$ symmetric arrangement are described by an exactly solvable worldsheet CFT~\cite{Giveon:1999px,Giveon:1999tq}, as we now review.  We will subsequently restore the parameter $k$ in section~\ref{sec:supertube} through a suitable generalization.

We choose to work in conventions where $\lstr=1$, so that T-duality is $R\to 1/R$.

%%%%%%%%%%%%%%%%%%%%%%%%%%%%%%%%%%%%%%%%%%
%%%%%%%%%%%%%%%%%%%%%%%%%%%%%%%%%%%%%%%%%%

\subsection{The \naive\ sigma model}
\label{sec:sigmodel} 

\begin{figure}[ht]
\centerline{\includegraphics[width=2.5in]{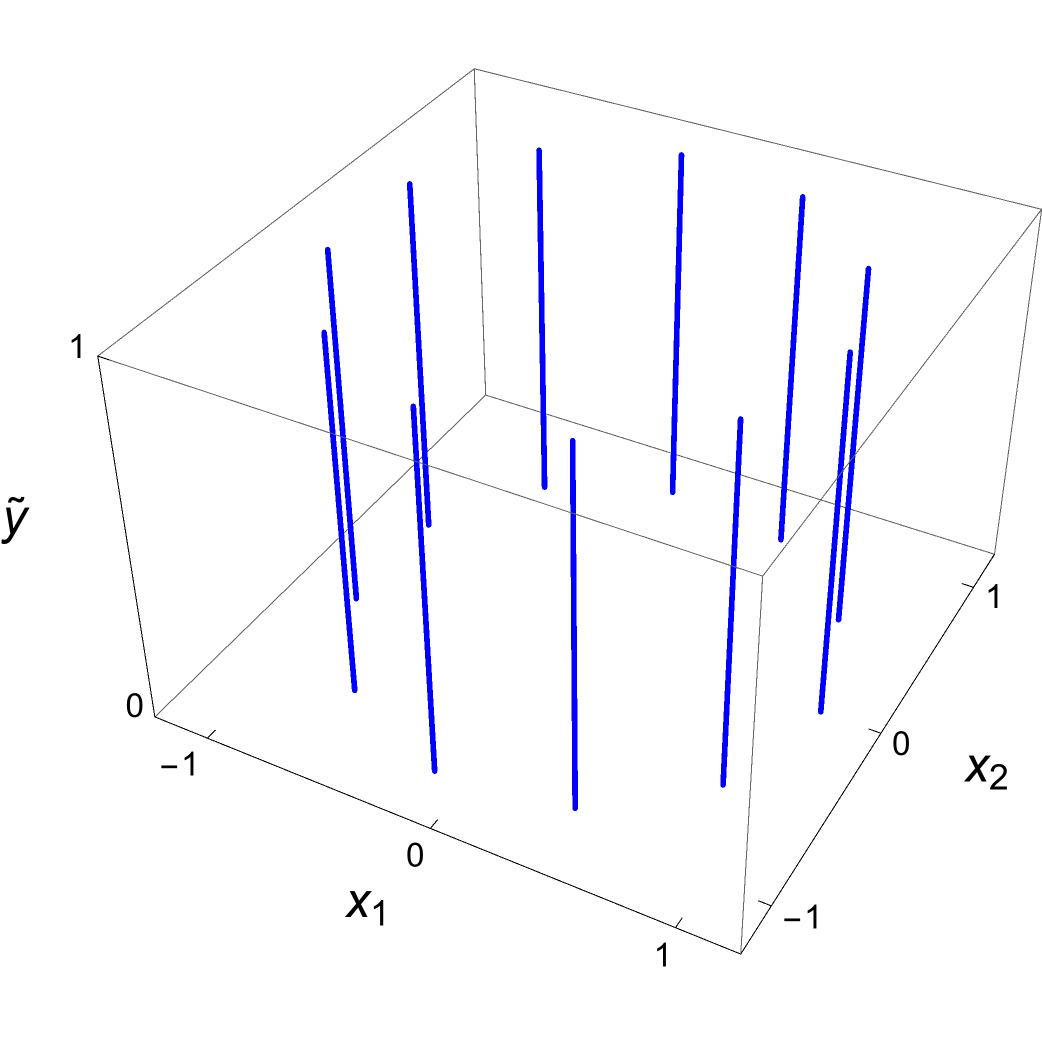}}
\setlength{\unitlength}{0.1\columnwidth}
\caption{\it 
Fivebranes separated onto the Coulomb branch at discrete points along a circle.
}
\label{fig:NS5_Coulomb.pdf}
\end{figure}

Consider the configuration~\eqref{Z5coulexact} for $k\tight=0$, which places NS5-branes in a $\IZ_{\nfive}$ symmetric arrangement around a (contractible) circle of radius $a$ in the transverse $x^1$-$x^2$ plane (see figure~\ref{fig:NS5_Coulomb.pdf}).
%
%%%%%%%%%%%%%
%%%%%%%%%%%%%

\noindent
For the analysis, it is useful to introduce the coordinates
\be
x_1+ix_2 = a\,\cosh\rho\,\sin\theta\, e^{i\phi}
~~,~~~~
x_3+ix_4 = a\,\sinh\rho\,\cos\theta\, e^{i\psi}  ~~;
\ee
carrying out the discrete sum over the source locations in~\eqref{Z5coulexact}, one finds
\begin{align}
\label{Z5circ}
Z_5 &= \frac{n_5 }{a^2(\cosh^2\rho-\sin^2\theta)_{\strut}}\,\Lambda_{n_5}
\nn\\
\Lambda_{n_5} &= \frac{\sinh(\nfive \chi)}{\cosh(\nfive \chi)-\cos(\nfive \phi)} 
=  1^{\strut} + \sum_{\ell\ne 0} e^{-n_5(|\ell|\chi+i\ell\phi)}
\end{align}
where $e^\chi = \frac{\cosh\rho}{\sin\theta}$.  The metric takes the form
\begin{align}
\label{Lambdametric}
ds^2 = -du\, dv + \nfive \Lambda_\nfive \bigl( d\rho^2 + d\theta^2\bigr)  + a^2 Z_5\Bigl(
\sin^2\!\theta\, \ch^2\!\rho \,d\phi^2 + \cos^2\!\theta\,\sh^2\!\rho\,d\psi^2\Bigr) ~.
\end{align}
In this \naive\ geometry, only the term $\Lambda_{n_5}$ is sensitive to the locations of the branes along the $\phi$ circle at $\rho=0$, $\theta=\pi/2$.  Well away from the fivebrane sources, the effects of $\Lambda_{n_5}$ are exponentially small and non-perturbative in $\alpha'$.  One might expect that it becomes important near the sources, however the curvature of the geometry is high here -- the branes are separated by a distance $2\pi/\sqrt{n_5}$ along the $\phi$ circle -- and there are thus large corrections to string propagation in this region.  So as far as supergravity (\ie\ the perturbative sigma model) is concerned, the throat of $n_5$ slightly separated NS5-branes is described by the geometry~\eqref{NS5Pmetric} smeared along the $\phi$ circle, with $\Lambda_\nfive\tight= 1$ in~\eqref{Z5circ} so that in equation~\eqref{NS5Pmetric}
\be
\label{Z5smeared}
Z_5 = \frac{n_5 }{a^2(\cosh^2\rho-\sin^2\theta)} ~,
\ee
and $Z_p \tight= Z_{ab} \tight= 0$.
This result does not mean that string theory is insensitive to the brane locations; indeed the \naive\ nonlinear sigma model based on the smeared geometry is too \naive.  But to see how this information is encoded in the worldsheet theory, we must understand the properties of the theory nonperturbatively in $\alpha'$.  For this we need an exact CFT description.%  
\footnote{In a closely related situation~\cite{Gregory:1997te,Tong:2002rq,Harvey:2005ab}, the locations of fivebranes on a {\it non-contractible} transverse circle were understood to be captured by worldsheet instanton effects.  The calculation that leads to this result is valid in the weak-coupling regime of the sigma model where the separations of the branes is much larger than the string scale.  In the decoupling limit of fivebranes on the Coulomb branch, the angular sphere has radius $\sqrt\nfive$, and the angular separation is $1/\sqrt\nfive$, and thus it is not clear that there is a reliable instanton calculation of this sort.
}
Besides the \naive\ sigma model above, three more presentations of the target space theory capture different aspects of the dynamics.  These presentations manifest the stringy physics near the branes, as they are indeed exactly solvable worldsheet CFT's.

%%%%%%%%%%%%%%%%%%%%%%%%%%%%%%%%%%%%%%%%%%
%%%%%%%%%%%%%%%%%%%%%%%%%%%%%%%%%%%%%%%%%%

\subsection{The WZW coset}
\label{sec:CosetModel}

It was shown in~\cite{Sfetsos:1998xd} that if one works with the smeared geometry~\eqref{Z5smeared}, a T-duality with respect to $\phi$ brings the metric on the transverse space to the form
\begin{align}
\label{cosetgeom}
ds^2 &= Q_5\Bigl[d\rho^2 + \tanh^2\!\rho \, \Bigl(\frac{d\tilde\phi}{\nfive}\Bigr)^2 + d\theta^2 + \cot^2\!\theta\Bigl(d\psi +\frac{d\tilde\phi}{\nfive}\Bigr)^{\!2}\,\Bigr]
\nn\\
e^{2\Phi} &= \frac{g_\eff^2}{\sin^2\!\theta\cosh^2\!\rho}
\end{align}
with $g_\eff = \sqrt{Q_5} \,\gstr/a$.  This is the target space geometry of the product of gauged WZW models
\be
\label{wzwcoset}
\left( \frac{SL(2,\IR)_{\nfive}}{U(1)} \times \frac{SU(2)_\nfive}{U(1)}\right)/\IZ_\nfive   
\ee

The locations of the fivebranes are seen in the spectrum of D-branes in the gauged WZW model%
~\cite{Elitzur:2000pq,Maldacena:2001ky,Israel:2005fn}.  
The A-type D-branes in $SU(2)/U(1)$ have an interpretation as straight lines between the points $\phi=\frac{2\pi m}{\nfive}$, a trajectory that extremizes the DBI action for D-branes stretching between NS5-branes (D2 and D4 for type IIA, D1 and D3 for type IIB).  The disk one-point function reproduces the semiclassical expectation for the tension of these branes.  D-branes resolve spacetime geometry down to the Planck scale~\cite{Shenker:1995xq} even though strings only resolve it down to the string scale, so it is not surprising that they ``know'' where the NS5-branes are located.  While the geometry~\eqref{cosetgeom} was derived by smearing the sources, the exact string description localizes the branes if the appropriate observables are considered.

Na\"ively it would still seem that the dilaton blows up at the $n_5$ distinct source locations of individual fivebranes where the full $Z_5$ in equation~\eqref{Z5coulexact} blows up.  When smeared, the second order poles smear out into a single pole along the ring at $\rho=0,\theta=\pi/2$ where the smeared $Z_5$ of~\eqref{Z5smeared} and thus the dilaton still diverges in the \naive\ sigma model.  After the above T-duality, the situation seems quite different~-- the dilaton now blows up at $\theta=0$ independent of the value of $\rho$, which can be arbitrarily far from the branes.  This feature is something of a fake however, as it arises from an application of the T-duality rules to an angular circle whose size shrinks to zero at the origin of polar coordinates, a region which is perfectly regular.  The $SU(2)/U(1)$ CFT is perfectly regular, with no large string coupling singularity; likewise for the $\sltwo/\uone$ CFT.

It appears that there is no remnant of the strong coupling singularity near any of the separated fivebrane sources, in contrast to the linearly growing dilaton of coincident fivebranes.  A likely explanation is that perturbative string worldsheets don't ``fit'' inside the throat of an individual fivebrane -- it takes a minimum  of two fivebranes to make a throat that perturbative strings can propagate into, down to an arbitrary depth at finite energy, because the angular directions of the transverse space in an NS5 throat involve
an $SU(2)$ current algebra whose level $\nfive$ 
has a lower bound of two (arising from the worldsheet fermions).  This bound is set by unitarity and supersymmetry of the worldsheet conformal field theory of perturbative strings.  In the worldsheet field theory corresponding to the background of figure~\ref{fig:NS5_Coulomb.pdf}, the transverse space is described by the coset CFT~\eqref{wzwcoset}, which is a perfectly sensible and regular theory with no obvious strong-coupling singularity.%
\footnote{String perturbation theory does break down at energies sufficient to create the {\it W-branes} that stretch between the separated NS5's~\cite{Giveon:1999px}.} 
When the fivebranes are spread out on their Coulomb branch, perturbative strings can't get close enough to them at low energies to see strong coupling, and so string perturbation theory at sufficiently low energy is self-consistent. 

Perhaps one lesson to draw from all of this is that some geometries that seem to be plagued by all sorts of pathologies, can nevertheless be quite benign when probed by perturbative strings at sufficiently low energies.  We see that separated fivebranes are an example of this phenomenon, one which we will make extensive use of in the following.

%%%%%%%%%%%%%%%%%%%%%%%%%%%%%%%%%%%%%%%%%%
%%%%%%%%%%%%%%%%%%%%%%%%%%%%%%%%%%%%%%%%%%

\subsection{The Landau-Ginsburg orbifold}
\label{sec:LGorb}

What seems to supplant the strong string coupling region near the fivebranes is an exponential tachyon wall~\cite{Giveon:1999px} formed by a condensate of winding strings in the $\frac{SL(2,R)}{U(1)}$ ``cigar'' CFT; the wall repels perturbative strings from the strong-coupling region.  It turns out that this coset sigma model is exactly equivalent to the $\cN=2$ Liouville CFT%
~\cite{Giveon:1999px,Hori:2001ax}%
\footnote{This equivalence goes by the name of {\it FZZ duality} and is based on the duality of the non-supersymmetric $\frac\sltwo\uone$ coset CFT and Sine-Liouville theory (the Liouville-dressed Sine-Gordon model) noted in unpublished work of V. Fateev, A. Zamolodchikov, and Al. Zamolodchikov.}
\be
\label{Sliou}
\cS = \frac1{2\pi}\int \!d^2z\, d^4\theta\, Y \bar Y + \mu \int \!d^2z \, d^2\theta\, \exp\bigl[-{\sqrt\nfive} \;Y\bigr]+ h.c.
\ee
(where here $\theta$ is the $\cN\tight= 2$ worldsheet supercoordinate).
This exact equivalence is a strong/weak coupling duality -- when $\nfive$ is large, so that the sigma model description is weakly coupled, the Liouville theory is strongly coupled; and when $\nfive$ is small, the Liouville theory is weakly coupled while the sigma model has string scale curvature and is strongly coupled.
Such behavior is familiar from other examples of stringy equivalence.  The Calabi-Yau/Landau-Ginsburg correspondence~\cite{Martinec:1988zu,Vafa:1988uu,Greene:1988ut} is a relation between (i) stringy sigma model geometries, and (ii) CFT's with a tachyon condensate and canonical kinetic term rather than a nontrivial metric.  Indeed, the basis of this correspondence is the exact equivalence between the $SU(2)_\nfive/U(1)$ coset model and the $\cN=2$ Landau-Ginsburg model with superpotential $\cW= X^\nfive$, and so the exact equivalence
\be
\left( \frac{SL(2,\IR)_{\nfive}}{U(1)} \!\times\! \frac{SU(2)_\nfive}{U(1)}\right)/\IZ_\nfive   
\;\equiv\;
\Bigl( \bigl[{\cN\!=\!2}~{\it Liouville}\bigr]\times \bigl[X^{\nfive}~{\it LG}\bigr]\Bigr)/\IZ_\nfive 
\ee
can be regarded as a non-compact version of the sigma model/Landau-Ginsburg correspondence~\cite{Giveon:1999zm,Giveon:1999px,Hori:2000kt}.  At the root of this equivalence is the fact that when geometrical curvatures are large, there is strong operator mixing, and no clean distinction between tachyon perturbations and metric perturbations.

The exponential Liouville wall is an example of stringy behavior not visible at the level of supergravity.    The fine-scale structure of the fivebrane source encoded in the function $\Lambda_{\nfive}$ in the sigma-model metric seems to be absent in the exact CFT, supplanted by this exponential wall and a Landau-Ginsburg potential.  The exponential wall in~\eqref{Sliou} represents the wavefunction of a winding string condensate concentrated in the region of the fivebrane sources.  Note that the imaginary part of the Liouville field $Y$ is a compact scalar parametrizing the $\tilde \phi$ circle of the coset model, and so a condensate of winding strings on this circle appears as a momentum condensate in the original variables of the \naive\ sigma model, and thus breaks the rotational symmetry of the smeared geometry to $\IZ_\nfive$ by a term that has the same asymptotic large distance scaling as the leading nontrivial correction to the geometry specified by $\Lambda$%
~\cite{Giveon:1999px,Giveon:2016dxe}.

With the exponential tachyon wall of the exact CFT, strings see structure near the sources quite different from the \naive\ geometry based on multicenter harmonic functions.  The exact scattering phase shift of the theory for probes carrying $SL(2)$ quantum numbers $(j,m,\mbar)$ is%
~\cite{Giveon:1999px,Giveon:2015cma},
\be
\label{twopoint}
\nu^{2j-1}
\biggl[\frac{\Gamma(1-\frac{2j-1}{\nfive})}{\Gamma(1+\frac{2j-1}{\nfive})}\biggr]
\biggl[\frac{\Gamma(-2j+2)\Gamma(j-m)\Gamma(j+\mbar)}{\Gamma(2j)\Gamma(-j-m+1)\Gamma(-j+\mbar+1)}\biggr]~,
\ee
where $\nu$ is a real constant which can depend on $k$ but not on $(j,m,\mbar)$.
When considering scattering states, $j=\frac12+ip$, this expression is a phase even at high radial momentum $p$; in contrast, scattering off the unsmeared fivebrane geometry would na\"ively lead to a nonzero transmission coefficient as the strings propagate down the `little throats' of the individual fivebranes;%  
\footnote{D. Kutasov, J. Lin and E. Martinec, unpublished.}
instead, an incident wave is entirely reflected.
The Liouville wall is an indicator that strings can't get into throats for individual fivebranes.

An intriguing feature of the exact expression~\eqref{twopoint} is that it factorizes into two pieces.  One piece is independent of $\nfive$, and is reproduced by minisuperspace QM approximation to string propagation in the geometry~\eqref{NS5Pmetric} with $Z_5$ the smeared harmonic function~\eqref{Z5smeared} (and $Z_p=Z_{ab}=0$); in this scattering problem, $\nfive$ is an overall scale in the metric and drops out.  The other factor depends on $p/\nfive$ and is reproduced by Liouville QM~\cite{Giveon:2015cma}.  The latter is a nonperturbative effect in the coset sigma model but perturbative in Liouville theory, while the former is a nonperturbative effect in Liouville theory and perturbative in the coset sigma model, thus illustrating the strong/weak coupling duality between the two.
Neither factor in~\eqref{twopoint} appears to be associated to scattering in the geometry~\eqref{Lambdametric}; rather, the factor associated to scattering in a geometry is reproduced using the smeared geometry~\eqref{Z5smeared}, and the effects of near-source structure seem better approximated by a tachyon wall rather than a metric with ``little throats'' for individual fivebranes.

The analysis of the $\nfive$-dependent factors in~\eqref{twopoint} by~\cite{Giveon:2015cma} in terms of an effective Liouville QM shows that strings with radial momentum $p$ can penetrate into the Liouville potential to a location of order $|Y|\sim\frac1{\sqrt\nfive}\log p\sim \sqrt{\nfive}\,\log\gstr^\eff$.
Our operating hypothesis is that the Liouville potential supplants the factor $\Lambda_\nfive$ in the geometry -- that the tachyon potential is a more accurate descriptor of the stringy structure near the sources.  Instead of the geometrical structure, a diagnostic of how close one is to the fivebrane sources is the effective value of the string coupling.  In Liouville theory this effective coupling is typically the value of the dilaton at the turning point of the classical trajectory of a probe string in the Liouville potential.
Thus one is led to a picture where high momentum strings are able to reach a point closer to the individual fivebranes than low momentum strings can reach, by an amount $\delta\rho\sim \frac1{\nfive}\delta\log p$, before being repelled back into the ambient spacetime.%
\footnote{In~\cite{Giveon:2015cma} this behavior was examined in the context of Euclidean dynamics in the $\tau$-$\rho$ plane.  The Liouville dynamics was interpreted as stringy structure near a black fivebrane horizon, and the growing time delay of returning perturbative strings as a function of momentum was seen as an indication that strings could probe a region beyond the geometrical horizon.  Here, we give a different interpretation of the same phenomenon -- as the stringy replacement for the near-source structure of slightly separated fivebranes, with the increasing time delay of scattered strings as a function of momentum (and the increasing effective string coupling at their turning point) as an indication of how close such strings get to the fivebrane sources before backscattering.}
As one gets closer to the fivebrane sources, the string coupling grows, and eventually the available energy can be used to create D-branes stretching between the fivebranes.  Such brane creation processes are effects beyond string perturbation theory, and so one may expect that in this regime, the perturbative expansion breaks down.

Both the coset sigma model and the Landau-Ginsburg description exhibit the discrete $\IZ_\nfive$ symmetry of the background; the Landau-Ginsburg presentation makes it explicit at the level of the Lagrangian, while one must dig a little deeper to see it in the coset sigma model.  The coset model does a good job of describing dynamics in the region sufficiently far from the sources, while the Landau-Ginsburg description is a more apt description close to the sources.  A similar mixed description was observed in the Calabi-Yau/Landau-Ginsburg correspondence%
\cite{Witten:1993yc,Aspinwall:1993nu,Bertolini:2013xga}.  
There, it was observed that some of the regions of Calabi-Yau moduli space where some cycles have been blown down, could be described by a `hybrid' model wherein a Landau-Ginsburg theory is fibered over a geometric base, with the LG potential varying over the base.  In this case one has a global factorization -- some directions in the target space are geometric and some are of stringy size, and the warping is not so severe that the cycle is blown down over some parts of the base and macroscopically blown up in others.  The present circumstance would seem to have more this latter sort of character -- a warping of geometry such that there is a region of space well described by geometry, and a complementary region well described by a Landau-Ginsburg model.

%%%%%%%%%%%%%%%%%%%%%%%%%%%%%%%%%%%%%%%%%%
%%%%%%%%%%%%%%%%%%%%%%%%%%%%%%%%%%%%%%%%%%

\subsection{The null-gauged WZW model}
\label{sec:nullGWZW}

A third exact formulation replaces the timelike gauging of the $SL(2,\IR)\cong SU(1,1)$ WZW model and the spacelike gauging of the $SU(2)$ WZW model, and the $\IZ_\nfive$ orbifold, instead rolling them all into the gauging of null currents in the product sigma model~\cite{Israel:2004ir}.  The target space geometry of this coset yields directly the smeared geometry~\eqref{Z5smeared}, and yet is fully equivalent as a CFT to~\eqref{wzwcoset}.  On the other hand, the presentation via null gauging leads to a simple modification that describes the supertube source~\eqref{profile1}, as we will see in section~\ref{sec:supertube}. 

In an Euler angle parametrization of group elements $(g,g')\in SU(1,1)\tight\times SU(2)$
\be
\label{Eulerangles}
g = e^{\frac i2(\tau+\sigma)\sigma_3} e^{\rho\sigma_1} e^{\frac i2(\tau-\sigma)\sigma_3}
~~,~~~~
g' = e^{\frac i2(\psi+\phi)\sigma_3} e^{i\theta\sigma_1} e^{\frac i2(\psi-\phi)\sigma_3}~~,
\ee
the level $\nfive$ WZW model is written in $\cN\!=\!1$ superspace as
\begin{align}
\label{Swzw}
\cS_{\sst \rm WZW} &= \frac{\nfive}{\pi}\int \! d^2\zhat \Bigl[ 
\Bigl(D\rho \Dbar\rho + \sh^2\!\rho \,D\sigma \Dbar\sigma - \ch^2\!\rho \,D\tau \Dbar\tau\Bigr) 
- \ch^2\!\rho\bigl(D\tau\Dbar\sigma \tight- D\sigma\Dbar\tau\bigr)
\nn\\ &\hskip 1cm +
\Bigl(D\theta \Dbar\theta + \sin^2\!\theta \,D\phi \Dbar\phi  + \cos^2\!\theta\,D\psi \Dbar\psi\Bigr)
- \cos^2\!\theta\bigl(D\phi\Dbar\psi \tight- D\psi\Dbar\phi\bigr)
\Bigr] ~,
\end{align}
where $d^2\zhat$ is the $\cN\tight=1$ superspace measure.
Following~\cite{Israel:2004ir}, we choose to gauge the $U(1)_L \!\times\! U(1)_R$ subgroup
\be
\bigl( g,g'\bigr) \longrightarrow \bigl( e^{\frac i2\lambda\sigma_3}\,g \,e^{\frac i2\xi \sigma_3},
e^{-\frac i2\lambda\sigma_3}\,g'\, e^{\frac i2\xi\sigma_3}\bigr) ~.
\ee
Because the individual $\sltwo$ and $\sutwo$ currents corresponding to this motion have equal level but opposite sign norm in their respective Killing metrics, the total currents on both left and right are null.
The gauging of a null subgroup is automatically anomaly-free, since the two-point function $\langle \cJ(z) \cJ(w)\rangle$ vanishes identically.  The gauge action for this subgroup is
\begin{align}
\label{Sgauge}
\cS_{\rm gauge} &= \frac{\nfive}{\pi}\int\!d^2\zhat\Bigl[ 
\cA\Bigl(\ch^2\!\rho \Dbar\tau + \sh^2\!\rho \Dbar\sigma \Bigr)
+\Bigl(\ch^2\!\rho D\tau - \sh^2\!\rho D\sigma \Bigr) \bar\cA 
-\frac{\ch2\rho}{2}\,\cA\bar\cA
\nn\\
&\hskip 1.5cm
-\cA\Bigl(\cos^2\!\theta \Dbar\psi - \sin^2\!\theta \Dbar\phi \Bigr)
+\Bigl(\cos^2\!\theta D\psi + \sin^2\!\theta D\phi \Bigr) \bar\cA 
-\frac{\cos2\theta}{2}\,\cA\bar\cA\Bigr] ~.
\end{align}
In the gauge $\tau=\sigma=0$, integrating out the gauge field leads to the action
\begin{align}
\label{S5Coul}
\cS &=  \frac{\nfive}{\pi}\int\!d^2\zhat \Bigl[ 
D\rho\Dbar\rho + D\theta\Dbar\theta + 
\frac{\nfive}{\Sigma}\Bigl(\sin^2\!\theta\,\ch^2\!\rho\,D\phi\Dbar\phi 
+\cos^2\!\theta\,\sh^2\!\rho\,D\psi\Dbar\psi\Bigr)
\nn\\
&\hskip 5cm + \frac{\nfive\cos^2\!\theta\,\ch^2\!\rho}{\Sigma}\Bigl(D\psi\Dbar\phi - D\phi\Dbar\psi\Bigr)\Bigr]
\end{align}
where 
\be
\label{NS5-Sigma}
\Sigma = \nfive\bigl(\cosh^2\!\rho - \sin^2\!\theta\bigr) ~.
\ee
Integrating out the gauge field also results in a nontrivial dilaton
\be
\label{NS5-Coul_dilaton}
\exp[-2\Phi] =  \frac{a^2\, \Sigma}{\gstr^2 Q_5^2} \, .
\ee
The null gauging approach thus leads directly to the NS5 geometry~\eqref{NS5Pmetric}, with $Z_5$ given by \eqref{Z5smeared} and $Z_p\!=\!Z_{ab}\!=\!0$; neither T-dualities nor orbifold quotients are required.  An analysis of the spectrum (see section~\ref{sec:spectrum} below and~\cite{Israel:2004ir}) shows agreement with that of the coset~\eqref{wzwcoset}.  Because this construction is completely equivalent to the coset orbifold~\eqref{cosetgeom}, and by extension the Landau-Ginsburg orbifold, the exact source structure is captured by this approach.

%%%%%%%%%%%%%%%%%%%%%%%%%%%%%%%%%%%%%%%%%%
%%%%%%%%%%%%%%%%%%%%%%%%%%%%%%%%%%%%%%%%%%

\section{Supertube CFTs}
\label{sec:supertube}

\begin{figure}[ht]
\centering
  \begin{subfigure}[b]{0.4\textwidth}
    \includegraphics[width=\textwidth]{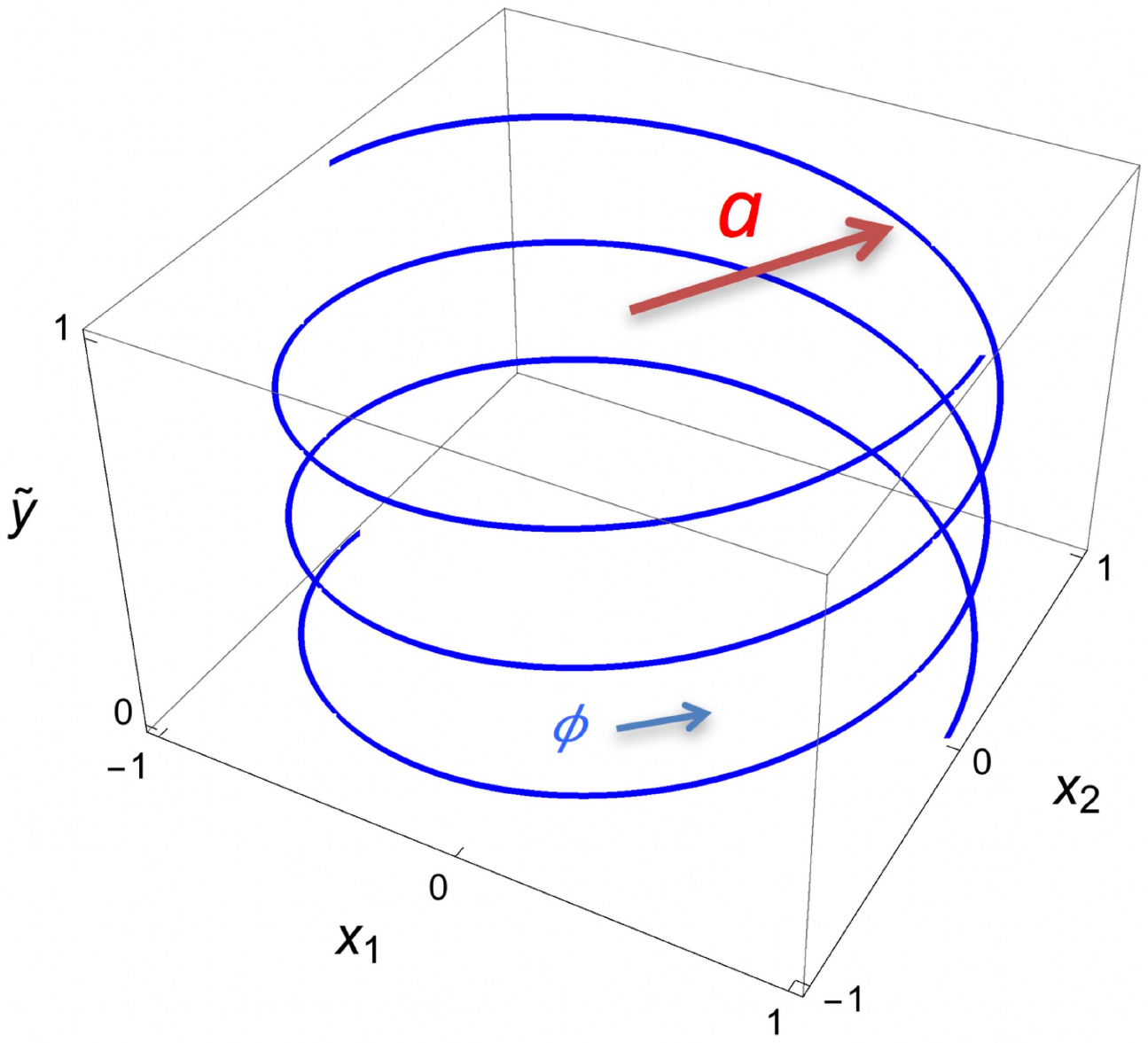}
    \caption{ }
    \label{fig:AdSchiralprimary3-1}
  \end{subfigure}
\qquad\qquad
  \begin{subfigure}[b]{0.4\textwidth}
    \includegraphics[width=\textwidth]{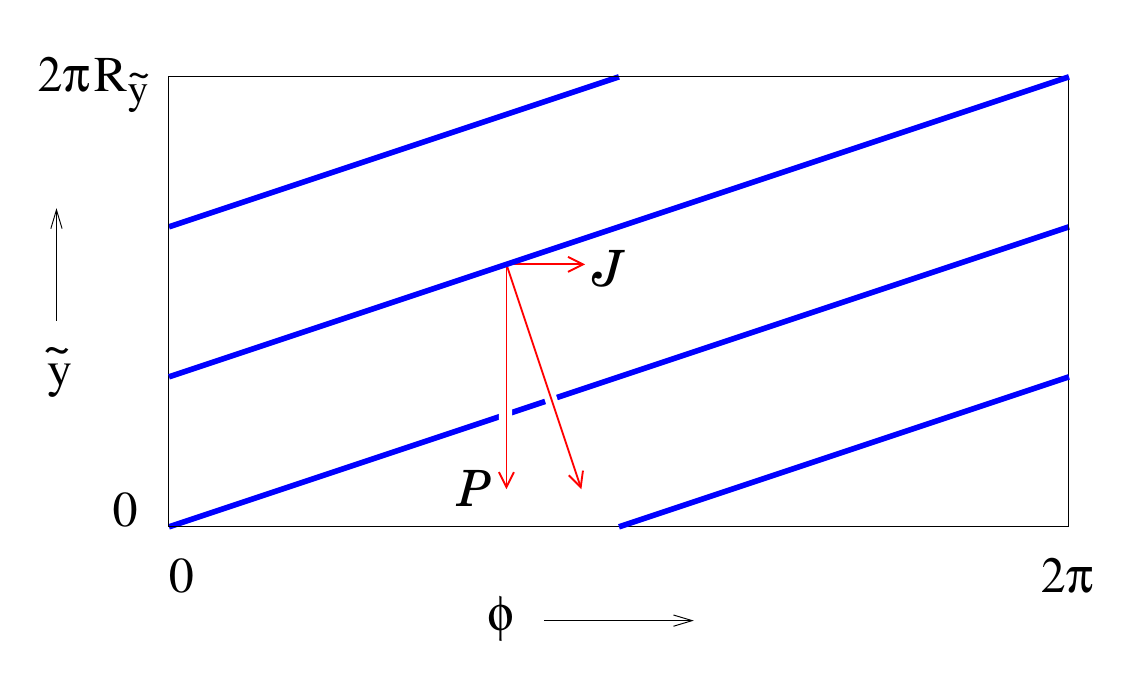}
    \vskip 0.8cm
    \caption{ }
    \label{fig:Supertube-alt3}
  \end{subfigure}
%
%\centerline{\includegraphics[width=5in]{AdSorbs-alt.pdf}}
%\setlength{\unitlength}{0.1\columnwidth}
\caption{\it 
Source for the NS5-P supertube.  (a) A single BPS source with $n_5=2$ and $k=3$; (b) Unrolling the circle of radius $\boldsymbol a$ reveals a fivebrane source moving transversely to its worldvolume on the $\ytil$-$\phi$ torus.  The fivebrane winds along the $(\nfive,k)$ cycle of this torus.
}
\label{fig:Supertube}
\end{figure}

So far, the fivebranes are not bound together, but rather are separated on the Coulomb branch of the NS5 worldvolume dynamics.  To construct a round supertube, we must spin them up into a helical profile~\eqref{profile1} compatible with the periodicity of the $\ytil$ circle and chosen to effect the desired monodromy, thereby re-introducing the discrete parameter $k$.  The result will be a set of sources wrapping a tilted cycle of the $(\ytil,\phi)$ torus, see figure~\ref{fig:Supertube}.

%%%%%%%%%%%%%%%%%%
%%%%%%%%%%%%%%%%%%
%
Having constructed the round NS5-P supertube as exact worldsheet CFT, we will then consider the T-dual description of the same CFT as an NS5-F1 supertube in section~\ref{sec:NS5F1gauging}, followed in section~\ref{sec:asymgauging} by a generalization to geometries involving spacetime spectral flow.

%%%%%%%%%%%%%%%%%%%%%%%%%%%%%%%%%%%%%%%%%%
%%%%%%%%%%%%%%%%%%%%%%%%%%%%%%%%%%%%%%%%%%

\subsection{The NS5-P supertube from tilted null gauging}
\label{sec:NS5Pgauging}

As for the NS5 branes on the Coulomb branch, we can obtain the na\"ive
geometry for the round NS5-P supertube by evaluating the
sum in \eqref{Z5coulexact} and the analogous superposition of the
harmonic functions in \eqref{NS5Pmetric}. We find
\begin{equation}
Z_5 = 1 +\frac{n_5}{a^2 \Sigma}\Lambda_{n_5} \, ,
\qquad Z_p = \frac{k^2}{n_5 R_{\tilde y}\Sigma} \Lambda_{n_5}
      \, ,  \qquad A_1 +i A_2 = \frac{i k
       }{ R_{\tilde y}\,a\,\Sigma}e^{i\phi -\chi}\Lambda_{n_5} \, ,
\end{equation}
where $\Lambda_{n_5}$ is given by \eqref{Z5circ} with the substitution
$\phi\rightarrow \phi + k (t+\tilde y)/(R_{\tilde y}n_5)$, and $\Sigma$ is given in~\eqref{NS5-Sigma}.
The geometry of smeared profiles was obtained
in~\cite{Maldacena:2000dr,Lunin:2001fv} and is given as before by setting $\Lambda_{n_5}
=1$.  Taking the decoupling limit leads to%
\footnote{We adopt a different convention as
compared to~\cite{Lunin:2001fv} in order that the momentum here and
in the following is a left-moving excitation; the two are related by
$y_{\rm there} = - y_{\rm here}$; also $\phi_{\rm there} = - \phi_{\rm here}$.}
\begin{align}
\label{smearedNS5Pmetric}
ds^2 &= -du\, dv + \nfive \bigl( d\rho^2 + d\theta^2\bigr)  + \frac{n_5^2}{\Sigma}\Bigl[
\sin^2\!\theta\, \cosh^2\!\rho \,d\phi^2 \nonumber\\
& \quad + \cos^2\!\theta\,\sinh^2\!\rho\,d\psi^2 +\frac{2 k}{n_5
  R_{\tilde y}} \sin^2\!\theta \,dv\,
  d\phi + \frac{k^2}{n_5^2 R_{\tilde y}^2} dv^2 \Bigr]+ dz_adz^a\, ,
\nonumber\\[8pt]
e^{2\Phi} & = \frac{n_5^2\gstr^2}{a^2\,\Sigma}\, \qquad B_{\psi \phi}  = \frac{n_5^2 \cos^2\theta \cosh^2\rho}{\Sigma} \,
                ,\qquad B_{\psi v} = \frac{n_5 k  \cos^2\theta}{\Rytil\,\Sigma}
            \, .
\end{align}

In order to construct the exact worldsheet CFT for this supertube, the
null-gauged WZW model discussed in section \ref{sec:nullGWZW} turns out to be the simplest description to work with.  One can of course transcribe the procedure to the traditional $G/H=\frac\sltwo\uone\!\times\! \frac\sutwo\uone$ coset description, where the deformation of the action by a current-current operator implements the spin-up.  However, it is known%
~\cite{Hassan:1992gi,Henningson:1992rn,Giveon:1993ph,Forste:2003km} 
that such current-current deformations often amount to a change in the choice of embedding of the gauge group $H$ in the WZW model for $G$.  For instance, spinning up fivebranes on the Coulomb branch -- giving them angular momentum on the transverse $\IS^3$ -- was described in the null gauging framework in~\cite{Itzhaki:2005zr}.
Such a description proves much more convenient in practice, as it leads directly to the spectrum through a modification of the BRST charge acting on the WZW Hilbert space of the numerator group $G$.  One also sees more clearly the modification of the target space geometry.

In the null-gauged WZW model above, brane sources are located at the points where the coefficient of the quadratic term $\cA\bar\cA$ vanishes and the gauging changes character, from an integration over a quadratic form to an integration over a Lagrange multiplier enforcing a constraint.  These points are the locus $\rho\!=\!0$, $\theta\!=\!\frac\pi2$, $\phi\!=\!\frac{2\pi m}{n_5}$.  Note that while these points end up looking singular from the point of view of the \naive\ target space geometry, they are perfectly regular in the full 2d worldsheet CFT.  Working on the covering space where $\ytil$ is noncompact, we want to tilt the fivebranes so that they spiral around the $\ytil$-$\phi$ cylinder at a helical pitch $\alpha$, see figure~\ref{fig:Supertube}.

%For future convenience, let us canonically normalize the $\cA\bar\cA$ term in the gauge action~\eqref{Sgauge} by rescaling $\cA\to \frac1{\sqrt\nfive}\cA$ and similarly for $\bar\cA$.
Adding a component $\alpha\, dv$ to the current being gauged on both left and right maintains the property that the current points in a null direction in $\IR^{1,1}\!\times\! SL(2,\IR)\!\times\! SU(2)$.  
The $\IR^{1,1}$ part of the action becomes
\be
\label{Stytil}
\cS_{t\ytil} = \frac{1}{2\pi} \int\!d^2\zhat \Bigl[ 
-(Du\Dbar v  + Dv \Dbar u)+ 2\alpha(\cA \, \Dbar v+\bar\cA \, Dv)\Bigr]
\ee
where $u\!=\!t\!-\!\ytil$, $v\!=\! t\!+\!\ytil$. 

Along the brane locus $\rho\!=\!0$, $\theta\!=\!\frac\pi2$, the gauge constraint now sets
\be
\nfive\, d\phi + \alpha \,dv = 0 ~,
\ee
and so we see that this shift in the gauging accomplishes our goal.
With this tilting of the null gauge direction, the quadratic term in gauge fields in~\eqref{Sgauge} is unchanged and so the location of the smeared supertube is unaffected.  The result of integrating out the gauge field is now
\begin{align}
\label{SNS5P}
\cS &=  \frac1{2\pi}\int\!d^2\zhat \Bigl[ -Du\Dbar v-Dv\Dbar u \Bigr] +
\frac{\nfive}{\pi}\int\!d^2\zhat \biggl[D\rho\Dbar\rho + D\theta\Dbar\theta 
\\[5pt]
&\hskip .5cm + 
\frac{\nfive}{\Sigma}\Bigl(\sin^2\!\theta\,\ch^2\!\rho\,D\phi\Dbar\phi +\cos^2\!\theta\,\sh^2\!\rho\,D\psi\Dbar\psi\Bigr) + \frac{\nfive\cos^2\!\theta\,\ch^2\!\rho}{\Sigma}\Bigl(D\psi\Dbar\phi - D\phi\Dbar\psi\Bigr)
\nn\\[5pt]
&\hskip 1cm + \frac{\alpha}{\Sigma}\Bigl(\sin^2\!\theta\bigl(D\phi\Dbar v+Dv\Dbar\phi\bigr) +\cos^2\!\theta\bigl(D\psi\Dbar v - Dv\Dbar \psi \bigr) \Bigr)+ \frac{\alpha^2}{\nfive\Sigma} Dv\Dbar v
\biggr]  ~,  \nn
\end{align}
with the denominator $\Sigma$ of the harmonic function in the solution
\be
\Sigma = \nfive\bigl(\cosh^2\rho-\sin^2\theta \bigr) 
\ee
again arising from the coefficient of the quadratic $\cA\bar\cA$ term in the gauged action~\eqref{Swzw}, \eqref{Sgauge}, and~\eqref{Stytil}.   In this duality frame for the supertube, this coefficient is unmodified from that of fivebranes on the Coulomb branch.  Consequently, the dilaton remains~\eqref{NS5-Coul_dilaton}.

% This result is precisely the supergravity solution for a round, noncompact NS5-P supertube~\cite{Maldacena:2000dr,Lunin:2001fv} in the fivebrane decoupling limit.%
% \footnote{We adopt a different convention as compared to~\cite{Lunin:2001fv} in order that the momentum here and in the following is a left-moving excitation; the two are related by $y_{\rm there} = - y_{\rm here}$.}

This result precisely matches the supergravity solution for the round
NS5-P supertube in the fivebrane decoupling limit
\eqref{smearedNS5Pmetric}.

So far the $\ytil$ direction is non-compact. We can now quotient by any $\ytil$ translation that is a symmetry of the source.
The choice of discrete translation amounts to a choice of how many units of discrete shift of the source around the $\phi$ circle are made before the identification along $\ytil$ (the parameter $k$ in~\eqref{profile1}).  Thus we have 
\be
\label{NS5P_params}
\alpha = \frac{k}{\Rytil}
\ee
and the quantum numbers determine the charges and other data of the solution as
\be
Q_5 =  n_5
~~,~~~~
Q_p = \frac{n_p\tilde\gstr^2}{V_4\Rytil^2} 
~~,~~~~
\alpha^2 = \frac{Q_pQ_5}{a^2} 
~~,~~~~
a = \frac{\sqrt{Q_p Q_5}\, \Rytil}{k}
~~,~~~~
J = \frac{n_pn_5}{2k} \, .
\ee
Thus the null gauging approach leads directly to the supertube geometry~\eqref{smearedNS5Pmetric}, via an almost trivial modification of the gauge group.

%%%%%%%%%%%%%%%%%%%%%%%%%%%%%%%%%%%%%%%%%%
%%%%%%%%%%%%%%%%%%%%%%%%%%%%%%%%%%%%%%%%%%

\subsection{The NS5-F1 supertube from axial gauging}
\label{sec:NS5F1gauging}

Having obtained the exact CFT for a class of NS5-P supertubes, T-duality leads to the corresponding NS5-F1 supertubes.  But T-duality is a CFT automorphism~-- a relabelling of the states that interchanges winding and momentum quantum numbers.  In the null gauged WZW model, T-duality of the $\ytil$ circle amounts to changing vector gauging to axial vector gauging.  

The change to axial gauging of the spatial direction of $\IR^{1,1}$ modifies the action~\eqref{Stytil} to
\be
\label{Sty}
\cS_{ty} = \frac{1}{2\pi} \int\!d^2\zhat \Bigl[ 
-(Du\Dbar v  + Dv \Dbar u)+ 2\alpha(\cA \, \Dbar u+\bar\cA \, Dv) - 2\alpha^2 \cA\bar\cA\Bigr] ~,
\ee
and one finds that the denominator of the harmonic function is modified to 
\be 
\Sigma = {\alpha^2}+ \nfive\bigl(\cosh^2\!\rho - \sin^2\!\theta\bigr) ~.
\ee
After eliminating the gauge field, the action becomes
\begin{align}
\label{SNS5F1}
\cS = &- \frac1{2\pi}\int\!d^2\zhat \Bigl[ Du\Dbar v+Dv\Dbar u \Bigr]+ \frac{n_5}{\pi}\int\!d^2\zhat \biggl[ 
 D\rho\Dbar\rho + D\theta\Dbar\theta \nn\\[5pt]
& 
+\frac{\cos^2\!\theta(\alpha^2 + \nfive
  \ch^2\!\rho)}{\Sigma}\Bigl(D\psi\Dbar\phi - D\phi\Dbar\psi\Bigr) 
+  \frac{\sin^2\!\theta}{\Sigma}\bigl(\alpha^2+\nfive \ch^2\!\rho\bigr)D\phi\Dbar\phi \nn\\[5pt]
& +
\frac{\cos^2\!\theta}{\Sigma}\bigl(\alpha^2+\nfive \sh^2\!\rho\bigr)D\psi\Dbar\psi
+ \frac{\alpha \cos^2\!\theta}{\Sigma}
\bigl(D\psi\Dbar u - Dv\Dbar \psi  \bigr) 
\nn\\[5pt]
& 
+\frac{\alpha \sin^2\!\theta}{\Sigma}\bigl(D\phi\Dbar u + Dv\Dbar \phi \bigr) +\frac{\alpha^2}{n_5\Sigma} Dv\Dbar u 
\biggr]  ~. 
\end{align}
This result agrees with the geometry of the NS5-F1 solution in~\cite{Lunin:2001fv} provided we identify
\be
\label{NS5F1_params}
Q_5 = n_5
~~,~~~~
Q_1 =  \frac{n_1 \gstr^2}{ V_4}
~~,~~~~
\alpha = {k R_y}
~~,~~~~
a = \frac{\sqrt{Q_1Q_5}}{k\, R_y}
~~,~~~~
J = \frac{\none\nfive}{2k}
~~,
\ee
which are just the T-dual versions of~\eqref{NS5P_params}.  
The dilaton that results from integrating out the gauge field is now
\be
\label{NS5F1_dilaton}
\exp[-2\Phi] = \frac{a^2\,\Sigma}{\gstr^2Q_5^2}  =  \frac{Q_1}{\gstr^2Q_5\alpha^2}\Bigl[{\alpha^2}+ {\nfive }\bigl(\cosh^2\rho-\sin^2\theta\bigr)\Bigr] ~~.
\ee

There are several key observations to make about the map from the CFT data to the supergravity solution data.  The first is that the parameter $\alpha$ is independent of the F1 charge $\none$.  In the NS5-F1 duality frame, the string coupling (relative to the volume $V_4$ of the compactification $\cM$ in string units) is a fixed scalar,
\be
\frac {\gstr^2}{V_4} = \frac{\nfive}{\none} ~.
\ee
The fact that $\none$ does not appear in the sigma model background, except as an additive constant in the dilaton, means that we are free to choose it arbitrarily.  Worldsheet string theory is an asymptotic expansion about vanishing string coupling, and we can achieve that while keeping the geometry fixed in string units by taking $\none\to\infty$.%
\footnote{There are subtleties in computing the fixed scalar value of the dilaton in the worldsheet formalism~\cite{Kutasov:1999xu,Porrati:2015eha}; the vertex operator $\cI$ associated to the zero mode of the dilaton is a rather subtle object requiring some care in its treatment.}

\begin{figure}[ht]
\centerline{\includegraphics[width=3.5in]{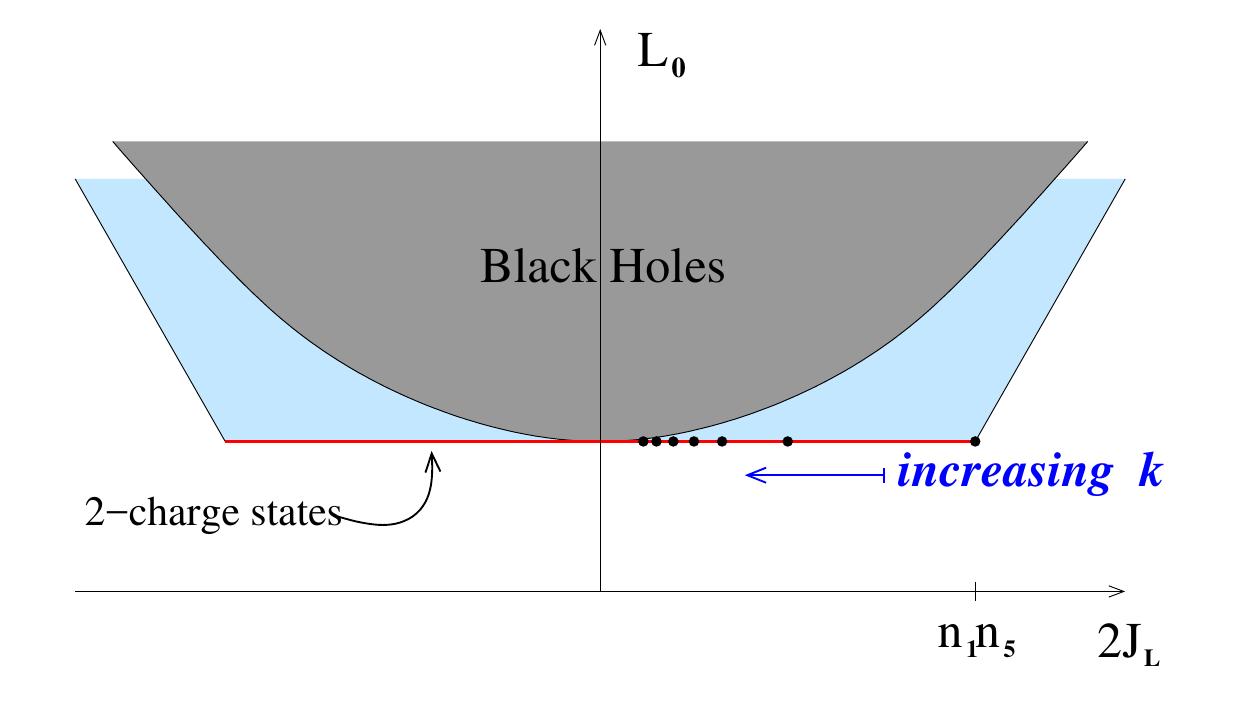}}
\setlength{\unitlength}{0.1\columnwidth}
\caption{\it 
Phase diagram for the spectrum, with the sequence of orbifold geometries indicated by black dots.  They are all 1/4-BPS states on the red line, converging toward the extremal black hole at $L_0\!=\!J_L\!=\!J_R=0$ in the limit $k\to\infty$.
}
\label{fig:Spectrum-orb}
\end{figure}

The second point is that the IR limit of the spacetime geometry in this duality frame is~\cite{Lunin:2001fv,Balasubramanian:2000rt}
\be
(AdS_3\times \IS^3)/\IZ_k \times \cM ~;
\ee
the `tilting' of the null gauging, which in the NS5-P frame determines the pitch of the fivebrane supertube spiral, in the NS5-F1 frame instead determines the order of an orbifold singularity in the target space geometry.  The orbifold decreases the angular momentum of the solution by a factor $k$, so that by increasing the integer $k$ to large values, one approaches the extremal black hole threshold along the BPS line (see figure~\ref{fig:Spectrum-orb}).
%
%%%%%%%%%%%%%
%%%%%%%%%%%%%

A third key point is that the tilt parameter $\alpha=k R_y$ controls the location of the crossover between the linear dilaton behavior of fivebrane throats and the constant dilaton of $AdS_3$; the crossover occurs at $\rho\sim \log\alpha$ when $\alpha$ is large.  The parameters $\nfive$ and $k$ are fixed for a given background, but $R_y$ is a modulus, and we can control the radial extent of the AdS region by dialing $R_y$.  The $\ads3$ region occupies the entire spacetime in the limit $R_y\to\infty$.

The fact that our construction allows such control should prove quite useful in investigating AdS dynamics; in particular, the observables in this background are not correlators of a dual CFT,%
\footnote{See~\cite{Giveon:2001up} for techniques for calculating such correlators in a string theory version of Witten diagrams~\cite{Witten:1998qj}.} 
but rather S-matrix elements of Little String Theory~\cite{Aharony:2004xn}.  Incoming scattering states from the linear dilaton region perturb scattering boundary conditions at an effective UV cutoff scale set by $\alpha$, where a matching of AdS and linear dilaton asymptotics takes place.
The perturbative string S-matrix in the NS5 background~\eqref{S5Coul} was considered in~\cite{Aharony:2004xn}, where it was shown that the amplitudes have poles associated to the discrete series representations of $\sltwo$.  It is natural to conjecture that the residues of these poles, in the decoupling limit $R_{y}\to\infty$ of the NS5-F1 supertube, describe the interactions of particles and strings in $\ads3$.

An intriguing feature is that the Virasoro symmetry which is such a powerful tool in the analysis of $AdS_3/CFT_2$ duality, is here not a spacetime symmetry of our model; rather it is an emergent symmetry in the IR, and only exact in the scaling limit $R_y\to\infty$ with the scaled energy $ER_y$ held fixed.  This latter quantity turns into energy in units of the AdS radius in the limit (times a redshift factor $1/k$ from the orbifold), which is the Virasoro conformal dimension.  It will be interesting to see if this approximate symmetry can be seen in the structure of the little string theory S-matrix.

Finally, we note that it was not in the end essential to proceed via the route we took~-- to begin with the Coulomb branch of fivebranes, spin them up to construct NS5-P supertubes, and then T-dualize the resulting CFT.  However, this approach makes manifest the geometry underlying the supertube, and the structure near the source which is hidden in the expectation values of winding operators in the NS5-F1 frame, but is easily understood as substringy structure of the brane profile in the NS5-P frame.

%%%%%%%%%%%%%%%%%%%%%%%%%%%%%%%%%%%%%%%%%%
%%%%%%%%%%%%%%%%%%%%%%%%%%%%%%%%%%%%%%%%%%

\subsection{Asymmetric null gauging and fractional spectral flow}
\label{sec:asymgauging}

Since null gauging is automatically anomaly free, there is no reason to choose vector or axial gauging for each of the factors in the numerator group $\IR^{1,1}\!\times\! SL(2,\IR)\!\times\! SU(2)$ separately.  Instead, we can allow for arbitrary coefficients
\begin{align}
\label{generalgauging}
\cJ &= l_1 J_3^{sl} + l_2 J_3^{su} + l_3\, \partial t + l_4\, \partial y
\nn\\
\bar\cJ &= r_1 \bar J_3^{sl} - r_2 \bar J_3^{su} + r_3\, \bar\partial t + r_4\, \bar\partial y
\end{align}
so long as the coefficient vectors are such that the current is null and the two-point function vanishes identically
\be
\label{nullcond}
\nfive(-l_1^2+l_2^2)-l_3^2+l_4^2 = 0 
~~,~~~~
\nfive(-r_1^2+r_2^2)-r_3^2+r_4^2 = 0 ~.
\ee
The general structure of asymmetric WZW gauging was worked out in~\cite{Bars:1991pt} (we follow the notation of~\cite{Quella:2002fk}) with the result
\be
\cS_{\rm gauge} = \frac{1}{4\pi}\int \tr\Bigl[ 2 \eL(\bar\cA) \cJ - 2\eR(\cA) \bar\cJ 
+ 2\eL(\bar\cA) g\;\! \eR(\cA) g^{-1} - \eL(\bar\cA)\eL(\cA) - \eR(\bar\cA)\eR(\cA)\Bigr] ~.
\ee
where $\eL,\eR$ are the embeddings of the left and right gauge groups into $G$.
For us, the $ \eL(\bar\cA)\eL(\cA)$ and $\eR(\bar\cA)\eR(\cA)$ terms vanish by the null condition~\eqref{nullcond}, and the coefficient of the $\cA\bar\cA$ term in the gauge action reduces to 
\be
\Sigma = \nfive\bigl(l_1r_1\cosh2\rho + l_2r_2\cos2\theta\bigr) + l_3r_3 - l_4r_4 ~.
\ee

In the NS5-P and NS5-F1 examples above, we solved the null condition~\eqref{nullcond} by having the embeddings in $\sltwo\!\times\!\sutwo$ and in $\IR^{1,1}$ be separately null.  More generally, one can have a solution such as 
\be
\label{specflowgauging}
l_1 = {\mu}\,\sinh\zeta
~~,~~~~
l_2 = {\mu}\,\cosh\zeta
~~,~~~~
l_3 = \sqrt\nfive\,\mu\,\cosh\xi
~~,~~~~
l_4 = \sqrt\nfive\,\mu\,\sinh\xi ~,
\ee
with the factor of $\sqrt\nfive$ arising from the relative normalization of the currents.
This more general gauging turns out to describe more general solutions involving fractional spectral flow in the $SU(2)$ $\cR$-symmetry of the spacetime theory~\cite{Giusto:2004id,Giusto:2012yz,Chakrabarty:2015foa}.  One has the freedom to perform independent spectral flows on left and right; however, if one wishes to preserve some supersymmetry one must restrict the spectral flow to one chirality.  Therefore, initially we will adopt the parametrization~\eqref{specflowgauging} for the left gauging, while preserving the double null form on the right
\be
\label{r_usual}
r_1=\beta ~~,~~~~ r_2=\beta ~~,~~~~ r_3 = \sqrt\nfive\, \alpha ~~,~~~~ r_4 = -\sqrt\nfive\,\alpha
\ee
that was used to generate the NS5-F1 background.

\begin{figure}[ht]
\centerline{\includegraphics[width=3.5in]{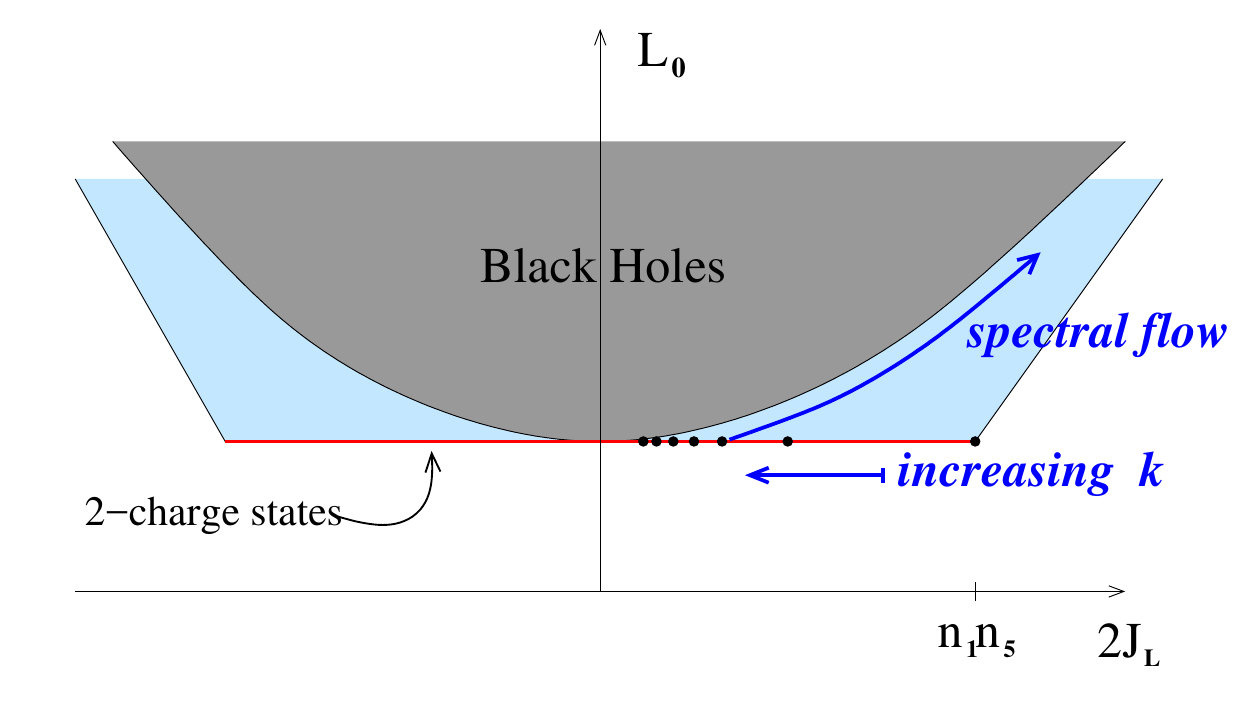}}
\setlength{\unitlength}{0.1\columnwidth}
\caption{\it 
Fractional spectral flow shifts the spectrum along the black hole threshold.
}
\label{fig:Spectrum-orb2}
\end{figure}

The spectrally flowed geometry of~\cite{Giusto:2012yz} can be written
in the fivebrane decoupling limit as follows
\begin{align}
\label{GLMTmetric}
ds^2 &= \frac{f_0}\Sigma \, \bigl(-\,dt^2 + dy^2\bigr)
+ \frac{Q_p}{\Sigma} \, \bigl(dt - dy\bigr)^2
+ Q_5 \bigl(d\rho^2 + d\theta^2 \bigr)
\nn\\
&\hskip 1cm
+ \frac{Q_5}{\Sigma} \Bigl( b^2\sinh^2\!\rho+(\coeff{\gamma_1}{\gamma_1+\gamma_2})b^2 + Q_1 \Bigr) \cos^2\!\theta\, d\psi^2
\nn\\ 
&\hskip 1cm
+ \frac{Q_5}{\Sigma} \Bigl( b^2\sinh^2\!\rho+(\coeff{\gamma_2}{\gamma_1+\gamma_2})b^2 + Q_1 \Bigr) \sin^2\!\theta\, d\phi^2
\\
&\hskip 2cm
- \frac{2Q_1Q_5}{R_y\,\Sigma}\Bigl(\gamma_1\cos^2\!\theta\,d\psi-\gamma_2\sin^2\!\theta\,d\phi\Bigr) \bigl(dt - dy\bigr)
\nn\\
&\hskip 1cm
+ \frac{2{Q_1Q_5}(\gamma_1+\gamma_2)\eta}{R_y\,\Sigma}\Bigl(\cos^2\!\theta\,d\psi+\sin^2\theta\,d\phi\Bigr) dy + dz_a\,dz^a
\nn\\[15pt]
%\end{align}
%
%\begin{align}
\label{GLMT_Bfield}
B_2 &= -\frac{Q_1}{\Sigma} \, dt\wedge dy
+\frac{Q_5\cos^2\!\theta}{\Sigma}\Bigl( b^2\sinh^2\!\rho + (\coeff{\gamma_2}{\gamma_1+\gamma_2})b^2 + Q_1 \Bigr) d\psi\wedge d\phi
\\\
&\hskip 1cm
+ \frac{ Q_pQ_5(\gamma_1+\gamma_2)\eta}{\Sigma} dy\wedge\bigl(\cos^2\!\theta\,d\psi - \sin^2\!\theta\, d\phi\bigr)
\nn\\
&\hskip 0.5cm
- \frac{{Q_1Q_5}\cos^2\!\theta}{R_y\,\Sigma}\bigl(\gamma_2\,dt + \gamma_1\,dy\bigr)\wedge d\psi
+ \frac{{Q_1Q_5}\sin^2\!\theta}{R_y\,\Sigma}\bigl(\gamma_1\,dt +\gamma_2\,dy\bigr)\wedge d\phi
\nn\\[15pt]
e^{2\Phi} &= {\frac{\gstr^2 Q_5}{\Sigma}} ~~,
\end{align}
where
\begin{align}
b^2 &= a^2\bigl(\gamma_1+\gamma_2\bigr)^2\eta
~~,~~~~
a = \frac{\sqrt{Q_1Q_5}}{R_y}
~~,~~~~
\eta = \frac{Q_1}{Q_1+Q_p}
~~,~~~~
Q_p = -a^2 \gamma_1\gamma_2
\nn\\[10pt]
f_0 &= b^2\Bigl[ \sinh^2\!\rho + \bigl(\coeff{\gamma_1}{\gamma_1+\gamma_2}\bigr)\sin^2\!\theta + \bigl(\coeff{\gamma_2}{\gamma_1+\gamma_2}\bigr)\cos^2\!\theta \Bigr]
~~,~~~~
\Sigma = f_0 + Q_1
%\\
%&= \frac{b^2}{2}\Bigl[\cosh 2\rho + \bigl(\coeff{\gamma_2 - \gamma_1}{\gamma_2 + \gamma_1}\bigr) \cos 2\theta \Bigr] 
~~.
\nn
\end{align}
With the parametrization of the gauge group
embedding~\eqref{specflowgauging}, \eqref{r_usual}, we match the above geometry via null gauging of $\IR^{1,1}\tight\times\sltwo\tight\times\sutwo$ provided the parameters map via%
\footnote{Note that the overall scale of the coefficients $l_i,r_i$ drops out of the final expression for the supergravity fields.  Until now we chose a convention where $l_1\tight=r_1\tight=1$; here and in the next subsection we switch to a convention where $l_1r_1=b^2/2$.}
\be
\label{param-map}
e^{2\xi} = \frac{Q_1}{Q_p}
~~,~~~~
e^{2\zeta} = -\frac{\gamma_2}{\gamma_1}
~~,~~~~
\alpha\mu = \frac{\sqrt{Q_1 Q_p}}{Q_5}
~~,~~~~
\beta\mu = \frac{\sqrt{Q_1Q_p}}{Q_5}\,\frac{\sinh\zeta}{\cosh\xi} ~~.
\ee
The parameters $\gamma_1$, $\gamma_2$ and the charge ratio $Q_p/Q_1$ were determined in~\cite{Giusto:2012yz} in terms of a spectral flow parameter $s$ and the orbifold order $k$ via%
\be
\label{specflow-params}
\gamma_1 = -\frac{s}{k} ~~,~~~~ \gamma_2 = \frac{s+1}{k}  
~~,~~~~
\frac{Q_p}{Q_1} = \frac{s(s+1)\nfive}{k^2 R_y^2}
~,
\ee
with the additional requirement that $k$ divides $s(s+1)$.  Note that
the parameter $\zeta$ of the gauging~\eqref{specflowgauging} encodes
the spectral flow parameter $s$, with the limit $s\to 0$ forcing
$Q_p\to 0$ as well through the null condition~\eqref{nullcond}, which
implies that $Q_p \propto s$; in this way we recover the unflowed
NS5-F1 geometry~\eqref{SNS5F1}.  Flow to nonzero $s$ moves the
spectrum along the black hole threshold in the $J_L$-$L_0$ plane, see
figure~\ref{fig:Spectrum-orb2}.  Note also that once again, the
geometry is independent of the F1 charge quantum $n_1$, which only
appears in the string coupling through the fixed scalar
condition~\eqref{NS5F1_dilaton}.  In Appendix \ref{sec:CTCs} we show
that there is a slightly more general choice of boost parameters that reproduces
 a one-parameter family of supersymmetric solutions
considered in \cite{Giusto:2004ip}. Generically these solutions
contain a region with closed time-like curves. The
map \eqref{param-map} selects the only solution among this family
which is free of pathologies.

%%%%%%%%%%%%%
%%%%%%%%%%%%%

\subsection{General spectral flow and non-supersymmetric solutions}

So far we restricted the more general choice~\eqref{specflowgauging} of null gauging to the left-movers in order to preserve spacetime supersymmetry; clearly if we had allowed the more general parametrization~\eqref{specflowgauging} of the null gauging for the right-movers as well, we would recover a solution with spectral flows in both chiralities of the spacetime $\cR$-symmetry.
This generalization has been considered in efforts to construct non-supersymmetric microstate geometries%
~\cite{Jejjala:2005yu,Chakrabarty:2015foa}.
In the fivebrane decoupling limit, the metric of the left/right flowed geometry takes the form
%~\cite{Jejjala:2005yu,Chakrabarty:2015foa}
\begin{align}
ds^2 &= \frac{f_0}{\Sigma}\bigl(-\,dt^2+dy^2\bigr) +\frac{M}{\Sigma}\bigl( c_p \,dt - s_p \,dy\bigr)^2
+ Q_5 \bigl( d\rho^2+d\theta^2\bigr)
\nn\\[5pt]
&\hskip .8cm 
+\frac{Q_5}{\Sigma}\Bigl( \bigl(r_+^2-r_-^2\bigr)\cosh^2\rho + r_-^2 + a_2^2 + Ms_1^2\Bigr)\sin^2\theta\, d\phi^2
\\[5pt]
&\hskip .8cm 
+\frac{Q_5}{\Sigma}\Bigl( \bigl(r_+^2-r_-^2\bigr)\sinh^2\rho + r_+^2 + a_1^2 + M s_1^2\Bigr)\cos^2\theta\, d\psi^2
\nn\\[5pt]
&\hskip .8cm 
+\frac{2\sqrt{M Q_5}\cos^2\theta}{\Sigma}\Bigl[
  \bigl(a_1c_1c_p-a_2s_1s_p\bigr) \, dt + \bigl(a_2s_1c_p-a_1c_1s_p\bigr)\, dy\Bigr]\,d\psi
\nn\\[5pt]
&\hskip .8cm 
-\frac{2\sqrt{M Q_5}\sin^2\theta}{\Sigma}\Bigl[ \bigl(a_2c_1c_p-a_1s_1s_p\bigr) \, dt+ \bigl(a_1s_1c_p-a_2c_1s_p\bigr)\, dy\Bigr]\,d\phi
+\bigl(dz_adz^a\bigr)
\nn
\end{align}
where  $c_i\tight=\cosh \delta_i$, $s_i\tight= \sinh\delta_i$, $Q_i = M s_ic_i$ for $i=1,p$; 
\begin{align}
r_\pm^2 &= \frac12 \[ \bigl(M-a_1^2-a_2^2\bigr)\pm \sqrt{\bigl(M-a_1^2-a_2^2\bigr)^2 - 4a_1^2a_2^2}\, \]
= -a_1a_2\Bigl(\frac{s_1s_p}{c_1c_p}\Bigr)^{\pm1}
\nn\\[5pt]
M &= a_1^2 + a_2^2 -a_1a_2\frac{c_1^2c_p^2+s_1^2s_p^2}{c_1c_ps_1s_p}
    ~; \label{rpmM}
\end{align}
and
\begin{align}
f_0 &=  \half \Bigl[ (r_+^2-r_-^2)\cosh 2\rho + (a_2^2 -  a_1^2)\cos 2\theta
+ r_+^2+r_-^2 + a_2^2+ a_1^2  \Bigr]  \, ,
\nn\\[8pt]
\Sigma &= f_0 + M s_1^2  ~.
\end{align}
The radial coordinate $\rho$ is related to the BTZ-like Schwarzschild coordinate $r$ used in~\cite{Jejjala:2005yu,Chakrabarty:2015foa} via
\be
\sinh^2\rho = \frac{r^2-r_+^2}{r_+^2-r_-^2} ~.
\ee
The parameters $r_\pm$ can formally be thought of as the locations of ``virtual'' inner and outer horizons (virtual because $r_\pm^2<0$).

Nonsingularity of the geometry at $\rho=0$ up to an orbifold singularity of order $k$ leads to the conditions
\be
\frac{s_pc_p\, k\Ry}{ a_1c_1c_p-a_2s_1s_p } = n\in \IZ
~~,~~~~
\frac{s_pc_p\, k\Ry}{ a_2c_1c_p-a_1s_1s_p } = - m\in \IZ  ~.
\ee
%where
%\be
%\Ry = \frac{M}{k}\,\frac{s_1c_1(s_1c_1s_pc_p)^{1/2}}{\sqrt{a_1a_2}\,(c_1^2c_p^2-s_1^2s_p^2)} 
%\ee
%is the radius of the $y$ circle.  
One also has the further relations
\be
M = mn \, a_1a_2\Bigl(\frac{c_1c_p}{s_1s_p}-\frac{s_1s_p}{c_1c_p}\Bigr)^2
~~,~~~~
a_1a_2 = \frac{M Q_5}{k^2\Ry^2} \, \frac{s_1^3c_1^3s_pc_p}{(c_1^2c_p^2-s_1^2s_p^2)^2} ~~.
\ee
It is useful to introduce dimensionless parameters
\be
\jhat^2 = \frac{a_2}{a_1} 
~~,~~~~ 
\shat^2 = \frac{s_1 s_p}{c_1 c_p} ~,
\ee
in terms of which $m$, $n$ can be expressed as follows
\be
\label{jsmn}
m+ n = \frac{\jhat  - \jhat^{-1}}{\shat - \shat^{-1}} 
~~,~~~~
m- n = \frac{\jhat + \jhat^{-1}}{\shat + \shat^{-1}} ~.
\ee

This modification is of course quite straightforward in our setup.  
In the null-gauged WZW model, we choose the generic parametrization of the gauge current~\eqref{specflowgauging} on both left and right 
\begin{align}
\label{generalparamsLR}
l_1 &= {\mu}\,\sinh\zeta\, , &
l_2 &= {\mu}\,\cosh\zeta\, ,&
l_3 &= \sqrt\nfive\,\mu\,\cosh\xi\, , &
l_4 &= \sqrt\nfive\,\mu\,\sinh\xi ~,&
\nn\\
r_1 &= {\mu}\,\sinh\bar\zeta \, , &
r_2& =- {\mu}\,\cosh\bar\zeta\, , &
r_3 &= \sqrt\nfive\,\mu\,\cosh\bar\xi \, , &
r_4 &= - \sqrt\nfive\,\mu\,\sinh\bar\xi \, .&
\end{align}
The expectation is that
$\zeta,\bar\zeta$ now encode the left and right spectral flows, and that the rapidity parameters $\xi,\bar\xi$ encode the shifts in the left- and right-moving momentum rather than the total momentum; and indeed this is the pattern.  
The coefficient of the $\cA\bar\cA$ term is $\Sigma$; one finds
\be
\Sigma_{\sst \rm GWZW} = Q_5\mu^2 \bigl[ 2\,\sh\zeta\,\sh\bar\zeta\cosh^2\!\rho
+ 2\,\ch\zeta\ch\bar\zeta\sin^2\theta - \ch(\zeta\tight+\bar\zeta)
+ \ch(\xi\tight+\bar\xi) \Bigr] ~,
\ee
and one also reads off the charges $Q_1,Q_p$ from $G_{ty}$ and $B_{ty}$ 
\be
Q_1 = Q_5\,\mu^2\sinh(\xi\tight+\bar\xi)
~~,~~~~
Q_p = Q_5\,\mu^2 \sinh(\bar \xi\tight-\xi) ~,
\ee
which yields the relation between the boost parameters $\delta_1,\delta_p$ for the charges $Q_1, Q_p$ and the boost parameters $\xi,\bar\xi$ in the null vectors
\be
\delta_1 = \frac{\xi+\bar\xi}2 
~~,~~~~
\delta_p = \frac{\bar \xi-\xi}2  
\ee
as well as the relation $M=2Q_5\mu^2$. 
Matching the various expressions, one also finds that
\begin{align}
r_+^2-r_-^2 = 2Q_5\mu^2 \sinh\zeta\,\sinh\bar\zeta
~~&,~~~~
a_1^2-a_2^2 = 2Q_5\mu^2\cosh\zeta\,\cosh\bar\zeta
%\nn\\[5pt]
%r_-^2+a_2^2 = \frac12 Q_5\,\mu^2 \cosh(\zeta\tight-\bar\zeta)
%~~&,~~~~
%r_+^2+a_1^2 = \frac12 Q_5\,\mu^2 \cosh(\zeta+\bar\zeta)
\end{align}
and the spectral flow parameters are determined as
\be
\coth\zeta = m+n = 2s+1
~~,~~~~
\coth\bar\zeta = m-n = 2\bar s+1  ~~,
\ee
which in the supersymmetric limit reproduces~\eqref{param-map}-\eqref{specflow-params}.  The angular momenta of the solution are
\be
J_L = \frac{n_1n_5}{2}\,\frac{2s+1}{k}
~~,~~~~
J_R = \frac{n_1n_5}{2}\,\frac{2\bar s+1}{k} ~~.
\ee

Thus we see that the non-supersymmetric solutions of~\cite{Jejjala:2005yu,Chakrabarty:2015foa} fit neatly into our general framework, and complete the set of models one can obtain from null gauging of $\IR^{1,1}\tight\times\sltwo\tight\times\sutwo$ using currents of the form~\eqref{generalgauging}, \eqref{generalparamsLR}.  Spectral flow in the spacetime CFT corresponds to a particular deformation of the null vector from a double-null form for the unflowed solution, to a generic null vector parametrized by rapidities $\xi,\zeta$ and $\bar\xi,\bar\zeta$ on the left and right, respectively.  The spectral flow parameters and charges are simple functions of the parameters of the gauging.

To summarize, the formalism of gauging a null current algebra $H$ in the WZW model for a group $G$ of signature $(10,2)$ -- or more generally $(9\!+\!d,1\!+\!d)$ -- yields a variety of interesting spacetime backgrounds that have been considered in the literature~\cite{Lunin:2001fv,Jejjala:2005yu,Giusto:2012yz,Chakrabarty:2015foa}, and suggests generalizations that connect to a much larger class of geometries, including%
~\cite{Elitzur:1998mm,Itzhaki:2005tu,Giveon:1999zm,Eguchi:2004ik,Eguchi:2003yy,Giveon:1999jg,Giveon:2003ku}.
The choice of embedding of the null gauging allows us to construct solvable worldsheet CFT's for supertube backgrounds that interpolate between little string theory in the UV and AdS geometries in the IR.  The fact that we have control over the string background at the nonperturbative level in $\alpha'$ allows many issues regarding AdS/CFT duality to be addressed fully in string theory rather than being limited to the supergravity approximation.  In particular, we have seen that the structure and effects of winding string condensates seen in fivebrane backgrounds on the Coulomb branch, carry over to the AdS limit and reveal properties of $\ads3$ string theory that were heretofore hidden by the smearing of sources used to construct the effective geometry in supergarvity.

%%%%%%%%%%%%%%%%%%%%%%%%%%%%%%%%%%%%%%%%%%
%%%%%%%%%%%%%%%%%%%%%%%%%%%%%%%%%%%%%%%%%%

\subsection{Variations on a theme}
\label{sec:variants}

Clearly the above models are but the first specimens in a large zoology of $G/H$ gauged WZW models where $G$ is noncompact, having several timelike directions, and $H$ gauges away all but one of them.  The variety of embeddings of $H$ into $G$ will lead to a variety of solutions to string theory, whose spacetime interpretation should be interesting to explore.  A few examples:

%%%%%%%%%%%%%%%%%%%%%%%%%%%%%%
\subsubsection{More general subgroups of $\IR^{1,1}\tight\times\sltwo\tight\times\sutwo$}

Our choice of left and right null generators of $\IR^{1,1}\tight\times\sltwo\tight\times\sutwo$ to gauge was certainly not the most general.  For instance, we could change vector gauging of $J_3^\susup,\bar J_3^\susup$ to axial vector gauging; this would orient the supertube circle in the $x^3$-$x^4$ plane instead of the $x^1$-$x^2$ plane.  One could choose entirely different $U(1)$ subgroups of $\sutwo$ on left and right, and similarly in $\sltwo$ one could make a different choice of generator within the elliptic conjugacy class (the timelike generators) on left and right.  These presumably correspond to supertubes having different angular orientations, or undergoing geodesic motion in the throat, though we have not checked the details.

%%%%%%%%%%%%%%%%%%%%%%%%%%%%%%
\subsubsection{Embedding $\ads3$ into Little String Theory}

Working from the AdS direction, all the $\ads3\times\cK$ examples of~\cite{Giveon:1999jg,Giveon:2003ku}, where $\cK$ can be decomposed as $U(1)\times(\cK/{U(1)})$ with $\cK/U(1)$ an $\cN\!=\!2$ SCFT, should be amenable to a treatment along the above lines.  Specifically,
\begin{enumerate}[(i)]
\item
Write the $\ads3$ factor as $\IR_t\times \sltwo/\uone$ (the parafermion decomposition of $\sltwo$ current algebra); the model thus is an orbifold of $\IR^{1,1}\tight\times \frac\sltwo\uone \tight \times \frac\cK\uone$.
\item
Rewrite the model as $(\IR^{1,1}\times\sltwo \times \cK)/(U(1)_L\times U(1)_R)$, where the denominator group is built out of null generators of the original denominator group $U(1)\times U(1)$.
\item
Tilt the gauging into another null direction in $\IR^{1,1}\tight\times \sltwo \tight \times \cK$.
\end{enumerate}
The choice of embedding of $H$ into $G$ leads to a family of supertubes with the original $\ads3$ spacetime as an IR limit.

%%%%%%%%%%%%%%%%%%%%%%%%%%%%%%
\subsubsection{Holographic RG flows of Little String Theory to $\ads3$ }

Working from the little string theory side, all the solvable examples in%
~\cite{Giveon:1999zm,Eguchi:2003yy,Eguchi:2004ik} 
fit into the above framework, since they all have $\IR^{1,1}\times\frac{\sltwo}{U(1)}\times \frac{G}{H}$ as part of the worldsheet CFT.  Replacing again the denominator group $U(1)\times H$ by a group $H'$, one of whose generators is a null generator in $U(1)\times H$, leads to a supertube state of little string theory compactified on the relevant background.  

For instance, the four-dimensional examples of~\cite{Giveon:1999zm} involve $\IR^{3,1}\times \frac{\sltwo}{U(1)}$ times a product of $\frac{\sutwo}{\uone}$ coset models, and were identified with NS5-branes wrapped on a Riemann surface (for instance the $\IS^2$ of the resolved conifold), leading to a low-energy dynamics on the branes which is the Seiberg-Witten solution to 4d $\cN\!=\!2$ gauge theory, near an Argyres-Douglas point.  One can then play the supertube game and compactify along the supertube spiral to generate a family of theories which have little string theory on NS5-branes wrapping the resolution of a Calabi-Yau singularity in the UV, and $\ads3$ dynamics in the IR.

%%%%%%%%%%%%%%%%%%%%%%%%%%%%%%
\subsubsection{I-brane supertubes }

An object known as the {\it I-brane} arises when two stacks of $\nfive$ and $\nfive'$ NS5-branes intersect over $\IR^{1,1}$~\cite{Itzhaki:2005tu}.  If one adds $\none$ fundamental strings along the intersection, the IR limit is described by string theory on $\ads3\!\times\! \IS^3\!\times\! \IS^3\!\times\! \IS^1$ with large $\cN\!=\!4$ superconformal symmetry in spacetime~\cite{Elitzur:1998mm}.  The I-brane intersection has a Coulomb branch where one separates each stack slightly in its transverse space.  There is a straighforward extension of the method of null gauging to this case, with $G = \IR^{1,1} \!\times\! \sltwo\!\times\!\sltwo\! \times\! \sutwo\!\times\! \sutwo$ and $H_{L,R}=U(1)\!\times\! U(1)$ where both factors $H_{L,R}$ have a null embedding in $G$.  In this way one will obtain an I-brane supertube whose geometry interpolates between the I-brane in the UV and $\ads3\!\times\! \IS^3\!\times\! \IS^3\!\times\! \IS^1$ in the IR.

%%%%%%%%%%%%%%%%%%%%%%%%%%%%%%
\subsubsection{Other groups }

Various related constructions using gauged WZW models for noncompact forms of Lie groups were considered in%
~\cite{Bars:1990rb,Balog:1990mu,Bars:1993jt,Klimcik:1994wp,Giveon:1995as}.  It would be interesting to understand their spacetime interpretation in terms of fivebranes, and to construct supertubes for them.

%%%%%%%%%%%%%%%%%%%%%%%%%%%%%%
\subsubsection{$\ads3$ orbifolds }

As noted in section~\ref{sec:Supertubes} and~\cite{Martinec:2001cf}, when $k$ and $n_5$ have a common divisor $p$, the prescription of~\cite{Lunin:2001fv} to construct a smeared source with a single strand on the $\nfive$-fold covering space of the $\ytil$ circle leads to $p$ supertube strands of $n_5/p$ coincident fivebranes.  This theory is singular, because now strings can propagate down the throats of any of the $p$ strands to a region of strong coupling.  As noted in~\cite{Martinec:2002xq}, a nonsingular theory separates the strands as in figure~\ref{fig:NS5-P_3strand-k15n6.pdf}, however there are $4p$ moduli for the relative separations of the $p$ strands, very analogous at the $(\ads3\times\IS^3)/\IZ_p$ orbifold point to the blowup modes of an ALE orbifold -- there is a triplet of modes that translate the branes in the three directions transverse to the ring of radius $a$, and a singlet mode that slides the branes along the ring.  And just like an ALE space, moving a finite distance in the singlet direction leads to coincident fivebranes and a singular worldsheet CFT.  In the NS5-F1 frame, this is because the local structure at the ring is that of $p$ nearly coincident KK monopoles, and this singlet modulus controls the stringy effective size of the homology two-sphere comprised of the KK circle fibered over the interval between the monopole centers.  The singlet moduli control the lengths of these intervals and hence the volumes of the two-spheres, which are T-dual to the separations of the fivebranes in the NS5-P frame.

The orbifold here is different from the one considered in~\cite{Martinec:2001cf,Martinec:2002xq}.  In those works, the angular directions $(\psi,\phi)$ on the $\IS^3$ are always transverse to the fivebranes, and so necessarily the order $p$ of a $\IZ_p$ quotient must divide $\nfive$ in order to maintain integrality of the magnetic $H_3$ flux of the solution.  For the supertube (say in the NS5-P frame), the orientation of the fivebrane is tilted into the $\ytil$ direction; the angular directions $(\psi,\phi)$ are no longer purely transverse to the branes, and so there is no such restriction on the order of the orbifold in the NS5-F1 frame after T-duality in $\ytil$.

%%%%%%%%%%%%%%%%%%%%%%%%%%%%%%
\subsubsection{The NS sector and the Rohm twist}

The $\ads3$ background that arises in the IR decoupling limit is naturally in the Ramond sector of the spacetime CFT, since there are unbroken global supersymmetries in the target spacetime.  The Neveu-Schwarz sector can be implemented through the use of twisted boundary conditions, whereby the identification $y\sim y+ 2\pi R_y$ is accompanied by a $2\pi$ rotation in a transverse direction.  This twist forces all the vertex operators for spacetime fermions to carry half-integer momentum/winding around the $y$ circle.  These boundary conditions were originally studied in~\cite{Rohm:1983aq}, where it was observed that for $R_y^2< \frac12$ there is a winding tachyon in the spectrum and the background is unstable.  Since we are interested in the large $R_y$ limit, this is not a problem; nor is there a potential generated at loop level that would cause $R_y$ to become dynamical and run away to either large or small values, because the asymptotic spacetime geometry is frozen by the vanishing dilaton at spatial infinity.  We might however expect to find a condensate of this winding tachyon in the core of the geometry where the $y$ circle pinches off.

%%%%%%%%%%%%%%%%%%%%%%%%%%%%%%%%%%%%%%%%%%
%%%%%%%%%%%%%%%%%%%%%%%%%%%%%%%%%%%%%%%%%%

\section{Perturbative string spectrum}
\label{sec:spectrum}

%%%%%%%%%%%%%%%%%%%%%%%%%%%%%%%%%%%%%%%%%%
%%%%%%%%%%%%%%%%%%%%%%%%%%%%%%%%%%%%%%%%%%

The spectrum of our model follows from by now standard results on the spectra of the supersymmetric $\sltwo$~\cite{Maldacena:2000hw} and $\sutwo$~\cite{Zamolodchikov:1986bd} WZW models, and the BRST formalism%
~\cite{Karabali:1988au,
Karabali:1989dk,
Hwang:1993nc} 
for gauging a subgroup~\cite{Goddard:1984vk}.  To begin, let us briefly recall the operator content of these current algebra conformal field theories.

%%%%%%%%%%%%%%%%%%%%%%%%%%%%%%
\subsection{Current algebra and parafermions}

As discussed in Appendix \ref{sec:WZWconventions}, the supersymmetric $\sutwo$ level $\nfive$ current algebra highest weight operators $\Phihat^\susup_{j'm'\mbar'}$ have a {\it parafermion} decomposition under the current $J^\susup_3$~\cite{Fateev:1985mm,Gepner:1986hr,Gepner:1987qi}%
\footnote{Our notation here largely follows~\cite{Martinec:2001cf}, see also~\cite{Giveon:2015raa}.  See the Appendix for details.}
\be
\label{supf}
\Phihat^\susup_{j'm'\mbar'} = \Psihat^\susup_{j'm'\mbar'} \,\exp\Bigl[i\frac2{\sqrt\nfive}\Bigl(m'\cY_{\!\susup}+\mbar'\bar\cY_{\!\susup}\Bigr)\Bigr] ~.
\ee
where $\cY_{\!\susup},\bar\cY_{\!\susup}$ bosonize the (total) left- and right-moving currents $J_3^\susup, \bar J_3^\susup$, and $m',\mbar'$ are the corresponding quantum numbers of the zero mode representation.  The conformal dimension $h\!=\!{j'(j'\!+\!1)}/\nfive$ of the $\sutwo$ primary $\Phihat_{j'm'\mbar'}$ decomposes as
\be
\label{supfspec}
h(\Psihat^\susup_{j'm'\mbar'}) = \frac{j'(j'+1)-(m')^2}{\nfive} 
~~,~~~~
\bar h(\Psihat^\susup_{j'm'\mbar'}) = \frac{j'(j'+1)-(\mbar')^2}{\nfive} 
\ee
with the rest made up by the dimension of the boson exponential.
Unitarity restricts the spins $j'$ of the underlying bosonic current
algebra to the allowed range
\be
\label{su2 reps}
j' = 0,\frac12,\dots,\frac{\nfive}{2}-1 ~.
\ee

The supersymmetric $\sltwo$ level $\nfive$ current algebra highest weight operators also have a super-parafermion decomposition.  Let 
$\cY_\slsup,\bar\cY_\slsup$ bosonize the (total) left- and right-moving timelike currents $J_3^\slsup, \bar J_3^\slsup$; then one can write
\be
\Phihat^\slsup_{jm\mbar} = \Psihat^\slsup_{jm\mbar} \,\exp\Bigl[i\frac2{\sqrt\nfive}\Bigl(m\cY_{\!\slsup}+\mbar\bar\cY_{\!\slsup}\Bigr)\Bigr] ~.
\ee
The conformal dimension $h\!=\!-{j(j-1)}/\nfive$ of the $\sltwo$ primary decomposes as
\be
\label{slpfspec}
h(\Psihat^\slsup_{jm\mbar}) = \frac{-j(j-1)+m^2}{\nfive} 
~~,~~~~
\bar h(\Psihat^\slsup_{jm\mbar}) = \frac{-j(j-1)+\bar m^2}{\nfive} 
\ee
with the rest made up by the dimension of the boson exponential.  

The dimensions of the fields transform under {\it spectral flow}, in which the $J_3$ charges are shifted by an amount 
\be
m \to m+\coeff12 \nfive w 
~~,~~~~
m' \to m'+\coeff12 \nfive w' 
\ee
and the dimensions of the primary fields change to
\begin{align}
h\bigl(\Psihat^{\slsup(w,\bar w)}_{jm\mbar}\bigr) &= -\frac{j(j-1)}{\nfive} - mw - \frac{\nfive}{4}  w^2
\nn\\
h\bigl(\Psihat^{\susup(w',\bar w')}_{j'm'\mbar'}\bigr) &= \frac{j'(j'+1)}{\nfive} + m'w' +\frac{\nfive}{4} (w')^2 ~;
\end{align}
and similarly for the right-movers.

%%%%%%%%%%%%%%%%%%%%%%%%%%%%%%
\subsection{Fivebranes on the Coulomb branch}

To begin, let us consider the null gauged WZW model that describes NS5-branes on the Coulomb branch, following~\cite{Israel:2004ir}.
Gauging the $\sltwo\!\times\!\sutwo$ subgroup 
\be
\label{NS5Coulgauging}
\cJ = J_3^\slsup + J_3^\susup~~,~~~~ \bar\cJ = \bar J_3^\slsup - \bar J_3^\susup
\ee
amounts to selecting from the affine $G=\sltwo\!\times\!\sutwo$ Hilbert space those states that are highest weight under the action of the currents $\cJ,\bar\cJ$ of the gauge group $H$.  This sets 
\begin{align}
\label{NS5coul constraint}
(m+\coeff12\nfive w) + (m'+\coeff12\nfive w') &= 0
\nn\\
(\mbar+\coeff12\nfive \bar w) - (\mbar'+\coeff12\nfive\bar w')  &= 0
\end{align}
and it was shown in~\cite{Israel:2004ir} 
(see also~\cite{Giveon:2015raa}) that the resulting spectrum~\eqref{supfspec}, \eqref{slpfspec} satisfying this constraint matches that of the coset orbifold~\eqref{wzwcoset}.  Since the null current has vanishing two-point function, it is a first class constraint and there are no subtleties when imposing it on the Hilbert space compared to usual gaugings of subgroups~\cite{Hwang:1993nc,Bjornsson:2007ha}.  The constraint imposes that all excitations in the conjugate null direction vanish; then one has the freedom to shift away any dependence on the null current itself via a gauge transformation.  The end result is that both scalars $\cY_\susup,\bar\cY_\susup$ and $\cY_\slsup,\bar\cY_\slsup$ are removed from the theory and one is left with the product of super-parafermion theories, with their quantum numbers correlated in the way that results from the $\IZ_\nfive$ orbifold. 

For example, consider a vertex operator that combines a primary field $\Phihat^\susup_{j'm'\mbar'}$ and spectrally flowed primary $\Phihat^{\slsup(\omega,\bar \omega)}_{jm\mbar}$; the dimension of this part of the operator is
\be
h(j,m,\omega;j',m') = \frac1\nfive\bigl[ j'(j'+1) - j(j-1)\bigr] -m\omega -\frac\nfive4 \omega^2 ~;
\ee
simultaneous spectral flow by an amount $w$ in $\sltwo$ and $-w$ in $\sutwo$ shifts the dimension to
\be
h(j,m,\omega;j',m') - w\bigl(m' + m + \frac\nfive2 \omega\bigr) ~.
\ee
States that satisfy the gauge condition~\eqref{NS5coul constraint} flow to equivalent states of the same dimension, and we are free to choose any representative of the equivalence class as a physical state.  We find the usual division of labor of a null constraint -- the gauge condition removes one null coordinate, and the gauge freedom removes the conjugate null coordinate.

%%%%%%%%%%%%%%%%%%%%%%%%%%%%%%
\subsection{NS5-P and NS5-F1 supertubes}

The tilted gauging that results in supertube geometries amounts to a simple modification of the gauge current~\eqref{NS5Coulgauging} to the more general choice~\eqref{generalgauging}.  This alters the conditions that quantum numbers of physical states must satisfy:
\begin{align}
\label{gauge_constraint}
l_1\, \bigl(2m + \nfive w\bigr) + l_2 \,\bigl(2m' + \nfive w'\bigr) - l_3\, E + l_4 \, P_y &= 0
\nn\\
r_1\, \bigl(2\mbar + \nfive\bar w\bigr) - r_2\, \bigl(2\mbar' + \nfive\bar w'\bigr)- r_3\, E + r_4\, P_y &= 0
\end{align}
To analyze these constraints, we proceed in stages.  Consider first the noncompact NS5-P supertube, for which 
\be
l_1=l_2=r_1=r_2 = 1~~,~~~~
l_3=l_4=r_3=r_4 = \alpha  ~.
\ee
The constraints are solved by
\begin{align}
0 &= 2(m+\mbar) + \nfive(w+\bar w) + 2(m'-\mbar') + \nfive(w'-\bar w') - 2\alpha(E - P_\ytil) 
\nn\\
0 &= 2(m-\mbar) + \nfive(w-\bar w) + 2(m'+\mbar') + \nfive(w'+\bar w')   ~.
\end{align}
The constraints can be used to solve for the $\sltwo$ quantum numbers $m,\mbar$, and then the spectral flow can be used to set the $\sltwo$ spectral flow parameters $w,\bar w$ to zero.  These are the analogues in the spectrum of the gauge choice made in sections~\ref{sec:nullGWZW} and~\ref{sec:supertube} that eliminates all dependence on $\tau,\rho$. 

One can now quotient by a $2\pi \Rytil$ translation along $\ytil$.
The momenta are quantized as $P_\ytil\!=\!n_\ytil/\Rytil$, and there is now a winding contribution to the left-right difference constraint
\begin{align}
\label{NS5F1constraint}
0 &= 2(m+\mbar) +\nfive(w+\bar w) +2(m'-\mbar') + \nfive(w'-\bar w') -
    2\alpha\Bigl(E - \frac{n_\ytil}{\Rytil}\Bigr) 
\nn\\
0 &= 2(m-\mbar ) + \nfive(w-\bar w) + 2(m'+\mbar') + \nfive(w'+\bar w') + 2\alpha\, w_\ytil \Rytil ~,
\end{align}
which requires $\alpha = k/\Rytil$ for some integer $k$ since all the other terms in the difference constraint are integral.  Note that the gauge constraints for the NS5-F1 supertube follow trivially -- one flips the sign of $r_4$ while sending $\Rytil\to 1/\Rytil=R_y$, and now the difference of left and right constraints requires the integer quantization of $\alpha/R_y$, \ie\ the condition $\alpha=k R_y$ of equation~\eqref{NS5F1_params}.

Of some interest are fundamental strings that have winding and momentum on the $\ytil$-$\phi$-$\psi$ torus.
The choices
\begin{align}
\label{windingstring}
j = j' + 1
~,~~~
w = \bar w = 0
~,~~~
E^2 = \Bigl(\frac{n_\ytil}{\Rytil} + w_\ytil  \Rytil\Bigr)^2 + 4m' w' + \nfive (w')^2
\end{align}
with $m,\mbar$ chosen to solve the constraints~\eqref{NS5F1constraint}, specify a physical state given suitably chosen oscillator excitations at level
\be
N_L = \frac12
~~,~~~~
N_R = \frac12 + n_\ytil w_\ytil +  (m' w'-\mbar' \bar w') + \frac\nfive4\bigl( (w')^2 - (\bar w')^2\bigr) ~.
\ee
The resulting set of string states are F1-P supertubes (typically not aligned with the NS5-P supertube as in~\cite{Bena:2011uw}).  Note that in the $\ads3$ limit $\Rytil\to 0$ the string becomes infinitely heavy and becomes part of the heavy background configuration, unless $n_\ytil=0$ \ie\ there is no winding on the T-dual $y$-circle, which is the angular direction of the $\ads3$ IR limit of the NS5-F1 duality frame.  In this limit, the standard convention is that the energy associated to the background charges are regulated and the static winding energy is subtracted off, and so one should perform this subtraction when analyzing the spectrum of strings that wind the $y$-circle.  For instance, in the Ramond sector of the spacetime CFT there is no gap to extracting a string from the system until one turns on the Ramond moduli of the theory~\cite{Seiberg:1999xz,Larsen:1999uk}, a result which assumes that one is measuring energy relative to that of the unexcited background.  The state~\eqref{windingstring} pulls winding number $w_y=n_\ytil$ out of the system; that winding has an associated energy $E_w = w_y \Ry = n_\ytil/\Rytil$ that has already been subtracted out in the ground state, and so one should continue to subtract it out in evaluating the energetics of such states.

%%%%%%%%%%%%%%%%%%%%%%%%%%%%%%
\subsection{Spectrally flowed supertubes}

Substituting the left and right gauge coefficients~\eqref{specflowgauging}, \eqref{r_usual} into judicious rescalings of the gauge constraints~\eqref{gauge_constraint}, one finds the following conditions on physical states for the spectrally flowed supertube of section~\ref{sec:asymgauging}:
\begin{align}
0 &= \bigl(2m+\nfive w\bigr) + (2s+1)\bigl(2m'+\nfive w'\bigr) -  k\Ry \Bigl(E-P_{y,L}\Bigr) - \frac{s(s+1)\nfive}{k\Ry}\Bigl( E + P_{y,L} \Bigr)
\nn\\
0 &= \bigl(2\mbar+\nfive \bar w\bigr) -\bigl(2\mbar'+\nfive \bar
    w'\bigr) - k\Ry \Bigl( E + P_{y,R} \Bigr) -
    \frac{s(s+1)\nfive}{k\Ry}\Bigl( E + P_{y,R} \Bigr) ~. \label{GLMTnullconstraints}
\end{align}
Note that these constraints reduce to the pure NS5-F1 constraints~\eqref{NS5F1constraint} when the spectral flow parameter $s$ vanishes.
The difference of these two constraints involves the left- and right-moving momenta $P_{y,L},P_{y,R}$ 
\begin{align}
\label{GLMTdiffconstraint}
0 = 2\Bigl(m-\mbar\Bigr) &+2\Bigl((2s\tight+1)m' + \mbar'\Bigr)  + \nfive\Bigl( w-\bar w +(2s\tight+1)w'+\bar w'\Bigr)
\\[5pt]
& + {k\Ry}\Bigl( P_{y,L} + P_{y,R} \Bigr) - \frac{s(s+1)\nfive}{k\Ry}\Bigl( P_{y,L} - P_{y,R} \Bigr) ~;
\nn
\end{align}
the usual zero mode quantization
\be 
P_{y,L} = \frac{n_y}{\Ry} + w_y \Ry
~~,~~~~
P_{y,R} = \frac{n_y}{\Ry} - w_y \Ry
\ee
implies that the $y$ zero mode contribution to the constraints is integral (recall that $k$ divides $s(s+1)$).  All contributions to the difference constraint~\eqref{GLMTdiffconstraint} are thus integral, and the constraint can be solved \eg\ by fixing the $\sutwo$ contribution in terms of the other contributions.

In~\cite{Giusto:2012yz}, the geometry~\eqref{GLMTmetric}, \eqref{GLMT_Bfield} was shown to be equivalent to global $\ads3\times\IS^3$ with the identification
\be
\Bigl( \frac y{R_y},\psi,\phi \Bigr) \sim \Bigl( \frac y{R_y},\psi,\phi \Bigr) + \frac{2\pi}{k} \Bigl(k,(s\!+\!1),s\Bigr)  ~.
\ee
In the Euler angle parametrization~\eqref{Eulerangles}, the left-handed angular coordinate $\psi\tight+\phi$ shifts by $(2s\tight+1)/k$ while the left $\ads3$ coordinate $(t\tight+y)/\Ry$ shifts by one; and at the same time the right-handed angular coordinate $\psi\tight+\phi$ shifts by $1/k$ while the right $\ads3$ coordinate $(t\tight-y)/\Ry$ shifts by minus one.  If we look at the spectrum of winding states of perturbative strings, the contributions of the winding on $\sutwo$ and on $y$ to the left- and right-moving constraints are precisely what one would expect as a result of these identifications.

%%%%%%%%%%%%%%%%%%%%%%%%%%%%%%%%%%%%%%%%%%
%%%%%%%%%%%%%%%%%%%%%%%%%%%%%%%%%%%%%%%%%%

\section{D-branes}
\label{sec:dbranes}

Stringy effects at the nonperturbative level in $\alpha'$ resolve the topology of the NS5 supertube source.  In the microstate geometries program, this sort of topology is conjectured to support excitations which account for black hole entropy.  Geometrical excitations sourcing non-singular solutions of the supergravity field equations have been found, leading to geometries which look like the BTZ black hole~\eqref{BTZgeom} to an arbitrary depth in redshift~\cite{Bena:2016ypk}; nevertheless, it is not clear how much entropy is to be found in this way, and several estimates%
~\cite{Bena:2006is,Bena:2008nh,deBoer:2009un} 
cast doubt on whether a substantial portion of the entropy comes from wiggling the geometry.  On the other hand, in bubbled geometries allowing an arbitrary depth of throat, so-called {\it W-branes} -- branes wrapping the topology down the throat -- account for a finite fraction of the black hole entropy in a crude quiver quantum mechanics truncation of the dynamics~\cite{Martinec:2015pfa}.  Finite depth of throat corresponds to the Coulomb branch of this dynamics, while the dominant contribution to the entropy comes from wrapped brane condensates on the Higgs branch~\cite{Denef:2007vg,Bena:2012hf} where the throat has become arbitrarily deep.  On the Coulomb branch, the wrapped branes are massive.  While one can see the constituents which will give rise to a vast increase in entropy on the Higgs branch when the throat reaches the depth of a `stretched horizon'~\cite{Susskind:1993if}, their contributions to the density of states are suppressed on the Coulomb branch.

These results motivate a search for such W-branes in supertube geometries.  In the NS5-F1 frame, the loop of KK monopoles provides the relevant topology.  The KK monopole cores are the poles of topological two-spheres where the $\IS^1$ fiber of the monopole geometry degenerates; there is such a sphere between any pair of cores.  While locally there seem to be $k$ independent two-spheres as we move around the $y$ circle at fixed $\phi$, the monodromy of the KK dipole loop around the $\phi$ circle relates them.  A D3-brane wrapping such a bubble will not come back to itself until it winds around the $\phi$ circle $k$ times.%
\footnote{When $k$ and $\nfive$ share a common divisor $\kappa$, there are multiple strands of fivebrane, so the W-brane need wrap the $\phi$ circle only $k/\kappa$ times before closing back on itself.}

\begin{figure}[ht]
\centerline{\includegraphics[width=3.0in]{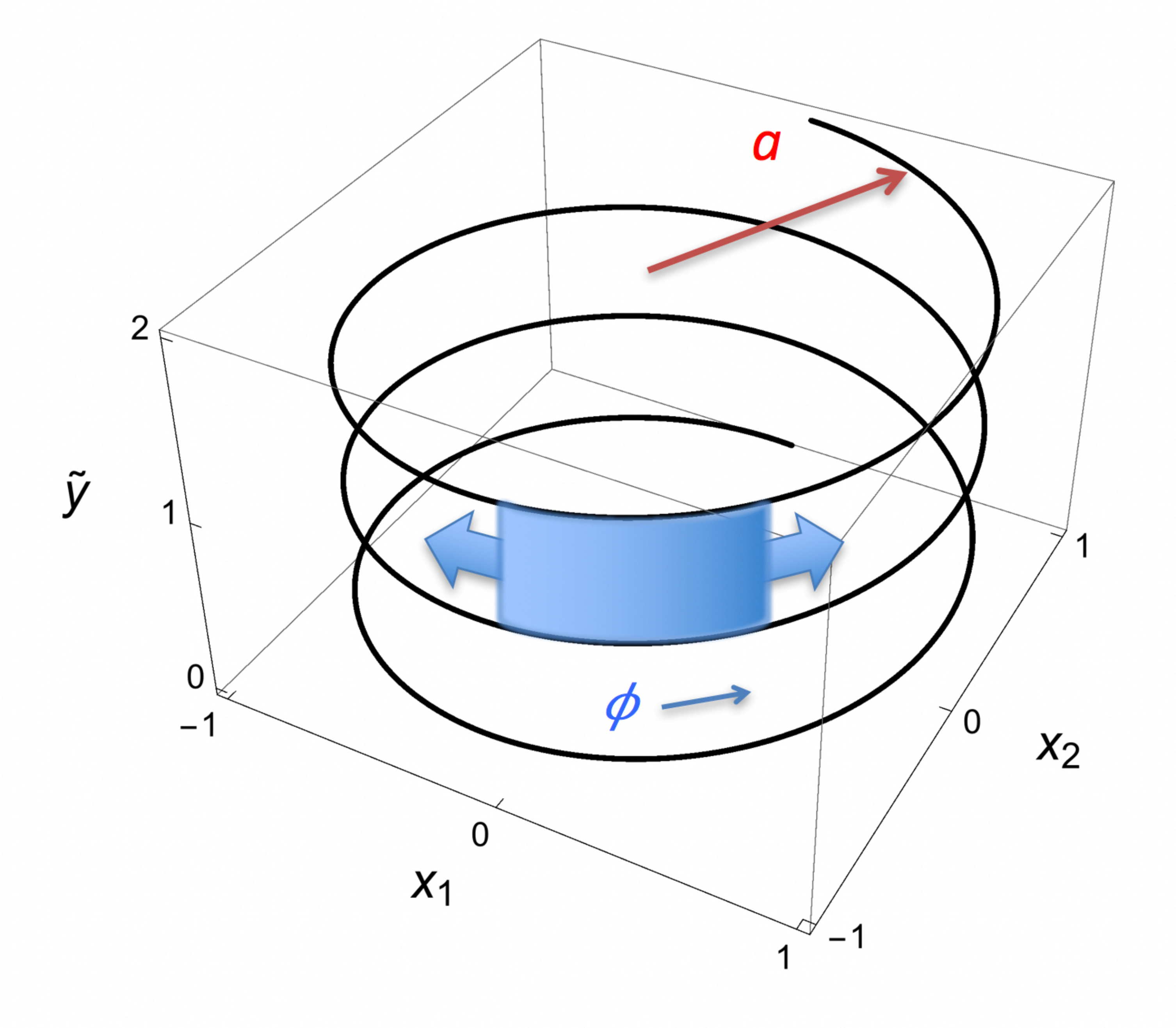}}
\setlength{\unitlength}{0.1\columnwidth}
\caption{\it 
A D2-brane can stretch between the windings of the NS5-P supertube, so long as the brane wraps around to close on itself along its other spatial dimension.
}
\label{fig:BraneStretch}
\end{figure}

The T-dual of such a brane is a D2-brane in the NS5-P frame stretching between the strands of the NS5-brane supertube, and as depicted in figure~\ref{fig:BraneStretch}, it is clear that such a brane must wrap the $\phi$ circle multiple times until it closes back on itself ($k$ times if $k$ and $\nfive$ are relatively prime).
%
%%%%%%%%%%%%%
%%%%%%%%%%%%%
This W-brane is closely related to the little string in that both source the antisymmetric tensor field $C_{ab}$ on type IIA five-branes.  In fact, one might regard this brane as being the manifestation of the little string on the Coulomb branch of little string theory, where it becomes massive due to the fivebrane separation.%
\footnote{A similar behavior is seen in the solvable examples of~\cite{Giveon:1999px,Giveon:1999tq}.  There, perturbative string theory is valid up to the energy scale set by the W-branes -- D1-branes stretched between the NS5 sources in type IIB, and D2-branes similarly stretched in type IIA.  When energies exceed this scale, one must account for D-brane pair creation processes, which are not describable in worldsheet perturbation theory.}
If we try to move closer to the black hole regime by dialing down the angular momentum carried by the supertube, the stretched W-brane becomes lighter (since its mass is proportional to the radius of the supertube $a\tight =\frac{\sqrt{Q_1Q_5}}{k\Ry}$, and $J_L \tight= 2k\Ry a^2$); the W-brane becomes massless as the fivebrane sources coalesce at the black hole threshold at $J_L=0$.

In this scenario, the passage to the black hole regime involves the condensation of stringy W-branes, much as in the quiver QM example analyzed in~\cite{Martinec:2015pfa}, and a phase transition lies between the two-charge supertube regime and the true black hole regime; it is natural to associate the phase boundary to the vanishing of $S_\bh$ in~\eqref{CardyJne0}.  

The $k$-wound W-brane admits open string excitations which are bound to the intersection of successive windings at the supertube source location.  Thus, once one pays the energetic cost of adding a W-brane to the system, there is a lot of entropy to be had from exciting it, since the branes wrap $k$ times around the $\phi$ circle, thus fractionating their momenta by a factor $k$.  This is just what one expects from the {\it long string sector} of the $\ads3$ spacetime CFT, which carries the softest excitations of the system and dominates the entropy at the threshold of the black hole regime.  Indeed, as we reviewed in section~\ref{sec:Supertubes}, the round supertube of winding $k$ has been identified with a state of the dual spacetime CFT built out of $k$-cycles in the symmetric group at the $\cM^N/S_N$ symmetric orbifold point, and having momentum excitations fractionated by a factor $k$.

\begin{figure}[ht]
\centerline{\includegraphics[width=2.5in]{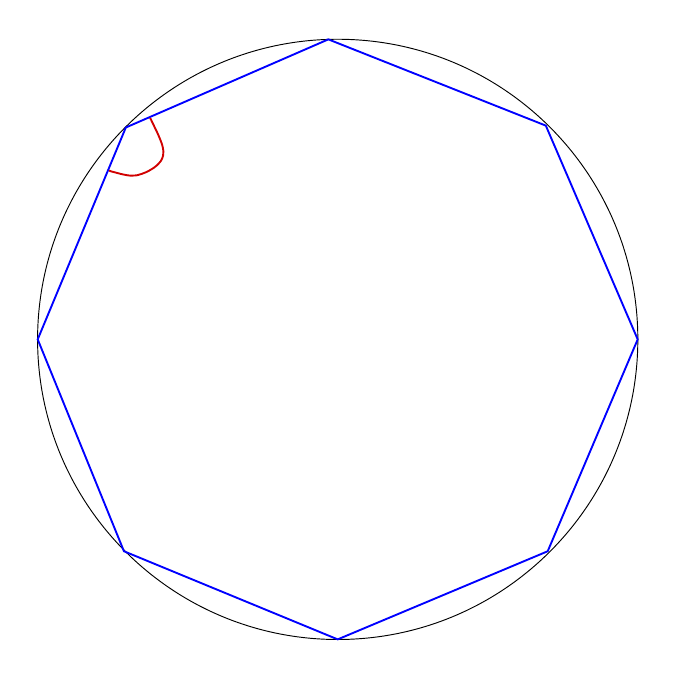}}
\setlength{\unitlength}{0.1\columnwidth}
\caption{\it 
A slice of the geometry in the NS5-P frame in the $\theta$-$\phi$ plane at fixed $\ytil$ and $\rho=0$, with a D2-brane stretching between strands of the multiwound fivebrane, for small $k$ and large $\Rytil$.  Due to monodromy, the D2-brane must recur on each interval between strands so that it closes back on itself after making enough circuits of the $\ytil$-$\phi$ torus. 
}
\label{fig:W-brane.pdf}
\end{figure}

While the above is a broad-brush picture of the W-branes, there are subtleties to be understood.  A cross-section of the NS5-P W-brane at fixed $\ytil$ in the $\theta$-$\phi$ plane of the NS5-brane sources at $\rho=0$ is depicted in figure~\ref{fig:W-brane.pdf}.  Here it is assumed that $k$ is small and $\Rytil$ is large, so that the brane slices at fixed $\phi$ are approximately those of the untilted fivebranes on the Coulomb branch.  The successive slices of the brane intersect at angles, and so break all spacetime supersymmetry (this feature was also observed~\cite{Tyukov:2016cbz} in the W-brane configurations of~\cite{Martinec:2015pfa}).
Moreover, the open strings bound to the intersections are tachyonic; one expects their condensation to yield a `floating brane' of the sort analyzed in~\cite{Bena:2008dw,Bena:2008nh}, where it was shown that such objects can be highly entropic.  But the open string instability is weaker and weaker as the orbifold order $k$ increases, or as $\Rytil$ decreases; the branes are pinned between neighboring strands of the supertube spiral and thus more and more closely follow the circular helix, with the angles between successive W-brane windings closer and closer to $\pi$, and the open strings bound to their intersection closer and closer to being massless.  Thus, even if these open strings are tachyonic, they are only very slightly so in the limit of interest, and one expects that the stable ground state is nearby in configuration space.
The energy threshold for the floating brane to fragment into fractional branes again will be rather low.  For instance, if one puts a sufficient energy density on the brane, the fragmented state will be restored, leading to a substantial increase in the W-brane entropy as the effective size of the box that the modes inhabit is increased by a factor $k$.  We thus regard the starting point of the wound W-brane to provide an accurate picture of the excitation spectrum of such D-branes.

%%%%%%%%%%%%%
%%%%%%%%%%%%%

A very rough estimate of the mass of the W-brane employs the DBI action; while this effective action is rather beyond its regime of applicability, it was seen to be reasonably accurate in an analysis of the D-brane spectrum in the $\frac\sutwo\uone$ coset model~\cite{Maldacena:2001ky}.  At sufficiently large $k$ and small $\Rytil$, the D-brane to a very good approximation simply wraps the $\ytil$-$\phi$ torus.  In the DBI action the dependence on the warp factor $\Sigma$ cancels between the dilaton and the volume element, and one has
\be
\label{Wbranemass}
E = \frac 1{ (2\pi)^2\lstr^3} \int e^{-\Phi} \sqrt{G+B} = \sqrt{\frac{\none\lstr^4}{\nfive V_4}}\,\frac{\Rytil \sqrt\nfive}{\lstr^2} = \sqrt{\frac{\none\lstr^4}{ V_4}}\,\frac{1}{\Ry}
\ee
Now recall that $E\Ry$ is the quantity being held fixed in the $\ads3$ decoupling limit, and is $k$ times the energy in units of the $\ads3$ radius (this factor of $k$ coming from the fact that we are on the $\IZ_k$ orbifold of $\ads3\tight\times\IS^3$).  In string units, one has $R_{\rm\sst  AdS}=\sqrt{\nfive}\,\lstr$, so
all told the mass of the W-brane in string units is
\be
\lstr M_w \sim \sqrt{\frac{\none\lstr^4}{ \nfive V_4}}\,\frac{1}{k} ~.
\ee

As one approaches the black hole threshold at large $k$, string perturbation theory breaks down when W-branes become as light as fundamental strings, and one passes to a phase where W-strings are condensed; much the same as in~\cite{Martinec:2015pfa}, the entropy is expected to consist of the multiplicity of degenerate ground states available to the condensate.  While a reliable treatment of this regime seems daunting, one at least has a handle on the relevant constituents via a controlled approach to the black hole threshold from below.
%

%%%%%%%%%%%%%%%%%%%%%%%%%%%%%%%%%%%%%%%%%%
%%%%%%%%%%%%%%%%%%%%%%%%%%%%%%%%%%%%%%%%%%

\section{Discussion}
\label{sec:discussion}

Supertube geometries have provided many insights into gauge-gravity duality, and inspired the fuzzball conjecture of black hole microstate structure.  In this work, we have developed a powerful tool for the construction of exact worldsheet CFT's for a broad class of such geometries.  Having control over effects nonperturbative in the string tension reveals an intricate topology of the fivebrane source that heretofore had been smeared away in the supergravity approximation.  Branes wrapping this topology appear to be a promising route to uncovering the long string sector of the spacetime theory, and relating this sector to the little strings associated to fivebranes, as foreseen in~\cite{Martinec:2014gka}.  

In the NS5-P duality frame, these {\it W-branes} are D2-branes stretching between strands of the fivebrane supertube.  In the NS5-F1 frame, the W-branes become D3-branes wrapping a cycle which is topologically $\IS^2\times\IS^1$, with the $\IS^2$ a collapsed sphere made by the structure of nearly coincident strands of the KK monopole ring, and the $\IS^1$ of the ring direction.  The bouquet of vanishing cycles has monodromy around the ring specified by the T-dual NS5-P picture, which causes the W-brane to wrap the ring direction many times.  These wrapped branes are the analogues of the W-branes in~\cite{Martinec:2015pfa}, whose Higgs branch states in a quiver QM truncation yield an entropy that grows with the charges in the appropriate way $\propto\sqrt{n_5n_1n_p}$ to account for a finite fraction of 3-charge BPS black hole entropy in a particular superselection sector.  There the branes were free to descend the AdS throat and approach the scale where a horizon should be forming; here the branes are held apart at a fixed separation by angular momentum and are massive on the Coulomb branch.  Nevertheless we see in the excitation spectrum the precursors of the degrees of freedom that are expected to form the long string sector associated to black hole physics.  Again these W-branes appear to support a great deal of entropy, and although the details here are different, the same sort of object is the main actor.

The picture that seems to be emerging is that smooth microstate geometries describe Coulomb branch configurations of the underlying sources, where strings see a smooth geometry because W-branes are massive in this phase.  Horizon formation is related to a Coulomb-Higgs transition where the W-branes become light and condense, similar to~\cite{Strominger:1995ac,Seiberg:1996vs,Witten:1996qb}.  The ground state of the condensate is highly degenerate and accounts for the BPS black hole entropy.  It also resolves the extremal horizon singularity that one is led to expect on the basis of the effective field theory arguments reviewed in the Introduction.  

This resolution of the horizon is somewhat different than the one envisioned in the original fuzzball proposal~\cite{Mathur:2005zp}, in that the geometrical mechanism that smoothly caps off the throat geometry at a finite redshift is supplanted by a new phase consisting of a brane condensate.  The geometry is still capped, but by a wall of branes rather than smooth geometry.  The properties of this wall lie at the heart of black hole horizon (and singularity) physics.  When a probe hits it, does it splatter on a hard wall, or is it caught in a soft net?  And for the nonextremal black hole, does this happen at the outer horizon or the inner horizon?  Do such geometrical notions even make sense in this stringy regime?  While we do not yet know the answers, we find it intriguing that there are multiple pictures of the family of Coulomb branch configurations of supertubes -- the same background has a `brane gas' description captured by the Landau-Ginsburg effective field theory, and a `coarse-grained geometry' description in the smeared sigma model.  Both are true at the same time, as one sees for instance in the scattering phase shift~\eqref{twopoint}, and it might not be too surprising if something similar persists in the physics of black holes near (and beyond) the horizon.

The dual manifestations of the background as a brane gas and as a geometry provides a rationale for why the W-branes can support themselves at the horizon scale without falling into the black hole, while ordinary matter cannot.  It is simply because the W-branes {\it are} the entropic black hole constituents.  In the present example, they are little strings at their correspondence point~\cite{Martinec:2014gka}; and like ordinary strings at their correspondence point~\cite{Horowitz:1996nw}, the degrees of freedom that carry black hole entropy live at the horizon scale because that is the scale of their quantum wavefunction.%
\footnote{The correspondence principle works somewhat differently in $\ads3$ and linear dilaton backgrounds~\cite{Giveon:2005mi}, see~\cite{Martinec:2014gka} for a discussion in the present context.}
Ordinary matter falls in simply because it is not responsible for gravitational entropy, and so is not coherent over the horizon scale, while the entropic degrees of freedom are coherent on this scale.

As an added bonus, our construction yields supertube configurations in the fivebrane decoupling limit, with no further approximations -- the decoupling limit of the second charge ($Q_1$ or $Q_p$) is {\it not} taken, and as a consequence the solutions describe holographic RG flows from little string theory in the UV to an AdS geometry in the IR.%
\footnote{Such flows were recently studied from a somewhat different point of view in~\cite{Giveon:2017nie}.}  
We thus have control over such flows at the nonperturbative level in $\alpha'$.  From the AdS perspective, we have access to scattering boundary conditions, where the AdS spacetime is not a closed system but interacts with perturbations entering and leaving across the UV boundary where it connects onto the linear dilaton throat of the fivebranes.  In this situation, $\ads3$ string theory is perhaps better thought of as the IR limit of a superselection sector of little string theory.   Virasoro symmetry, which is a key organizing principle of the spacetime CFT of the IR limit, is an emergent symmetry in little string theory.  It would be interesting to see how it arises as an approximate symmetry and what effects it might have.

There are many generalizations and extensions of our work that would be interesting to explore.  We discussed a few related to fivebrane physics in section~\ref{sec:variants}.
In addition, other cosets of the SL(2,R) WZW model have been considered in the context of black hole physics and cosmology~\cite{Witten:1991yr,Kiritsis:1994np,Elitzur:2002rt}, and null gauging has been considered in~\cite{Klimcik:1994wp,Horowitz:1994ei,Tseytlin:1995fh,Giveon:1995as}, to list but a few examples from an extensive literature.  The resulting geometries often exhibit regions of closed timelike curves, some hidden behind horizons and some not.  Usually such pathologies are regarded as cause to reject the solution.  While we have restricted our attention to null gaugings that only involve the compact timelike generator of $\sltwo$ in order to obtain spacetime backgrounds with a self-consistent perturbative expansion, one could consider other possibilities.  It might be interesting to revisit some of these constructions with the advantage of hindsight and a new perspective.

A general class of backgrounds of interest are nonlinear sigma models with Wess-Zumino term, which admit a null current%
~\cite{Horowitz:1994rf,Tseytlin:1995fh}
that can be gauged%
~\cite{Hull:1989jk,Hull:1991uw,Hull:1993ct}
on left and right, in a spacetime of signature $(10,2)$.  As we have seen, the left and right null currents can be different.  For us, the rather symmetric $\IR^{1,1}\tight\times\sltwo\tight\times\sutwo\tight\times\cM$ (with $\cM=\IT^4$ or $K3$) has already yielded several families of solutions.  
More generally~\cite{Hull:1989jk,Hull:1991uw}, in a background with nontrivial NS flux $H^{\sst(3)}$ and global Killing vectors $\xi_a$, motion along $\xi_a$ can be gauged provided
\be
\xi^i_a H^{\sst(3)}_{ijk} = \partial_{[j} \omega_{k],a}
\ee
for some globally defined one-forms $\omega_a$.  Defining
\be
c_{ab} = \omega_{i,a}\xi^i_b ~,
\ee
the action
\be
\cS = \frac1{2\pi}\int d^2\!\zhat \,\Bigl[ \bigl(g_{ij}+b_{ij}\bigr)Dx^i\Dbar x^j
+ \alpha\,\omega_{i,a}\Bigl(  \bar\cA^a Dx^i + \cA^a\Dbar x^i \Bigr)
- \alpha^2 \, c_{[ab]} \cA^a\bar\cA^b \Bigr]
\ee
is gauge invariant if $c_{(ab)}=0$ (this latter condition can be relaxed, but then the Wess-Zumino coupling cannot be written in a gauge-invariant way as a local two-dimensional action~\cite{Hull:1989jk,Hull:1991uw}).  We are then interested in gauging a null Killing vector.  The locations of fivebranes are naturally associated to the vanishing locus of $c_{[ab]}$.
One may expect a further generalization to asymmetric gaugings.
Needless to say, it would be interesting to explore the range of possibilities afforded by our construction.  A natural class of models to try to embed in this framework are the general supertubes~\eqref{NS5Pmetric}.

We have seen however that one can be misled by effective geometry when considering the near-source region.  Supergravity tends to smear over near-coincident fivebrane sources, and unless one has sufficient control over stringy effects, important details of the source structure will be missed.  Of course, it is too much to ask for an exact solution for the general fivebrane profile; rather one seeks an appropriate effective theory that captures the essential physics.  For this reason it would also be useful to develop the Landau-Ginsburg approach.  It seems natural to consider hybrid models~\cite{Witten:1993yc,Aspinwall:1993nu,Bertolini:2013xga}, with a Landau-Ginsburg description of the transverse space to the fivebranes fibered over a manifold of signature $(2,2)$ with a null Killing vector that can be gauged.

Finally, the notion that spacetime in the presence of NS5-branes seems best described by null gauging of a target manifold of signature (10,2) recalls a very similar structure that arises in $\cN=2$ heterotic strings~\cite{Kutasov:1996fp,Kutasov:1996zm}, as well as other suggestions of a twelve-dimensional structure in string theory~\cite{Vafa:1996xn,Tseytlin:1996it,Bars:1996dz,Hewson:1996yh,Tseytlin:1996ne}.
Is this phenomenon just formalism, or is it an indicator of some deeper structure underlying string theory?  
Are fivebranes better thought of as nonsingular flux geometries in a space of dimension 10+2?  In this setting, perturbative strings see the locations of fivebranes as places where the gauge action degenerates.  Is the notion of time more flexible in fivebrane dynamics?  If so, one might have the seeds of a resolution of the Hawking paradox, wherein the usual causal structure of spacetime is substantially modified in the cores of the brane bound states that underlie black hole microstates.  

%%%%%%%%%%%%%%%%%%%%%%%%%%%%%%%%%%%%%%%%%%

%\newpage
%\appendix

%%%%%%%%%%%%%%%%%%%%%%%%%%%%%%%%%%%%%
%%%%%%%%%%%%%%%%%%%%%%%%%%%%%%%%%%%%%

%%%%%%%%%%%%%
\vskip 2cm
\section*{Acknowledgements}
%%%%%%%%%%%%%

We thank 
Iosif Bena,
David Kutasov, 
David Turton,
and Nicholas Warner
for useful discussions; and 
Stefano Giusto 
for a helpful correspondence.
%.
This work is supported in part by DOE grant DE-SC0009924.  

%%%%%%%%%%%%%%%%%%%%%%%%%%%%%%%%%%%%%
%%%%%%%%%%%%%%%%%%%%%%%%%%%%%%%%%%%%%

%\newpage
\vskip 3cm
\appendix

\section{Solutions with closed timelike curves}
\label{sec:CTCs}

The spectrally flowed supersymmetric solution \eqref{GLMTmetric}
arises as the extremal limit of the general rotating 3-charge solutions
of \cite{Cvetic:1996xz,Cvetic:1998xh}. In fact, this limit gives a one-parameter
family of solutions which in general admit closed time-like curves. As
shown in \cite{Giusto:2004ip}, requiring absence of these pathologies precisely
selects the metric \eqref{GLMTmetric}. We now show
that this more general family of supersymmetric solutions can be
obtained from the null gauging \eqref{specflowgauging}, with the
following modification of the parameters \eqref{param-map}:
\begin{equation}
e^{2\xi} = \frac{Q_1}{Q_p} \,  , \quad e^{2\zeta} = 1 -
\frac{2\gamma_{+}}{\gamma_{+} + \kappa \,\gamma_{-}} \, , \quad \alpha \mu
= \frac{\sqrt{Q_1 Q_p}}{Q_5} \, , \quad \beta \mu = \kappa \frac{\sqrt{Q_1
  Q_p}}{Q_5}\frac{\sinh \zeta}{\cosh \xi} \, .
\end{equation}
This leads to the following metric and B-field:
\begin{align}
ds^2 &= \frac{f_0}\Sigma \, \bigl(-\,dt^2 + dy^2\bigr)
+ \frac{Q_p}{\Sigma} \, \bigl(dt - dy\bigr)^2
+Q_5 \left(d\rho^2 + d\theta^2 \right)
\label{extremalfull}\\
&
+\frac{Q_5}{\Sigma}\Bigl(b^2 \sinh^2 \rho + a_1^2+ Q_1 \Bigr) \cos^2\!\theta\, d\psi^2
\nn\\ 
&
+ \frac{Q_5}{\Sigma}\Bigl(b^2 \sinh^2 \rho + a_2^2+ Q_1 \Bigr) \sin^2\!\theta\, d\phi^2\nn
\\
&
- \frac{2Q_1Q_5}{R_y\,\Sigma}\Bigl(\gamma_1\cos^2\!\theta\,d\psi-\gamma_2\sin^2\!\theta\,d\phi\Bigr) \bigl(dt - dy\bigr)
\nn\\
&
+
  \frac{2{Q_1Q_5}(\gamma_1+\gamma_2)\eta}{R_y\,\Sigma}\Bigl(\cos^2\!\theta\,d\psi+\sin^2\theta\,d\phi\Bigr)
  dy + dz_a\,dz^a \, ,
\nn\\[15pt]
%\end{align}
%
%\begin{align}
\label{extremalfull_Bfield}
B_2 &= -\frac{Q_1}{\Sigma} \, dt\wedge dy
+\frac{Q_5\cos^2\!\theta}{\Sigma}\Bigl(b^2\sinh^2\rho+a_2^2+Q_1 \Bigr) d\psi\wedge d\phi
\\\
&
+\frac{ Q_p Q_5(\gamma_1+\gamma_2)\eta}{R_y \Sigma}dy\wedge\bigl(\cos^2\!\theta\,d\psi - \sin^2\!\theta\, d\phi\bigr)
\nn\\
&
- \frac{{Q_1Q_5}\cos^2\!\theta}{R_y\,\Sigma}\bigl(\gamma_2\,dt + \gamma_1\,dy\bigr)\wedge d\psi
+ \frac{{Q_1Q_5}\sin^2\!\theta}{R_y\,\Sigma}\bigl(\gamma_1\,dt +
  \gamma_2\,dy\bigr)\wedge d\phi \, ,
\nn \\[15pt]
e^{2\Phi} &= {\frac{\gstr^2 Q_5}{\Sigma}} ,
\end{align}
where
\begin{align}
f_0  &=
  b^2\sinh^2 \rho  +a_1^2 \sin^2\theta+ a_2^2\cos^2\theta \, ,\quad \Sigma  = 
  f_0 + Q_1\, , \\
 b^2 &= \kappa^{-1} a^2
       (\gamma_1+\gamma_2)^2 \eta\, , \quad a = \frac{\sqrt{Q_1 Q_5}}{R_y} \, ,\quad \eta = \frac{Q_1}{Q_1 +Q_p} \, , \nn \\
a_1^2 &= \frac{a^2 (\gamma_1 +\gamma_2) \eta}{2\kappa}
  \left[(1+\kappa)\gamma_1 + (1-\kappa)\gamma_2\right] \, , \nn \\
a_2^2 &= \frac{a^2 (\gamma_1 +\gamma_2) \eta}{2\kappa}
  \left[(1-\kappa)\gamma_1 + (1+\kappa)\gamma_2\right] \, .
\end{align}
The parameter $\kappa$ determines the $Q_p$ charge in terms of the
angular momenta
\begin{equation}\label{QpCTC}
Q_p = \frac{a^2}{4\kappa^2}\left[\kappa^2 (\gamma_1-\gamma_2)^2 -
  (\gamma_1+\gamma_2)^2\right] \, .
\end{equation}
Setting $\kappa = 1$, one recovers the solution \eqref{GLMTmetric}, 
\eqref{GLMT_Bfield}. In this case the  parameters $\gamma_1$, $\gamma_2$ are
related to the spectral flow and orbifold order in the dual CFT as in
\eqref{specflow-params}. The solution \eqref{extremalfull},
\eqref{extremalfull_Bfield} matches precisely the five-brane
decoupling limit of the extremal solution derived in \cite{Giusto:2004ip}.

The null gauging constrains the perturbative string spectrum by
imposing the linear relations \eqref{gauge_constraint}. For general
$\kappa$ these constraints become
\begin{align}
0 & = Q_1 \gamma_{+}  (2m + n_5 w) - \kappa Q_1 \gamma_{-}(2m'+n_5
    w') - \kappa Q_{+} R_y E + \kappa Q_{-} R_y P_{y,L} \, ,\\
0 & = Q_1\gamma_{+} (2\bar m + n_5 \bar w)-Q_1\gamma_{+} (2\bar m'+ n_5 \bar w') - Q_{+} R_y E - Q_{+} R_y P_{y,R} \, , \nn
\end{align}
where we defined
\begin{equation}
\gamma_{\pm} = \gamma_1 \pm \gamma_2 \, , \qquad Q_{\pm} = Q_1 \pm Q_p
\, .
\end{equation}
Note that by using the relation \eqref{QpCTC} the $Q_1$ charge drops out
from the equations.
For $\kappa = 1$, using the relations \eqref{specflow-params}, these constraints
reduce to \eqref{GLMTnullconstraints}. It is interesting to see that the
parameter $\kappa$ enters asymmetrically in the left and right
constraints. Thus, it is not clear how to solve the difference
constraint with the appropriate quantization for the
$SL(2,\mathbb{R})$ and $SU(2)$ quantum numbers, unless $\kappa = 1$.

\section{Current algebra properties}
\label{sec:WZWconventions}

%%%%%%%%%%%%%%%%%%%%%%%%%%%%%%%%%%%%%%%%%%
%%%%%%%%%%%%%%%%%%%%%%%%%%%%%%%%%%%%%%%%%%

\subsection{$\sutwo$}
\label{sec:sutwocft}

The supersymmetric $\sutwo$ level $\nfive$ current algebra consists of currents $j_\susup^a$ and their fermionic superpartners $\psi_\susup^a$ having the OPE structure
\begin{align}
J_\susup^a(z)\,J_\susup^b(0) &\sim \frac{\frac12 \nfive \, \delta^{ab}}{z^2} + \frac{i\epsilon^{abc} J^\susup_c(0)}{z}
\nn\\
J_\susup^a(z)\, \psi_\susup^b(0) &\sim i\epsilon^{abc}\frac{\psi_c^\susup(0)}{z}
\\
\psi_\susup^a(z)\, \psi_\susup^b(0) &\sim \frac{\delta^{ab}}{z}
\nn
\end{align}
with the Killing metric $\delta^{ab} = {\rm diag}(+1,+1,+1)$.  One can define a set of ``bosonic'' $\sutwo$ level $\nfivetil=\nfive-2$ currents that commute with the fermions
\be
j_\susup^a = J_\susup^a + \frac i2\epsilon^{abc}\psi^\susup_b \psi^\susup_c  ~.
\ee
The primary fields $\Phihat_{j'm'\mbar'}$ of the current algebra have conformal dimensions
\be
h = \bar h = \frac{j'(j'+1)}{\nfive} ~;
\ee
unitarity restricts the allowed spins $j'$ of the underlying bosonic
current algebra to the allowed range
\be
\label{su2 reps app}
j' = 0,\frac12,\dots,\frac{\nfive}{2}-1 ~.
\ee

These operators have a {\it parafermion} decomposition under the current $J^\susup_3$%
~\cite{Fateev:1985mm,Gepner:1986hr,Gepner:1987qi}
\footnote{Our notation here largely follows~\cite{Martinec:2001cf}, see also~\cite{Giveon:2015raa}, except that we work in conventions where $\alpha'=1$, so that T-duality is $R\to 1/R$, instead of the convention $\alpha'=2$ of those works.}
obtained by extracting the dependence on $J_3^\susup, \bar J_3^\susup$.  To this end, one bosonizes the currents
\begin{align}
j_3^\susup=i\sqrt{\nfivetil}\,\partial Y_\susup 
~~&,~~~~
\psi^+_\susup \psi^-_\susup = i\sqrt2\,\partial H_\susup
\nn\\
J_3^\susup=i\sqrt{\nfive}\,\partial \cY _\susup
~~&,~~~~
J_\cR^\susup = \frac\nfivetil\nfive \psi^+_\susup\psi^-_\susup - \frac2\nfive j_3^\susup
= i\sqrt{\frac{2\nfivetil}{\nfive}} \, \partial \cH_\susup ~.
\end{align}
and similarly for the right-movers.  The $\sutwo$ primary field $\Phihat_{j'm'\mbar'}$ can then be decomposed as
\be
\label{supf app}
\Phihat^\susup_{j'm'\mbar'} = \Psihat^\susup_{j'm'\mbar'} \,\exp\Bigl[i\frac2{\sqrt\nfive}\Bigl(m'\cY_{\!\susup}+\mbar'\bar\cY_{\!\susup}\Bigr)\Bigr] ~.
\ee
The conformal dimension of the $\sutwo$ primary $\Phihat_{j'm'\mbar'}$ decomposes as
\be
\label{supfspec-app}
h(\Psihat^\susup_{j'm'\mbar'}) = \frac{j'(j'+1)-(m')^2}{\nfive} 
~~,~~~~
\bar h(\Psihat^\susup_{j'm'\mbar'}) = \frac{j'(j'+1)-(\mbar')^2}{\nfive} 
\ee
with the rest made up by the dimension of the $\cY,\bar\cY$ exponentials.  The point of this exercise is that the fields $\Psihat^\susup_{j'm'\mbar'}$ commute with the current $J_3^\susup$, and so are the natural building blocks for representations of the gauged theory; constraints on linear combinations of this and other currents will not involve the parafermion fields.

The superparafermion theory is a representation of the $\cN\!=\!2$ superconformal algebra generated by the supersymmetry currents
\be
G_\susup^\pm = \sqrt{\frac{2}{\nfive}} \, j_\susup^\mp\psi_\susup^\pm
\ee
together with the stress tensor and the $\cR$-symmetry current $J_\cR^\susup$.

The super-parafermion fields are in turn built out of parafermions of the bosonic $\sutwo$ subalgebra of level $\nfivetil= \nfive -2$, and the additional scalar $\cH_\susup$, $\bar\cH_\susup$ that bosonizes the $\cR$-current
\be
\Psihat^\susup_{j'm'\mbar'} = \Psi^\susup_{j'm'\mbar'} \,\exp\Bigl[-i\sqrt{\frac{8}{\nfive\nfivetil}}\Bigl(m'\cH_\susup+\mbar'\bar\cH_\susup \Bigr)\Bigr] ~;
\ee
There are thus two notions of {\it spectral flow} in this theory: Spectral flow in the $\cR$-symmetry $J_\cR^\susup$ (see for instance~\cite{Giveon:2015raa}), and spectral flow in $J_3^\susup$.

The shift $m'\to (m'\!+\!\frac12\nfive w')$ in the $\cY$ exponential in $\Phihat^{\susup}_{j'm'\mbar'}$ in~\eqref{supf app} defines the left spectral flow of interest to us here.  The states flowed in this way have a shifted exponential but the same underlying superparafermion state; their conformal dimensions are
\be
h\bigl(\Psihat^{(w',\bar w')}_{j'm'\mbar'}\bigr) = \frac{j'(j'+1)}{\nfive} + m'w' +\frac{\nfive}{4} (w')^2
\ee
and similarly for the right-handed spectral flow.

%%%%%%%%%%%%%%%%%%%%%%%%%%%%%%%%%%%%%%%%%%
%%%%%%%%%%%%%%%%%%%%%%%%%%%%%%%%%%%%%%%%%%

\subsection{$\sltwo$}
\label{sec:sltwocft}

The supersymmetric $\sltwo$ level $\nfive$ current algebra consists of currents $j_\slsup^a$ and their fermionic superpartners $\psi_\slsup^a$ having the OPE structure
\begin{align}
J_\slsup^a(z)\,J_\slsup^b(0) &\sim \frac{\frac12 \nfive \,h^{ab}}{z^2} + \frac{i\epsilon^{abc} J^\slsup_c(0)}{z}
\nn\\
J_\slsup^a(z)\, \psi_\slsup^b(0) &\sim i\epsilon^{abc}\frac{\psi_c^\slsup(0)}{z}
\\
\psi_\slsup^a(z)\, \psi_\slsup^b(0) &\sim \frac{h^{ab}}{z}
\nn
\end{align}
with the Killing metric $h^{ab} = {\rm diag}(+1,+1,-1)$.  One can define a set of ``bosonic'' $\sltwo$ level $\nfivehat\!=\!\nfive\!+\!2$ currents that commute with the fermions
\be
j_\slsup^a = J_\slsup^a + \frac i2\epsilon^{abc}\psi^\slsup_b \psi^\slsup_c  ~.
\ee
The primary fields $\Phihat^{\slsup}_{jm\mbar}$ of the current algebra have conformal dimensions
\be
h = \bar h = -\frac{j(j-1)}{\nfive} ~.
\ee

These operators also have a superparafermion decomposition under the current $J^\slsup_3$%
~\cite{Dixon:1989cg,Griffin:1990fg,Dijkgraaf:1991ba}%
\footnote{Again our notation here largely follows~\cite{Martinec:2001cf}, see also~\cite{Giveon:2015raa}.}
obtained by extracting the dependence on $J_3^\slsup, \bar J_3^\slsup$.  To this end, one bosonizes the currents
\begin{align}
j_3^\slsup=- i\sqrt{\nfivehat}\,\partial Y_\slsup 
~~&,~~~~
\psi^+_\slsup \psi^-_\slsup = i\sqrt2\,\partial H_\slsup
\nn\\
J_3^\slsup=- i\sqrt{\nfive}\,\partial \cY _\slsup
~~&,~~~~
J_\cR^\slsup = \frac\nfivehat\nfive \psi^+_\slsup\psi^-_\slsup + \frac2\nfive j_3^\slsup
= i\sqrt{\frac{2\nfivehat}{\nfive}} \, \partial \cH_\slsup ~.
\end{align}
and similarly for the right-movers.  Note that the boson $\cY$, $\bar\cY$ is timelike.  The $\sltwo$ primary field $\Phihat_{jm\mbar}$ can then be decomposed as
\be
\label{slpf app}
\Phihat^\slsup_{jm\mbar} = \Psihat^\slsup_{jm\mbar} \,\exp\Bigl[i\frac2{\sqrt\nfive}\Bigl(m\cY_{\!\slsup}+\mbar\bar\cY_{\!\slsup}\Bigr)\Bigr] ~.
\ee
The conformal dimension of the $\sltwo$ primary $\Phihat^{\slsup}_{jm\mbar}$ decomposes as
\be
\label{slpfspec-app}
h(\Psihat^\slsup_{jm\mbar}) = \frac{-j(j-1)+m^2}{\nfive} 
~~,~~~~
\bar h(\Psihat^\slsup_{jm\mbar}) = \frac{-j(j-1)+\mbar^2}{\nfive} 
\ee
with the rest made up by the dimension of the $\cY,\bar\cY$ exponentials.  Again the fields $\Psihat^\slsup_{jm\mbar}$ commute with the current $J_3^\slsup$, and so are the natural building blocks for representations of the gauged theory.

The superparafermion theory is a representation of the $\cN\!=\!2$ superconformal algebra generated by the supersymmetry currents
\be
G_\slsup^\pm = \sqrt{\frac{2}{\nfive}} \, j_\slsup^\mp\psi_\slsup^\pm
\ee
together with the stress tensor and the $\cR$-symmetry current $J_\cR^\slsup$.

The super-parafermion can again be written in terms of an ordinary parafermion for the bosonic $\sltwo$ subalgebra of level $\nfivehat\!=\!\nfive\!+\!2$, together with another free boson $\cH_\slsup$, $\bar\cH_\slsup$
\be
\Psihat^\slsup_{jm\mbar} = \Psi^\slsup_{jm\mbar} \,\exp\Bigl[-i\sqrt{\frac{8}{\nfive\nfivehat}}\Bigl(m\cH_\slsup+\mbar\bar\cH_\slsup \Bigr)\Bigr] ~;
\ee
one can spectral flow either in the $\cR$-symmetry current as in~\cite{Giveon:2015raa}, or in the group current.  The shift of the $J_3^\susup$ charge $m\to (m\!+\!\frac12 \nfive w)$ leads to the flowed conformal dimension 
\be
h\bigl(\Psihat^{(w,\bar w)}_{jm\mbar}\bigr) = -\frac{j(j-1)}{\nfive} - mw - \frac{\nfive}{4}  w^2
\ee

Unitary representations of bosonic $\sltwo$ current algebra lie in either the principal discrete series (on both left and right)
\be
\cD_j^+ = \bigl\{ \ket{j,m}~\bigl| ~  j\in\IR_+\, ;~~ m\!=\! j+n\, ,~~n\in\IN \bigr\}
\ee
restricted to the range
\be
\frac 12 \le j < \frac{\nfive+1}{2} ~;
\ee
in addition one has the conjugate representations $\cD_j^-$; and the continuous series representations $\cC_j^\alpha$ (again on both left and right)
\be 
\cC_j^\alpha = \bigl\{ \ket{j,m}~\bigl| ~  j\!=\! \coeff12+is\, ,~~s\in\IR\, ;~~m\!=\! \alpha+n\, ,~~n\in\IZ\, ,~~0\!\le\alpha\!<1\in\IR \bigr\} ~.
\ee
In the application to coset models, one is often interested in more general representations where $j\in\IR$ but $j-m$ and $j-\mbar$ are allowed to take non-integer values, and $m,\mbar$ are quantized to the values appropriate to the constant radius of the asymptotic angular circle of the $\frac\sltwo\uone$ coset geometry.  These non-normalizable operators are dual to off-shell operators in the dual little string theory~\cite{Aharony:2004xn}.

An important property of the representations results from the duality between the supersymmetric $\frac{\sltwo}{\uone}$ coset sigma model and $\cN\!=\!2$ Liouville theory (see~\cite{Giveon:2016dxe} and references therein).  A general version of this duality posits an isomorphism between the discrete series affine representations of the underlying bosonic WZW model
\be
\label{fieldident}
\bigl(\cD^-_j\bigr)^{w=\bar w=0} \equiv \bigl(\cD^+_{ \jdual }\bigr)^{w=\bar w= -1} ~,~~~~~~ \jdual \equiv \frac \nfive2-j +1~,
\ee 
which can be extended to the supersymmetric theory.
The quantum numbers of states in these two representations are equal, and embody the nature of the duality.  In particular, the isomorphism relates the two conformal operators
\be
(J_{-1}^+) (\bar J_{-1}^+)  \Phi_{1, -1, -1}^{\sst(0,0)}
~~~\longleftrightarrow~~~
\Phi_{\frac \nfive2,\frac\nfive2+1,\frac\nfive2+1}^{\sst(-1,-1)}
\ee
in the bosonic $\sltwo$ theory (where we have dropped the `$\slsup$' decoration to reduce clutter).  One thus has an equivalence between the vertex operator that has the leading large $\rho$ asymptotics of the metric on the $\frac\sltwo\uone$ coset, and a winding tachyon condensate.  In the supersymmetric theory, the corresponding dual operators are the background metric of the supersymmetric coset, and the superpotential of $\cN=2$ Liouville theory.  This equivalence means that if one of these operators is condensed in the background of the model, then so is the other, since they are dual versions of the same object. 
% An important property of the representations results from the duality
% between the $\frac{\sltwo}{\uone}$ coset sigma model and $\cN\!=\!2$
% Liouville theory (see~\cite{Giveon:2016dxe} and references therein).
% A general version of this duality posits an isomorphism between the
% discrete series affine representations of the underlying bosonic WZW model
% \be
% \bigl(\cD^-_j\bigr)^{w=\bar w=0} \equiv \bigl(\cD^+_{ \jdual }\bigr)^{w=\bar w=1} ~~,~~~~ \jdual = \frac \nfive2-j+1 ~,
% \ee 
% which can be extended to the supersymmetric theory.
% The quantum numbers of states in these two representations are equal, and embody the nature of the duality.  In particular, the isomorphism relates the two conformal operators
% \be
% (J_{-1}^+)^\ell (\bar J_{-1})^\ell  \Phihat_{\ell, -\ell, -\ell}^{\sst(0,0)}
% ~~\longleftrightarrow~~
% \Phihat_{\frac \nfive2-\ell,\frac\nfive2,\frac\nfive2}^{\sst(1,1)}
% \ee
% in the $\sltwo$ theory, where we have dropped the `$\it sl$' decoration.  For $\ell\!=\!1$, one has the equivalence between the vertex operator which has the leading large $\rho$ asymptotics of the metric on the $\frac\sltwo\uone$ coset, and the winding tachyon condensate which has the same leading asymptotic as the $\ell=1$ term in $\Lambda_\nfive$, equation~\eqref{Z5circ}.  This equivalence means that if one of these operators is condensed in the background of the model, then so is the other, since they are dual versions of the same object.

%%%%%%%%%%%%%%%%%%%%%%%%%%%%%%%%%%%%%
%%%%%%%%%%%%%%%%%%%%%%%%%%%%%%%%%%%%%

\newpage

%\begin{adjustwidth}{-.5mm}{-0.5mm} % to adjust the L and R margins 

%%%%%%%%%%%%%%%%%%%%%%%%%%%%%%%%%%%%
%%%%%%%%%%%%%%%%%%%%%%%%%%%%%%%%%%%%

%\bibliographystyle{utphysM}      
\bibliographystyle{JHEP}      

% Uses file "utphysM.bst", a modification of utphys.bst 
% which is compatible with the `mciteplus' package for combined citations.
% Available at www.physics.ohio-state.edu/~turton.7/bibtex.html

\bibliography{microstates}       % calls file "microstates.bib"

\providecommand{\href}[2]{#2}\begingroup\raggedright\begin{thebibliography}{100}

\bibitem{Hawking:1976ra}
S.~W. Hawking, \emph{{Breakdown of Predictability in Gravitational Collapse}},
  \href{http://dx.doi.org/10.1103/PhysRevD.14.2460}{\emph{Phys. Rev.}
  {\bfseries D14} (1976) 2460--2473}.

\bibitem{Seiberg:1999xz}
N.~Seiberg and E.~Witten, \emph{{The D1/D5 system and singular CFT}},
  {\emph{JHEP} {\bfseries 04} (1999) 017},
  [\href{https://arxiv.org/abs/hep-th/9903224}{{\ttfamily hep-th/9903224}}].

\bibitem{Mathur:2009hf}
S.~D. Mathur, \emph{{The information paradox: A pedagogical introduction}},
  \href{http://dx.doi.org/10.1088/0264-9381/26/22/224001}{\emph{Class. Quant.
  Grav.} {\bfseries 26} (2009) 224001},
  [\href{https://arxiv.org/abs/0909.1038}{{\ttfamily 0909.1038}}].

\bibitem{Almheiri:2012rt}
A.~Almheiri, D.~Marolf, J.~Polchinski and J.~Sully, \emph{{Black Holes:
  Complementarity or Firewalls?}},
  \href{http://dx.doi.org/10.1007/JHEP02(2013)062}{\emph{JHEP} {\bfseries 1302}
  (2013) 062}, [\href{https://arxiv.org/abs/1207.3123}{{\ttfamily 1207.3123}}].

\bibitem{Marolf:2011dj}
D.~Marolf and A.~Ori, \emph{{Outgoing gravitational shock-wave at the inner
  horizon: The late-time limit of black hole interiors}},
  \href{http://dx.doi.org/10.1103/PhysRevD.86.124026}{\emph{Phys. Rev.}
  {\bfseries D86} (2012) 124026},
  [\href{https://arxiv.org/abs/1109.5139}{{\ttfamily 1109.5139}}].

\bibitem{Murata:2013daa}
K.~Murata, H.~S. Reall and N.~Tanahashi, \emph{{What happens at the horizon(s)
  of an extreme black hole?}},
  \href{http://dx.doi.org/10.1088/0264-9381/30/23/235007}{\emph{Class. Quant.
  Grav.} {\bfseries 30} (2013) 235007},
  [\href{https://arxiv.org/abs/1307.6800}{{\ttfamily 1307.6800}}].

\bibitem{Maldacena:1997ih}
J.~M. Maldacena and A.~Strominger, \emph{{Universal low-energy dynamics for
  rotating black holes}},
  \href{http://dx.doi.org/10.1103/PhysRevD.56.4975}{\emph{Phys. Rev.}
  {\bfseries D56} (1997) 4975--4983},
  [\href{https://arxiv.org/abs/hep-th/9702015}{{\ttfamily hep-th/9702015}}].

\bibitem{Cvetic:1997uw}
M.~Cvetic and F.~Larsen, \emph{{General rotating black holes in string theory:
  Grey body factors and event horizons}},
  \href{http://dx.doi.org/10.1103/PhysRevD.56.4994}{\emph{Phys.Rev.} {\bfseries
  D56} (1997) 4994--5007},
  [\href{https://arxiv.org/abs/hep-th/9705192}{{\ttfamily hep-th/9705192}}].

\bibitem{Cvetic:1998xh}
M.~Cvetic and F.~Larsen, \emph{{Near horizon geometry of rotating black holes
  in five dimensions}},
  \href{http://dx.doi.org/10.1016/S0550-3213(98)00604-X}{\emph{Nucl. Phys.}
  {\bfseries B531} (1998) 239--255},
  [\href{https://arxiv.org/abs/hep-th/9805097}{{\ttfamily hep-th/9805097}}].

\bibitem{KeskiVakkuri:1998nw}
E.~Keski-Vakkuri, \emph{{Bulk and boundary dynamics in BTZ black holes}},
  \href{http://dx.doi.org/10.1103/PhysRevD.59.104001}{\emph{Phys. Rev.}
  {\bfseries D59} (1999) 104001},
  [\href{https://arxiv.org/abs/hep-th/9808037}{{\ttfamily hep-th/9808037}}].

\bibitem{Balasubramanian:2004zu}
V.~Balasubramanian and T.~S. Levi, \emph{{Beyond the veil: Inner horizon
  instability and holography}},
  \href{http://dx.doi.org/10.1103/PhysRevD.70.106005}{\emph{Phys. Rev.}
  {\bfseries D70} (2004) 106005},
  [\href{https://arxiv.org/abs/hep-th/0405048}{{\ttfamily hep-th/0405048}}].

\bibitem{Ooguri:2006in}
H.~Ooguri and C.~Vafa, \emph{{On the Geometry of the String Landscape and the
  Swampland}},
  \href{http://dx.doi.org/10.1016/j.nuclphysb.2006.10.033}{\emph{Nucl. Phys.}
  {\bfseries B766} (2007) 21--33},
  [\href{https://arxiv.org/abs/hep-th/0605264}{{\ttfamily hep-th/0605264}}].

\bibitem{Skenderis:1999bs}
K.~Skenderis, \emph{{Black holes and branes in string theory}},
  \href{http://dx.doi.org/10.1007/3-540-46634-7_12}{\emph{Lect. Notes Phys.}
  {\bfseries 541} (2000) 325--364},
  [\href{https://arxiv.org/abs/hep-th/9901050}{{\ttfamily hep-th/9901050}}].

\bibitem{Lunin:2001fv}
O.~Lunin and S.~D. Mathur, \emph{{Metric of the multiply wound rotating
  string}}, \href{http://dx.doi.org/10.1016/S0550-3213(01)00321-2}{\emph{Nucl.
  Phys.} {\bfseries B610} (2001) 49--76},
  [\href{https://arxiv.org/abs/hep-th/0105136}{{\ttfamily hep-th/0105136}}].

\bibitem{Lunin:2002iz}
O.~Lunin, J.~M. Maldacena and L.~Maoz, \emph{{Gravity solutions for the D1-D5
  system with angular momentum}},
  \href{https://arxiv.org/abs/hep-th/0212210}{{\ttfamily hep-th/0212210}}.

\bibitem{Polchinski:2000uf}
J.~Polchinski and M.~J. Strassler, \emph{{The String dual of a confining
  four-dimensional gauge theory}},
  \href{https://arxiv.org/abs/hep-th/0003136}{{\ttfamily hep-th/0003136}}.

\bibitem{Lin:2004nb}
H.~Lin, O.~Lunin and J.~M. Maldacena, \emph{{Bubbling AdS space and 1/2 BPS
  geometries}}, {\emph{JHEP} {\bfseries 10} (2004) 025},
  [\href{https://arxiv.org/abs/hep-th/0409174}{{\ttfamily hep-th/0409174}}].

\bibitem{Mathur:2005zp}
S.~D. Mathur, \emph{{The fuzzball proposal for black holes: An elementary
  review}}, \href{http://dx.doi.org/10.1002/prop.200410203}{\emph{Fortsch.
  Phys.} {\bfseries 53} (2005) 793--827},
  [\href{https://arxiv.org/abs/hep-th/0502050}{{\ttfamily hep-th/0502050}}].

\bibitem{Bena:2007kg}
I.~Bena and N.~P. Warner, \emph{{Black holes, black rings and their
  microstates}}, \href{http://dx.doi.org/10.1007/978-3-540-79523-0}{\emph{Lect.
  Notes Phys.} {\bfseries 755} (2008) 1--92},
  [\href{https://arxiv.org/abs/hep-th/0701216}{{\ttfamily hep-th/0701216}}].

\bibitem{Skenderis:2008qn}
K.~Skenderis and M.~Taylor, \emph{{The fuzzball proposal for black holes}},
  \href{http://dx.doi.org/10.1016/j.physrep.2008.08.001}{\emph{Phys. Rept.}
  {\bfseries 467} (2008) 117--171},
  [\href{https://arxiv.org/abs/0804.0552}{{\ttfamily 0804.0552}}].

\bibitem{Bena:2013dka}
I.~Bena and N.~P. Warner, \emph{{Resolving the Structure of Black Holes:
  Philosophizing with a Hammer}},
  \href{https://arxiv.org/abs/1311.4538}{{\ttfamily 1311.4538}}.

\bibitem{Martinec:2014gka}
E.~J. Martinec, \emph{{The Cheshire Cap}},
  \href{http://dx.doi.org/10.1007/JHEP03(2015)112}{\emph{JHEP} {\bfseries 03}
  (2015) 112}, [\href{https://arxiv.org/abs/1409.6017}{{\ttfamily 1409.6017}}].

\bibitem{Martinec:2015pfa}
E.~J. Martinec and B.~E. Niehoff, \emph{{Hair-brane Ideas on the Horizon}},
  \href{http://dx.doi.org/10.1007/JHEP11(2015)195}{\emph{JHEP} {\bfseries 11}
  (2015) 195}, [\href{https://arxiv.org/abs/1509.00044}{{\ttfamily
  1509.00044}}].

\bibitem{Bena:2011zw}
I.~Bena, B.~D. Chowdhury, J.~de~Boer, S.~El-Showk and M.~Shigemori,
  \emph{{Moulting Black Holes}},
  \href{http://dx.doi.org/10.1007/JHEP03(2012)094}{\emph{JHEP} {\bfseries 1203}
  (2012) 094}, [\href{https://arxiv.org/abs/1108.0411}{{\ttfamily 1108.0411}}].

\bibitem{Balasubramanian:2007qv}
V.~Balasubramanian et~al., \emph{{Typicality versus thermality: An analytic
  distinction}}, \href{http://dx.doi.org/10.1007/s10714-008-0606-8}{\emph{Gen.
  Rel. Grav.} {\bfseries 40} (2008) 1863--1890},
  [\href{https://arxiv.org/abs/hep-th/0701122}{{\ttfamily hep-th/0701122}}].

\bibitem{Balasubramanian:2008da}
V.~Balasubramanian, J.~de~Boer, S.~El-Showk and I.~Messamah, \emph{{Black Holes
  as Effective Geometries}},
  \href{http://dx.doi.org/10.1088/0264-9381/25/21/214004}{\emph{Class. Quant.
  Grav.} {\bfseries 25} (2008) 214004},
  [\href{https://arxiv.org/abs/0811.0263}{{\ttfamily 0811.0263}}].

\bibitem{Mateos:2001qs}
D.~Mateos and P.~K. Townsend, \emph{{Supertubes}},
  \href{http://dx.doi.org/10.1103/PhysRevLett.87.011602}{\emph{Phys. Rev.
  Lett.} {\bfseries 87} (2001) 011602},
  [\href{https://arxiv.org/abs/hep-th/0103030}{{\ttfamily hep-th/0103030}}].

\bibitem{Emparan:2001ux}
R.~Emparan, D.~Mateos and P.~K. Townsend, \emph{{Supergravity supertubes}},
  {\emph{JHEP} {\bfseries 07} (2001) 011},
  [\href{https://arxiv.org/abs/hep-th/0106012}{{\ttfamily hep-th/0106012}}].

\bibitem{Gibbons:2013tqa}
G.~Gibbons and N.~Warner, \emph{{Global structure of five-dimensional
  fuzzballs}},
  \href{http://dx.doi.org/10.1088/0264-9381/31/2/025016}{\emph{Class.Quant.Grav.}
  {\bfseries 31} (2014) 025016},
  [\href{https://arxiv.org/abs/1305.0957}{{\ttfamily 1305.0957}}].

\bibitem{Bena:2015bea}
I.~Bena, S.~Giusto, R.~Russo, M.~Shigemori and N.~P. Warner, \emph{{Habemus
  Superstratum! A constructive proof of the existence of superstrata}},
  \href{http://dx.doi.org/10.1007/JHEP05(2015)110}{\emph{JHEP} {\bfseries 05}
  (2015) 110}, [\href{https://arxiv.org/abs/1503.01463}{{\ttfamily
  1503.01463}}].

\bibitem{Giusto:2015dfa}
S.~Giusto, E.~Moscato and R.~Russo, \emph{{AdS$_{3}$ holography for 1/4 and 1/8
  BPS geometries}},
  \href{http://dx.doi.org/10.1007/JHEP11(2015)004}{\emph{JHEP} {\bfseries 11}
  (2015) 004}, [\href{https://arxiv.org/abs/1507.00945}{{\ttfamily
  1507.00945}}].

\bibitem{Bena:2016agb}
I.~Bena, E.~Martinec, D.~Turton and N.~P. Warner, \emph{{Momentum Fractionation
  on Superstrata}},  \href{https://arxiv.org/abs/1601.05805}{{\ttfamily
  1601.05805}}.

\bibitem{Bena:2016ypk}
I.~Bena, S.~Giusto, E.~J. Martinec, R.~Russo, M.~Shigemori, D.~Turton et~al.,
  \emph{{Smooth horizonless geometries deep inside the black-hole regime}},
  \href{https://arxiv.org/abs/1607.03908}{{\ttfamily 1607.03908}}.

\bibitem{Kanitscheider:2007wq}
I.~Kanitscheider, K.~Skenderis and M.~Taylor, \emph{{Fuzzballs with internal
  excitations}}, {\emph{JHEP} {\bfseries 06} (2007) 056},
  [\href{https://arxiv.org/abs/0704.0690}{{\ttfamily 0704.0690}}].

\bibitem{Harvey:1995rn}
J.~A. Harvey and A.~Strominger, \emph{{The heterotic string is a soliton}},
  \href{http://dx.doi.org/10.1016/0550-3213(95)00604-4,
  10.1016/0550-3213(95)00310-O}{\emph{Nucl. Phys.} {\bfseries B449} (1995)
  535--552}, [\href{https://arxiv.org/abs/hep-th/9504047}{{\ttfamily
  hep-th/9504047}}].

\bibitem{Alvarez:1987wg}
O.~Alvarez, T.~P. Killingback, M.~L. Mangano and P.~Windey, \emph{{String
  Theory and Loop Space Index Theorems}},
  \href{http://dx.doi.org/10.1007/BF01239011}{\emph{Commun. Math. Phys.}
  {\bfseries 111} (1987) 1}.

\bibitem{Pilch:1986en}
K.~Pilch, A.~N. Schellekens and N.~P. Warner, \emph{{Path Integral Calculation
  of String Anomalies}},
  \href{http://dx.doi.org/10.1016/0550-3213(87)90109-X}{\emph{Nucl. Phys.}
  {\bfseries B287} (1987) 362--380}.

\bibitem{Giveon:1994fu}
A.~Giveon, M.~Porrati and E.~Rabinovici, \emph{{Target space duality in string
  theory}}, \href{http://dx.doi.org/10.1016/0370-1573(94)90070-1}{\emph{Phys.
  Rept.} {\bfseries 244} (1994) 77--202},
  [\href{https://arxiv.org/abs/hep-th/9401139}{{\ttfamily hep-th/9401139}}].

\bibitem{Kanitscheider:2006zf}
I.~Kanitscheider, K.~Skenderis and M.~Taylor, \emph{{Holographic anatomy of
  fuzzballs}},
  \href{http://dx.doi.org/10.1088/1126-6708/2007/04/023}{\emph{JHEP} {\bfseries
  04} (2007) 023}, [\href{https://arxiv.org/abs/hep-th/0611171}{{\ttfamily
  hep-th/0611171}}].

\bibitem{Giveon:1999px}
A.~Giveon and D.~Kutasov, \emph{{Little string theory in a double scaling
  limit}}, \href{http://dx.doi.org/10.1088/1126-6708/1999/10/034}{\emph{JHEP}
  {\bfseries 10} (1999) 034},
  [\href{https://arxiv.org/abs/hep-th/9909110}{{\ttfamily hep-th/9909110}}].

\bibitem{Giveon:1999tq}
A.~Giveon and D.~Kutasov, \emph{{Comments on double scaled little string
  theory}}, \href{http://dx.doi.org/10.1088/1126-6708/2000/01/023}{\emph{JHEP}
  {\bfseries 01} (2000) 023},
  [\href{https://arxiv.org/abs/hep-th/9911039}{{\ttfamily hep-th/9911039}}].

\bibitem{Gregory:1997te}
R.~Gregory, J.~A. Harvey and G.~W. Moore, \emph{{Unwinding strings and t
  duality of Kaluza-Klein and h monopoles}}, {\emph{Adv. Theor. Math. Phys.}
  {\bfseries 1} (1997) 283--297},
  [\href{https://arxiv.org/abs/hep-th/9708086}{{\ttfamily hep-th/9708086}}].

\bibitem{Tong:2002rq}
D.~Tong, \emph{{NS5-branes, T duality and world sheet instantons}},
  \href{http://dx.doi.org/10.1088/1126-6708/2002/07/013}{\emph{JHEP} {\bfseries
  07} (2002) 013}, [\href{https://arxiv.org/abs/hep-th/0204186}{{\ttfamily
  hep-th/0204186}}].

\bibitem{Harvey:2005ab}
J.~A. Harvey and S.~Jensen, \emph{{Worldsheet instanton corrections to the
  Kaluza-Klein monopole}},
  \href{http://dx.doi.org/10.1088/1126-6708/2005/10/028}{\emph{JHEP} {\bfseries
  10} (2005) 028}, [\href{https://arxiv.org/abs/hep-th/0507204}{{\ttfamily
  hep-th/0507204}}].

\bibitem{Sfetsos:1998xd}
K.~Sfetsos, \emph{{Branes for Higgs phases and exact conformal field
  theories}},
  \href{http://dx.doi.org/10.1088/1126-6708/1999/01/015}{\emph{JHEP} {\bfseries
  01} (1999) 015}, [\href{https://arxiv.org/abs/hep-th/9811167}{{\ttfamily
  hep-th/9811167}}].

\bibitem{Elitzur:2000pq}
S.~Elitzur, A.~Giveon, D.~Kutasov, E.~Rabinovici and G.~Sarkissian,
  \emph{{D-branes in the background of NS five-branes}},
  \href{http://dx.doi.org/10.1088/1126-6708/2000/08/046}{\emph{JHEP} {\bfseries
  08} (2000) 046}, [\href{https://arxiv.org/abs/hep-th/0005052}{{\ttfamily
  hep-th/0005052}}].

\bibitem{Maldacena:2001ky}
J.~M. Maldacena, G.~W. Moore and N.~Seiberg, \emph{{Geometrical interpretation
  of D-branes in gauged WZW models}},
  \href{http://dx.doi.org/10.1088/1126-6708/2001/07/046}{\emph{JHEP} {\bfseries
  07} (2001) 046}, [\href{https://arxiv.org/abs/hep-th/0105038}{{\ttfamily
  hep-th/0105038}}].

\bibitem{Israel:2005fn}
D.~Israel, A.~Pakman and J.~Troost, \emph{{D-branes in little string theory}},
  \href{http://dx.doi.org/10.1016/j.nuclphysb.2005.05.027}{\emph{Nucl. Phys.}
  {\bfseries B722} (2005) 3--64},
  [\href{https://arxiv.org/abs/hep-th/0502073}{{\ttfamily hep-th/0502073}}].

\bibitem{Shenker:1995xq}
S.~H. Shenker, \emph{{Another length scale in string theory?}},
  \href{https://arxiv.org/abs/hep-th/9509132}{{\ttfamily hep-th/9509132}}.

\bibitem{Hori:2001ax}
K.~Hori and A.~Kapustin, \emph{{Duality of the fermionic 2-D black hole and N=2
  liouville theory as mirror symmetry}},
  \href{http://dx.doi.org/10.1088/1126-6708/2001/08/045}{\emph{JHEP} {\bfseries
  08} (2001) 045}, [\href{https://arxiv.org/abs/hep-th/0104202}{{\ttfamily
  hep-th/0104202}}].

\bibitem{Martinec:1988zu}
E.~J. Martinec, \emph{{Algebraic Geometry and Effective Lagrangians}},
  \href{http://dx.doi.org/10.1016/0370-2693(89)90074-9}{\emph{Phys. Lett.}
  {\bfseries B217} (1989) 431--437}.

\bibitem{Vafa:1988uu}
C.~Vafa and N.~P. Warner, \emph{{Catastrophes and the Classification of
  Conformal Theories}},
  \href{http://dx.doi.org/10.1016/0370-2693(89)90473-5}{\emph{Phys. Lett.}
  {\bfseries B218} (1989) 51--58}.

\bibitem{Greene:1988ut}
B.~R. Greene, C.~Vafa and N.~P. Warner, \emph{{Calabi-Yau Manifolds and
  Renormalization Group Flows}},
  \href{http://dx.doi.org/10.1016/0550-3213(89)90471-9}{\emph{Nucl. Phys.}
  {\bfseries B324} (1989) 371}.

\bibitem{Giveon:1999zm}
A.~Giveon, D.~Kutasov and O.~Pelc, \emph{{Holography for noncritical
  superstrings}},
  \href{http://dx.doi.org/10.1088/1126-6708/1999/10/035}{\emph{JHEP} {\bfseries
  10} (1999) 035}, [\href{https://arxiv.org/abs/hep-th/9907178}{{\ttfamily
  hep-th/9907178}}].

\bibitem{Hori:2000kt}
K.~Hori and C.~Vafa, \emph{{Mirror symmetry}},
  \href{https://arxiv.org/abs/hep-th/0002222}{{\ttfamily hep-th/0002222}}.

\bibitem{Giveon:2016dxe}
A.~Giveon, N.~Itzhaki and D.~Kutasov, \emph{{Stringy Horizons II}},
  \href{http://dx.doi.org/10.1007/JHEP10(2016)157}{\emph{JHEP} {\bfseries 10}
  (2016) 157}, [\href{https://arxiv.org/abs/1603.05822}{{\ttfamily
  1603.05822}}].

\bibitem{Giveon:2015cma}
A.~Giveon, N.~Itzhaki and D.~Kutasov, \emph{{Stringy Horizons}},
  \href{http://dx.doi.org/10.1007/JHEP06(2015)064}{\emph{JHEP} {\bfseries 06}
  (2015) 064}, [\href{https://arxiv.org/abs/1502.03633}{{\ttfamily
  1502.03633}}].

\bibitem{Witten:1993yc}
E.~Witten, \emph{{Phases of N=2 theories in two-dimensions}},
  \href{http://dx.doi.org/10.1016/0550-3213(93)90033-L}{\emph{Nucl. Phys.}
  {\bfseries B403} (1993) 159--222},
  [\href{https://arxiv.org/abs/hep-th/9301042}{{\ttfamily hep-th/9301042}}].

\bibitem{Aspinwall:1993nu}
P.~S. Aspinwall, B.~R. Greene and D.~R. Morrison, \emph{{Calabi-Yau moduli
  space, mirror manifolds and space-time topology change in string theory}},
  \href{http://dx.doi.org/10.1016/0550-3213(94)90321-2}{\emph{Nucl. Phys.}
  {\bfseries B416} (1994) 414--480},
  [\href{https://arxiv.org/abs/hep-th/9309097}{{\ttfamily hep-th/9309097}}].

\bibitem{Bertolini:2013xga}
M.~Bertolini, I.~V. Melnikov and M.~R. Plesser, \emph{{Hybrid conformal field
  theories}}, \href{http://dx.doi.org/10.1007/JHEP05(2014)043}{\emph{JHEP}
  {\bfseries 05} (2014) 043},
  [\href{https://arxiv.org/abs/1307.7063}{{\ttfamily 1307.7063}}].

\bibitem{Israel:2004ir}
D.~Israel, C.~Kounnas, A.~Pakman and J.~Troost, \emph{{The Partition function
  of the supersymmetric two-dimensional black hole and little string theory}},
  \href{http://dx.doi.org/10.1088/1126-6708/2004/06/033}{\emph{JHEP} {\bfseries
  06} (2004) 033}, [\href{https://arxiv.org/abs/hep-th/0403237}{{\ttfamily
  hep-th/0403237}}].

\bibitem{Maldacena:2000dr}
J.~M. Maldacena and L.~Maoz, \emph{{De-singularization by rotation}},
  {\emph{JHEP} {\bfseries 12} (2002) 055},
  [\href{https://arxiv.org/abs/hep-th/0012025}{{\ttfamily hep-th/0012025}}].

\bibitem{Hassan:1992gi}
S.~F. Hassan and A.~Sen, \emph{{Marginal deformations of WZNW and coset models
  from O(d,d) transformation}},
  \href{http://dx.doi.org/10.1016/0550-3213(93)90429-S}{\emph{Nucl. Phys.}
  {\bfseries B405} (1993) 143--165},
  [\href{https://arxiv.org/abs/hep-th/9210121}{{\ttfamily hep-th/9210121}}].

\bibitem{Henningson:1992rn}
M.~Henningson and C.~R. Nappi, \emph{{Duality, marginal perturbations and
  gauging}}, \href{http://dx.doi.org/10.1103/PhysRevD.48.861}{\emph{Phys. Rev.}
  {\bfseries D48} (1993) 861--868},
  [\href{https://arxiv.org/abs/hep-th/9301005}{{\ttfamily hep-th/9301005}}].

\bibitem{Giveon:1993ph}
A.~Giveon and E.~Kiritsis, \emph{{Axial vector duality as a gauge symmetry and
  topology change in string theory}},
  \href{http://dx.doi.org/10.1016/0550-3213(94)90460-X}{\emph{Nucl. Phys.}
  {\bfseries B411} (1994) 487--508},
  [\href{https://arxiv.org/abs/hep-th/9303016}{{\ttfamily hep-th/9303016}}].

\bibitem{Forste:2003km}
S.~Forste and D.~Roggenkamp, \emph{{Current current deformations of conformal
  field theories, and WZW models}},
  \href{http://dx.doi.org/10.1088/1126-6708/2003/05/071}{\emph{JHEP} {\bfseries
  05} (2003) 071}, [\href{https://arxiv.org/abs/hep-th/0304234}{{\ttfamily
  hep-th/0304234}}].

\bibitem{Itzhaki:2005zr}
N.~Itzhaki, D.~Kutasov and N.~Seiberg, \emph{{Non-supersymmetric deformations
  of non-critical superstrings}},
  \href{http://dx.doi.org/10.1088/1126-6708/2005/12/035}{\emph{JHEP} {\bfseries
  12} (2005) 035}, [\href{https://arxiv.org/abs/hep-th/0510087}{{\ttfamily
  hep-th/0510087}}].

\bibitem{Kutasov:1999xu}
D.~Kutasov and N.~Seiberg, \emph{{More comments on string theory on AdS(3)}},
  \href{http://dx.doi.org/10.1088/1126-6708/1999/04/008}{\emph{JHEP} {\bfseries
  04} (1999) 008}, [\href{https://arxiv.org/abs/hep-th/9903219}{{\ttfamily
  hep-th/9903219}}].

\bibitem{Porrati:2015eha}
J.~Kim and M.~Porrati, \emph{{On the central charge of spacetime current
  algebras and correlators in string theory on AdS$_{3}$}},
  \href{http://dx.doi.org/10.1007/JHEP05(2015)076}{\emph{JHEP} {\bfseries 05}
  (2015) 076}, [\href{https://arxiv.org/abs/1503.07186}{{\ttfamily
  1503.07186}}].

\bibitem{Balasubramanian:2000rt}
V.~Balasubramanian, J.~de~Boer, E.~Keski-Vakkuri and S.~F. Ross,
  \emph{{Supersymmetric conical defects: Towards a string theoretic description
  of black hole formation}},
  \href{http://dx.doi.org/10.1103/PhysRevD.64.064011}{\emph{Phys. Rev.}
  {\bfseries D64} (2001) 064011},
  [\href{https://arxiv.org/abs/hep-th/0011217}{{\ttfamily hep-th/0011217}}].

\bibitem{Giveon:2001up}
A.~Giveon and D.~Kutasov, \emph{{Notes on AdS(3)}},
  \href{http://dx.doi.org/10.1016/S0550-3213(01)00573-9}{\emph{Nucl. Phys.}
  {\bfseries B621} (2002) 303--336},
  [\href{https://arxiv.org/abs/hep-th/0106004}{{\ttfamily hep-th/0106004}}].

\bibitem{Witten:1998qj}
E.~Witten, \emph{{Anti-de Sitter space and holography}}, {\emph{Adv. Theor.
  Math. Phys.} {\bfseries 2} (1998) 253--291},
  [\href{https://arxiv.org/abs/hep-th/9802150}{{\ttfamily hep-th/9802150}}].

\bibitem{Aharony:2004xn}
O.~Aharony, A.~Giveon and D.~Kutasov, \emph{{LSZ in LST}},
  \href{http://dx.doi.org/10.1016/j.nuclphysb.2004.05.015}{\emph{Nucl. Phys.}
  {\bfseries B691} (2004) 3--78},
  [\href{https://arxiv.org/abs/hep-th/0404016}{{\ttfamily hep-th/0404016}}].

\bibitem{Bars:1991pt}
I.~Bars and K.~Sfetsos, \emph{{Generalized duality and singular strings in
  higher dimensions}},
  \href{http://dx.doi.org/10.1142/S0217732392000963}{\emph{Mod. Phys. Lett.}
  {\bfseries A7} (1992) 1091--1104},
  [\href{https://arxiv.org/abs/hep-th/9110054}{{\ttfamily hep-th/9110054}}].

\bibitem{Quella:2002fk}
T.~Quella and V.~Schomerus, \emph{{Asymmetric cosets}},
  \href{http://dx.doi.org/10.1088/1126-6708/2003/02/030}{\emph{JHEP} {\bfseries
  02} (2003) 030}, [\href{https://arxiv.org/abs/hep-th/0212119}{{\ttfamily
  hep-th/0212119}}].

\bibitem{Giusto:2004id}
S.~Giusto, S.~D. Mathur and A.~Saxena, \emph{{Dual geometries for a set of
  3-charge microstates}},
  \href{http://dx.doi.org/10.1016/j.nuclphysb.2004.09.001}{\emph{Nucl. Phys.}
  {\bfseries B701} (2004) 357--379},
  [\href{https://arxiv.org/abs/hep-th/0405017}{{\ttfamily hep-th/0405017}}].

\bibitem{Giusto:2012yz}
S.~Giusto, O.~Lunin, S.~D. Mathur and D.~Turton, \emph{{D1-D5-P microstates at
  the cap}}, \href{http://dx.doi.org/10.1007/JHEP02(2013)050}{\emph{JHEP}
  {\bfseries 1302} (2013) 050},
  [\href{https://arxiv.org/abs/1211.0306}{{\ttfamily 1211.0306}}].

\bibitem{Chakrabarty:2015foa}
B.~Chakrabarty, D.~Turton and A.~Virmani, \emph{{Holographic description of
  non-supersymmetric orbifolded D1-D5-P solutions}},
  \href{http://dx.doi.org/10.1007/JHEP11(2015)063}{\emph{JHEP} {\bfseries 11}
  (2015) 063}, [\href{https://arxiv.org/abs/1508.01231}{{\ttfamily
  1508.01231}}].

\bibitem{Giusto:2004ip}
S.~Giusto, S.~D. Mathur and A.~Saxena, \emph{{3-charge geometries and their CFT
  duals}}, \href{http://dx.doi.org/10.1016/j.nuclphysb.2005.01.009}{\emph{Nucl.
  Phys.} {\bfseries B710} (2005) 425--463},
  [\href{https://arxiv.org/abs/hep-th/0406103}{{\ttfamily hep-th/0406103}}].

\bibitem{Jejjala:2005yu}
V.~Jejjala, O.~Madden, S.~F. Ross and G.~Titchener, \emph{{Non-supersymmetric
  smooth geometries and D1-D5-P bound states}},
  \href{http://dx.doi.org/10.1103/PhysRevD.71.124030}{\emph{Phys. Rev.}
  {\bfseries D71} (2005) 124030},
  [\href{https://arxiv.org/abs/hep-th/0504181}{{\ttfamily hep-th/0504181}}].

\bibitem{Elitzur:1998mm}
S.~Elitzur, O.~Feinerman, A.~Giveon and D.~Tsabar, \emph{{String theory on
  AdS(3) x S**3 x S**3 x S**1}},
  \href{http://dx.doi.org/10.1016/S0370-2693(99)00101-X}{\emph{Phys. Lett.}
  {\bfseries B449} (1999) 180--186},
  [\href{https://arxiv.org/abs/hep-th/9811245}{{\ttfamily hep-th/9811245}}].

\bibitem{Itzhaki:2005tu}
N.~Itzhaki, D.~Kutasov and N.~Seiberg, \emph{{I-brane dynamics}},
  \href{http://dx.doi.org/10.1088/1126-6708/2006/01/119}{\emph{JHEP} {\bfseries
  01} (2006) 119}, [\href{https://arxiv.org/abs/hep-th/0508025}{{\ttfamily
  hep-th/0508025}}].

\bibitem{Eguchi:2004ik}
T.~Eguchi and Y.~Sugawara, \emph{{Conifold type singularities, N=2 Liouville
  and SL(2:R)/U(1) theories}},
  \href{http://dx.doi.org/10.1088/1126-6708/2005/01/027}{\emph{JHEP} {\bfseries
  01} (2005) 027}, [\href{https://arxiv.org/abs/hep-th/0411041}{{\ttfamily
  hep-th/0411041}}].

\bibitem{Eguchi:2003yy}
T.~Eguchi, Y.~Sugawara and S.~Yamaguchi, \emph{{Supercoset CFT's for string
  theories on noncompact special holonomy manifolds}},
  \href{http://dx.doi.org/10.1016/S0550-3213(03)00148-2}{\emph{Nucl. Phys.}
  {\bfseries B657} (2003) 3--52},
  [\href{https://arxiv.org/abs/hep-th/0301164}{{\ttfamily hep-th/0301164}}].

\bibitem{Giveon:1999jg}
A.~Giveon and M.~Rocek, \emph{{Supersymmetric string vacua on AdS(3) x N}},
  \href{http://dx.doi.org/10.1088/1126-6708/1999/04/019}{\emph{JHEP} {\bfseries
  04} (1999) 019}, [\href{https://arxiv.org/abs/hep-th/9904024}{{\ttfamily
  hep-th/9904024}}].

\bibitem{Giveon:2003ku}
A.~Giveon and A.~Pakman, \emph{{More on superstrings in AdS(3) x N}},
  \href{http://dx.doi.org/10.1088/1126-6708/2003/03/056}{\emph{JHEP} {\bfseries
  03} (2003) 056}, [\href{https://arxiv.org/abs/hep-th/0302217}{{\ttfamily
  hep-th/0302217}}].

\bibitem{Bars:1990rb}
I.~Bars and D.~Nemeschansky, \emph{{String Propagation in Backgrounds With
  Curved Space-time}},
  \href{http://dx.doi.org/10.1016/0550-3213(91)90223-K}{\emph{Nucl. Phys.}
  {\bfseries B348} (1991) 89--107}.

\bibitem{Balog:1990mu}
J.~Balog, L.~Feher, L.~O'Raifeartaigh, P.~Forgacs and A.~Wipf, \emph{{Toda
  Theory and $W$ Algebra From a Gauged {WZNW} Point of View}},
  \href{http://dx.doi.org/10.1016/0003-4916(90)90029-N}{\emph{Annals Phys.}
  {\bfseries 203} (1990) 76--136}.

\bibitem{Bars:1993jt}
I.~Bars, \emph{{Curved space-time geometry for strings and affine noncompact
  algebras}},  in \emph{{Quantum Aspects of Black Holes Santa Barbara,
  California, June 21-26, 1993}}, pp.~51--76, 1993,
  \href{https://arxiv.org/abs/hep-th/9309042}{{\ttfamily hep-th/9309042}}.

\bibitem{Klimcik:1994wp}
C.~Klimcik and A.~A. Tseytlin, \emph{{Exact four-dimensional string solutions
  and Toda like sigma models from 'null gauged' WZNW theories}},
  \href{http://dx.doi.org/10.1016/0550-3213(94)90089-2}{\emph{Nucl. Phys.}
  {\bfseries B424} (1994) 71--96},
  [\href{https://arxiv.org/abs/hep-th/9402120}{{\ttfamily hep-th/9402120}}].

\bibitem{Giveon:1995as}
A.~Giveon, O.~Pelc and E.~Rabinovici, \emph{{WZNW models and gauged WZNW models
  based on a family of solvable Lie algebras}},
  \href{http://dx.doi.org/10.1016/0550-3213(95)00663-X}{\emph{Nucl. Phys.}
  {\bfseries B462} (1996) 53--98},
  [\href{https://arxiv.org/abs/hep-th/9509013}{{\ttfamily hep-th/9509013}}].

\bibitem{Martinec:2001cf}
E.~J. Martinec and W.~McElgin, \emph{{String theory on AdS orbifolds}},
  \href{http://dx.doi.org/10.1088/1126-6708/2002/04/029}{\emph{JHEP} {\bfseries
  04} (2002) 029}, [\href{https://arxiv.org/abs/hep-th/0106171}{{\ttfamily
  hep-th/0106171}}].

\bibitem{Martinec:2002xq}
E.~J. Martinec and W.~McElgin, \emph{{Exciting AdS orbifolds}},
  \href{http://dx.doi.org/10.1088/1126-6708/2002/10/050}{\emph{JHEP} {\bfseries
  10} (2002) 050}, [\href{https://arxiv.org/abs/hep-th/0206175}{{\ttfamily
  hep-th/0206175}}].

\bibitem{Rohm:1983aq}
R.~Rohm, \emph{{Spontaneous Supersymmetry Breaking in Supersymmetric String
  Theories}}, \href{http://dx.doi.org/10.1016/0550-3213(84)90007-5}{\emph{Nucl.
  Phys.} {\bfseries B237} (1984) 553--572}.

\bibitem{Maldacena:2000hw}
J.~M. Maldacena and H.~Ooguri, \emph{{Strings in AdS(3) and SL(2,R) WZW model
  1.: The Spectrum}}, \href{http://dx.doi.org/10.1063/1.1377273}{\emph{J. Math.
  Phys.} {\bfseries 42} (2001) 2929--2960},
  [\href{https://arxiv.org/abs/hep-th/0001053}{{\ttfamily hep-th/0001053}}].

\bibitem{Zamolodchikov:1986bd}
A.~B. Zamolodchikov and V.~A. Fateev, \emph{{Operator Algebra and Correlation
  Functions in the Two-Dimensional Wess-Zumino SU(2) x SU(2) Chiral Model}},
  {\emph{Sov. J. Nucl. Phys.} {\bfseries 43} (1986) 657--664}.

\bibitem{Karabali:1988au}
D.~Karabali, Q.-H. Park, H.~J. Schnitzer and Z.~Yang, \emph{{A GKO Construction
  Based on a Path Integral Formulation of Gauged Wess-Zumino-Witten Actions}},
  \href{http://dx.doi.org/10.1016/0370-2693(89)91120-9}{\emph{Phys. Lett.}
  {\bfseries B216} (1989) 307--312}.

\bibitem{Karabali:1989dk}
D.~Karabali and H.~J. Schnitzer, \emph{{BRST Quantization of the Gauged WZW
  Action and Coset Conformal Field Theories}},
  \href{http://dx.doi.org/10.1016/0550-3213(90)90075-O}{\emph{Nucl. Phys.}
  {\bfseries B329} (1990) 649--666}.

\bibitem{Hwang:1993nc}
S.~Hwang and H.~Rhedin, \emph{{The BRST Formulation of G/H WZNW models}},
  \href{http://dx.doi.org/10.1016/0550-3213(93)90165-L}{\emph{Nucl. Phys.}
  {\bfseries B406} (1993) 165--186},
  [\href{https://arxiv.org/abs/hep-th/9305174}{{\ttfamily hep-th/9305174}}].

\bibitem{Goddard:1984vk}
P.~Goddard, A.~Kent and D.~I. Olive, \emph{{Virasoro Algebras and Coset Space
  Models}}, \href{http://dx.doi.org/10.1016/0370-2693(85)91145-1}{\emph{Phys.
  Lett.} {\bfseries B152} (1985) 88--92}.

\bibitem{Fateev:1985mm}
V.~A. Fateev and A.~B. Zamolodchikov, \emph{{Parafermionic Currents in the
  Two-Dimensional Conformal Quantum Field Theory and Selfdual Critical Points
  in Z(n) Invariant Statistical Systems}}, {\emph{Sov. Phys. JETP} {\bfseries
  62} (1985) 215--225}.

\bibitem{Gepner:1986hr}
D.~Gepner and Z.-a. Qiu, \emph{{Modular Invariant Partition Functions for
  Parafermionic Field Theories}},
  \href{http://dx.doi.org/10.1016/0550-3213(87)90348-8}{\emph{Nucl. Phys.}
  {\bfseries B285} (1987) 423}.

\bibitem{Gepner:1987qi}
D.~Gepner, \emph{{Space-Time Supersymmetry in Compactified String Theory and
  Superconformal Models}},
  \href{http://dx.doi.org/10.1016/0550-3213(88)90397-5}{\emph{Nucl. Phys.}
  {\bfseries B296} (1988) 757}.

\bibitem{Giveon:2015raa}
A.~Giveon, J.~Harvey, D.~Kutasov and S.~Lee, \emph{{Three-Charge Black Holes
  and Quarter BPS States in Little String Theory}},
  \href{http://dx.doi.org/10.1007/JHEP12(2015)145}{\emph{JHEP} {\bfseries 12}
  (2015) 145}, [\href{https://arxiv.org/abs/1508.04437}{{\ttfamily
  1508.04437}}].

\bibitem{Bjornsson:2007ha}
J.~Bjornsson and S.~Hwang, \emph{{On the unitarity of gauged non-compact WZNW
  strings}},
  \href{http://dx.doi.org/10.1016/j.nuclphysb.2007.11.027}{\emph{Nucl. Phys.}
  {\bfseries B797} (2008) 464--498},
  [\href{https://arxiv.org/abs/0710.1050}{{\ttfamily 0710.1050}}].

\bibitem{Bena:2011uw}
I.~Bena, J.~de~Boer, M.~Shigemori and N.~P. Warner, \emph{{Double, Double
  Supertube Bubble}},
  \href{http://dx.doi.org/10.1007/JHEP10(2011)116}{\emph{JHEP} {\bfseries 10}
  (2011) 116}, [\href{https://arxiv.org/abs/1107.2650}{{\ttfamily 1107.2650}}].

\bibitem{Larsen:1999uk}
F.~Larsen and E.~J. Martinec, \emph{{U(1) charges and moduli in the D1-D5
  system}}, {\emph{JHEP} {\bfseries 06} (1999) 019},
  [\href{https://arxiv.org/abs/hep-th/9905064}{{\ttfamily hep-th/9905064}}].

\bibitem{Bena:2006is}
I.~Bena, C.-W. Wang and N.~P. Warner, \emph{{The foaming three-charge black
  hole}}, \href{http://dx.doi.org/10.1103/PhysRevD.75.124026}{\emph{Phys. Rev.}
  {\bfseries D75} (2007) 124026},
  [\href{https://arxiv.org/abs/hep-th/0604110}{{\ttfamily hep-th/0604110}}].

\bibitem{Bena:2008nh}
I.~Bena, N.~Bobev, C.~Ruef and N.~P. Warner, \emph{{Entropy Enhancement and
  Black Hole Microstates}},
  \href{http://dx.doi.org/10.1103/PhysRevLett.105.231301}{\emph{Phys. Rev.
  Lett.} {\bfseries 105} (2010) 231301},
  [\href{https://arxiv.org/abs/0804.4487}{{\ttfamily 0804.4487}}].

\bibitem{deBoer:2009un}
J.~de~Boer, S.~El-Showk, I.~Messamah and D.~Van~den Bleeken, \emph{{A bound on
  the entropy of supergravity?}},
  \href{http://dx.doi.org/10.1007/JHEP02(2010)062}{\emph{JHEP} {\bfseries 02}
  (2010) 062}, [\href{https://arxiv.org/abs/0906.0011}{{\ttfamily 0906.0011}}].

\bibitem{Denef:2007vg}
F.~Denef and G.~W. Moore, \emph{{Split states, entropy enigmas, holes and
  halos}}, \href{http://dx.doi.org/10.1007/JHEP11(2011)129}{\emph{JHEP}
  {\bfseries 11} (2011) 129},
  [\href{https://arxiv.org/abs/hep-th/0702146}{{\ttfamily hep-th/0702146}}].

\bibitem{Bena:2012hf}
I.~Bena, M.~Berkooz, J.~de~Boer, S.~El-Showk and D.~Van~den Bleeken,
  \emph{{Scaling BPS Solutions and pure-Higgs States}},
  \href{http://dx.doi.org/10.1007/JHEP11(2012)171}{\emph{JHEP} {\bfseries 1211}
  (2012) 171}, [\href{https://arxiv.org/abs/1205.5023}{{\ttfamily 1205.5023}}].

\bibitem{Susskind:1993if}
L.~Susskind, L.~Thorlacius and J.~Uglum, \emph{{The Stretched horizon and black
  hole complementarity}},
  \href{http://dx.doi.org/10.1103/PhysRevD.48.3743}{\emph{Phys. Rev.}
  {\bfseries D48} (1993) 3743--3761},
  [\href{https://arxiv.org/abs/hep-th/9306069}{{\ttfamily hep-th/9306069}}].

\bibitem{Tyukov:2016cbz}
A.~Tyukov and N.~P. Warner, \emph{{Supersymmetry and Wrapped Branes in
  Microstate Geometries}},  \href{https://arxiv.org/abs/1608.04023}{{\ttfamily
  1608.04023}}.

\bibitem{Bena:2008dw}
I.~Bena, N.~Bobev, C.~Ruef and N.~P. Warner, \emph{{Supertubes in Bubbling
  Backgrounds: Born-Infeld Meets Supergravity}},
  \href{http://dx.doi.org/10.1088/1126-6708/2009/07/106}{\emph{JHEP} {\bfseries
  07} (2009) 106}, [\href{https://arxiv.org/abs/0812.2942}{{\ttfamily
  0812.2942}}].

\bibitem{Strominger:1995ac}
A.~Strominger, \emph{{Open p-branes}},
  \href{http://dx.doi.org/10.1016/0370-2693(96)00712-5}{\emph{Phys. Lett.}
  {\bfseries B383} (1996) 44--47},
  [\href{https://arxiv.org/abs/hep-th/9512059}{{\ttfamily hep-th/9512059}}].

\bibitem{Seiberg:1996vs}
N.~Seiberg and E.~Witten, \emph{{Comments on string dynamics in
  six-dimensions}},
  \href{http://dx.doi.org/10.1016/0550-3213(96)00189-7}{\emph{Nucl. Phys.}
  {\bfseries B471} (1996) 121--134},
  [\href{https://arxiv.org/abs/hep-th/9603003}{{\ttfamily hep-th/9603003}}].

\bibitem{Witten:1996qb}
E.~Witten, \emph{{Phase transitions in M theory and F theory}},
  \href{http://dx.doi.org/10.1016/0550-3213(96)00212-X}{\emph{Nucl. Phys.}
  {\bfseries B471} (1996) 195--216},
  [\href{https://arxiv.org/abs/hep-th/9603150}{{\ttfamily hep-th/9603150}}].

\bibitem{Horowitz:1996nw}
G.~T. Horowitz and J.~Polchinski, \emph{{A correspondence principle for black
  holes and strings}},
  \href{http://dx.doi.org/10.1103/PhysRevD.55.6189}{\emph{Phys. Rev.}
  {\bfseries D55} (1997) 6189--6197},
  [\href{https://arxiv.org/abs/hep-th/9612146}{{\ttfamily hep-th/9612146}}].

\bibitem{Giveon:2005mi}
A.~Giveon, D.~Kutasov, E.~Rabinovici and A.~Sever, \emph{{Phases of quantum
  gravity in AdS(3) and linear dilaton backgrounds}},
  \href{http://dx.doi.org/10.1016/j.nuclphysb.2005.04.015}{\emph{Nucl. Phys.}
  {\bfseries B719} (2005) 3--34},
  [\href{https://arxiv.org/abs/hep-th/0503121}{{\ttfamily hep-th/0503121}}].

\bibitem{Giveon:2017nie}
A.~Giveon, N.~Itzhaki and D.~Kutasov, \emph{{$T\bar T$ and LST}},
  \href{https://arxiv.org/abs/1701.05576}{{\ttfamily 1701.05576}}.

\bibitem{Witten:1991yr}
E.~Witten, \emph{{On string theory and black holes}},
  \href{http://dx.doi.org/10.1103/PhysRevD.44.314}{\emph{Phys. Rev.} {\bfseries
  D44} (1991) 314--324}.

\bibitem{Kiritsis:1994np}
E.~Kiritsis and C.~Kounnas, \emph{{Dynamical topology change in string
  theory}}, \href{http://dx.doi.org/10.1016/0370-2693(94)90942-3}{\emph{Phys.
  Lett.} {\bfseries B331} (1994) 51--62},
  [\href{https://arxiv.org/abs/hep-th/9404092}{{\ttfamily hep-th/9404092}}].

\bibitem{Elitzur:2002rt}
S.~Elitzur, A.~Giveon, D.~Kutasov and E.~Rabinovici, \emph{{From big bang to
  big crunch and beyond}},
  \href{http://dx.doi.org/10.1088/1126-6708/2002/06/017}{\emph{JHEP} {\bfseries
  06} (2002) 017}, [\href{https://arxiv.org/abs/hep-th/0204189}{{\ttfamily
  hep-th/0204189}}].

\bibitem{Horowitz:1994ei}
G.~T. Horowitz and A.~A. Tseytlin, \emph{{On exact solutions and singularities
  in string theory}},
  \href{http://dx.doi.org/10.1103/PhysRevD.50.5204}{\emph{Phys. Rev.}
  {\bfseries D50} (1994) 5204--5224},
  [\href{https://arxiv.org/abs/hep-th/9406067}{{\ttfamily hep-th/9406067}}].

\bibitem{Tseytlin:1995fh}
A.~A. Tseytlin, \emph{{Exact solutions of closed string theory}},
  \href{http://dx.doi.org/10.1088/0264-9381/12/10/003}{\emph{Class. Quant.
  Grav.} {\bfseries 12} (1995) 2365--2410},
  [\href{https://arxiv.org/abs/hep-th/9505052}{{\ttfamily hep-th/9505052}}].

\bibitem{Horowitz:1994rf}
G.~T. Horowitz and A.~A. Tseytlin, \emph{{A New class of exact solutions in
  string theory}},
  \href{http://dx.doi.org/10.1103/PhysRevD.51.2896}{\emph{Phys. Rev.}
  {\bfseries D51} (1995) 2896--2917},
  [\href{https://arxiv.org/abs/hep-th/9409021}{{\ttfamily hep-th/9409021}}].

\bibitem{Hull:1989jk}
C.~M. Hull and B.~J. Spence, \emph{{The Gauged Nonlinear $\sigma$ Model With
  {Wess-Zumino} Term}},
  \href{http://dx.doi.org/10.1016/0370-2693(89)91688-2}{\emph{Phys. Lett.}
  {\bfseries B232} (1989) 204--210}.

\bibitem{Hull:1991uw}
C.~M. Hull, G.~Papadopoulos and B.~J. Spence, \emph{{Gauge symmetries for (p,q)
  supersymmetric sigma models}},
  \href{http://dx.doi.org/10.1016/0550-3213(91)80035-K}{\emph{Nucl. Phys.}
  {\bfseries B363} (1991) 593--621}.

\bibitem{Hull:1993ct}
C.~M. Hull, G.~Papadopoulos and P.~K. Townsend, \emph{{Potentials for (p,0) and
  (1,1) supersymmetric sigma models with torsion}},
  \href{http://dx.doi.org/10.1016/0370-2693(93)90327-E}{\emph{Phys. Lett.}
  {\bfseries B316} (1993) 291--297},
  [\href{https://arxiv.org/abs/hep-th/9307013}{{\ttfamily hep-th/9307013}}].

\bibitem{Kutasov:1996fp}
D.~Kutasov and E.~J. Martinec, \emph{{New principles for string / membrane
  unification}},
  \href{http://dx.doi.org/10.1016/0550-3213(96)00302-1}{\emph{Nucl. Phys.}
  {\bfseries B477} (1996) 652--674},
  [\href{https://arxiv.org/abs/hep-th/9602049}{{\ttfamily hep-th/9602049}}].

\bibitem{Kutasov:1996zm}
D.~Kutasov, E.~J. Martinec and M.~O'Loughlin, \emph{{Vacua of M theory and N=2
  strings}}, \href{http://dx.doi.org/10.1016/0550-3213(96)00303-3}{\emph{Nucl.
  Phys.} {\bfseries B477} (1996) 675--700},
  [\href{https://arxiv.org/abs/hep-th/9603116}{{\ttfamily hep-th/9603116}}].

\bibitem{Vafa:1996xn}
C.~Vafa, \emph{{Evidence for F theory}},
  \href{http://dx.doi.org/10.1016/0550-3213(96)00172-1}{\emph{Nucl. Phys.}
  {\bfseries B469} (1996) 403--418},
  [\href{https://arxiv.org/abs/hep-th/9602022}{{\ttfamily hep-th/9602022}}].

\bibitem{Tseytlin:1996it}
A.~A. Tseytlin, \emph{{Selfduality of Born-Infeld action and Dirichlet
  three-brane of type IIB superstring theory}},
  \href{http://dx.doi.org/10.1016/0550-3213(96)00173-3}{\emph{Nucl. Phys.}
  {\bfseries B469} (1996) 51--67},
  [\href{https://arxiv.org/abs/hep-th/9602064}{{\ttfamily hep-th/9602064}}].

\bibitem{Bars:1996dz}
I.~Bars, \emph{{Supersymmetry, p-brane duality and hidden space-time
  dimensions}}, \href{http://dx.doi.org/10.1103/PhysRevD.54.5203}{\emph{Phys.
  Rev.} {\bfseries D54} (1996) 5203--5210},
  [\href{https://arxiv.org/abs/hep-th/9604139}{{\ttfamily hep-th/9604139}}].

\bibitem{Hewson:1996yh}
S.~Hewson and M.~Perry, \emph{{The Twelve-dimensional super (2+2)-brane}},
  \href{http://dx.doi.org/10.1016/S0550-3213(97)00120-X,
  10.1016/S0550-3213(97)80035-1}{\emph{Nucl. Phys.} {\bfseries B492} (1997)
  249--277}, [\href{https://arxiv.org/abs/hep-th/9612008}{{\ttfamily
  hep-th/9612008}}].

\bibitem{Tseytlin:1996ne}
A.~A. Tseytlin, \emph{{Type IIB instanton as a wave in twelve-dimensions}},
  \href{http://dx.doi.org/10.1103/PhysRevLett.78.1864}{\emph{Phys. Rev. Lett.}
  {\bfseries 78} (1997) 1864--1867},
  [\href{https://arxiv.org/abs/hep-th/9612164}{{\ttfamily hep-th/9612164}}].

\bibitem{Cvetic:1996xz}
M.~Cvetic and D.~Youm, \emph{{General Rotating Five Dimensional Black Holes of
  Toroidally Compactified Heterotic String}},
  \href{http://dx.doi.org/10.1016/0550-3213(96)00355-0}{\emph{Nucl. Phys.}
  {\bfseries B476} (1996) 118--132},
  [\href{https://arxiv.org/abs/hep-th/9603100}{{\ttfamily hep-th/9603100}}].

\bibitem{Dixon:1989cg}
L.~J. Dixon, M.~E. Peskin and J.~D. Lykken, \emph{{N=2 Superconformal Symmetry
  and SO(2,1) Current Algebra}},
  \href{http://dx.doi.org/10.1016/0550-3213(89)90459-8}{\emph{Nucl. Phys.}
  {\bfseries B325} (1989) 329--355}.

\bibitem{Griffin:1990fg}
P.~A. Griffin and O.~F. Hernandez, \emph{{Feigin-Fuchs derivation of SU(1,1)
  parafermion characters}},
  \href{http://dx.doi.org/10.1016/0550-3213(91)90150-V}{\emph{Nucl. Phys.}
  {\bfseries B356} (1991) 287--298}.

\bibitem{Dijkgraaf:1991ba}
R.~Dijkgraaf, H.~L. Verlinde and E.~P. Verlinde, \emph{{String propagation in a
  black hole geometry}},
  \href{http://dx.doi.org/10.1016/0550-3213(92)90237-6}{\emph{Nucl. Phys.}
  {\bfseries B371} (1992) 269--314}.

\end{thebibliography}\endgroup

%\end{adjustwidth}

\end{document}